\documentclass[12pt]{article}

\advance\voffset by -2.0cm \advance\hoffset by -1.25cm
\usepackage[colorlinks,linkcolor=.]{hyperref}

\usepackage{cite}
\usepackage{amsmath}
\usepackage{amssymb}
\usepackage{amsthm}

\usepackage{lscape}
\usepackage{booktabs}
\usepackage{tabularx}
\usepackage{longtable}
\usepackage{ltablex}
\usepackage{caption}

\usepackage{authblk}

\usepackage{multirow}

\usepackage[table]{xcolor}

\newcommand{\CC}{\mathbb{C}} 
\newcommand{\RR}{\mathbb{R}} 
\newcommand{\QQ}{\mathbb{Q}} 
 
\newcommand{\HH}{\mathbb{H}}
\newcommand{\ZZ}{\mathbb{Z}} 
 
\newcommand{\PP}{\mathbb{P}}

\newcommand{\half}{{\frac{1}{2}}}

\newcommand{\A}{\mathcal{A}}

\newcommand{\C}{\mathcal{C}}
\newcommand{\E}{\mathcal{E}}
\newcommand{\F}{\mathcal{F}}

\newcommand{\Hh}{\mathcal{H}}

\newcommand{\M}{\mathcal{M}}
\newcommand{\N}{\mathcal{N}}

\newcommand{\calS}{\mathcal{S}}
\newcommand{\T}{\mathcal{T}}

\newcommand{\Z}{\mathcal{Z}}

\newcommand{\cc}{\mathsf{c}}

\newcommand{\modu}[1]{\,(\mbox{mod }#1)}

\usepackage{adjustbox}

\newtheorem{theorem}{Theorem}

\newtheorem{corollary}[theorem]{Corollary}
\newtheorem{claim}[theorem]{Claim}

\DeclareMathOperator{\Sym}{Sym}

\DeclareMathOperator{\Tr}{Tr}
\DeclareMathOperator{\Real}{Re}

\newcommand{\avg}[1]{\left\langle#1\right\rangle}

\newcommand{\floor}[1]{\left\lfloor#1\right\rfloor}

\newcommand{\parens}[1]{\!\left(#1\right)}
\newcommand{\braces}[1]{\!\left\{#1\right\}}
\newcommand{\brackets}[1]{\!\left[#1\right]}
\DeclareRobustCommand{\beginProtected}[1]{\begin{#1}}
\DeclareRobustCommand{\endProtected}[1]{\end{#1}}
\newcommand{\piecewise}[4]{\left\{\beginProtected{array}{rl}#1&:#2\\#3&:#4\endProtected{array}\right.}
\newcommand{\column}[2]{
\begin{pmatrix}
#1 \\
#2
\end{pmatrix}}

\newcommand{\smcolumn}[2]{
\left(\begin{smallmatrix}
#1 \\
#2
\end{smallmatrix}\right)}

\newcommand{\twoMatrix}[4]{
\begin{pmatrix}
#1 & #2 \\
#3 & #4
\end{pmatrix}}

\newcommand{\smtwoMatrix}[4]{
\left(\begin{smallmatrix}
#1 & #2 \\
#3 & #4
\end{smallmatrix}\right)}

\newcommand{\gmo}[1]{\Gamma_{\langle-1\rangle}(#1)}

\numberwithin{equation}{section}

\def\be{\begin{equation}}
\def\ee{\end{equation}}

\allowdisplaybreaks[3]
\title{
No More Walls!\\ {\large A Tale of Modularity, Symmetry, and Wall Crossing for $1/4$ BPS Dyons}}
\author[1]{Natalie M. Paquette}
\author[2]{Roberto Volpato} 
\author[1]{Max Zimet}
\affil[1]{Stanford Institute for Theoretical Physics, Stanford University, Stanford, CA 94305 USA}
\affil[2]{Dipartimento di Fisica e Astronomia `Galileo Galilei', Universit\`a di Padova \& INFN sez. di Padova, Via Marzolo 8, 35131 Padova, Italy}
\date{}
\newcommand{\finishRow}[1]{\multicolumn{\the\numexpr 14-#1\relax}{C}{}}

\begin{document}
\maketitle

\begin{abstract}
We determine the generating functions of $1/4$ BPS dyons in a class of 4d $\mathcal{N}=4$ string vacua arising as CHL orbifolds of $K3 \times T^2$, a classification of which has been recently completed. We show that all such generating functions obey some simple physical consistency conditions that are very often sufficient to fix them uniquely. The main constraint we impose is the absence of unphysical walls of marginal stability: discontinuities of 1/4 BPS degeneracies can only occur when 1/4 BPS dyons decay into pairs of 1/2 BPS states.  Formally, these generating functions in  spacetime can be described as multiplicative lifts of certain supersymmetric indices (twining genera) on the worldsheet of the corresponding nonlinear sigma model on K3. As a consequence, our procedure also leads to an explicit derivation of almost all of these twining genera. The worldsheet indices singled out in this way match precisely a set of functions of interest in moonshine, as predicted by a recent conjecture.
\end{abstract}

\tableofcontents

\hypersetup{linkcolor=blue}
\section{Introduction}

Half-maximal ($\N=4$) supersymmetric string models in four dimensions constitute a rich class of theories where many exact results can be derived. The prototypical example of such theories is given by type IIA on K3$\times T^2$ or, dually, by heterotic on $T^6$. 
In this model, the geometry of the moduli space, as well as various terms in the low energy effective action, are known exactly, including all perturbative and non-perturbative corrections.  Even more remarkably, this is the framework of the first successful attempts of matching the Bekenstein-Hawking-Wald black hole entropy formula  with a precise counting of the corresponding microstates in string theory \cite{StromingerVafa}. More precisely, the generating functions for the multiplicities of $1/2$- and $1/4$-BPS states have been determined \cite{DVV}.

Compactifications of string theory on K3 are also the arena of many interesting open conjectures in mathematical physics. It was noticed long ago that the generating function for the multiplicities of 1/4 BPS dyons matches exactly the square of the denominator of a generalized (Borcherds) Kac-Moody algebra \cite{DVV}. This observation was one of the motivations behind the attempts to construct an algebra of BPS states in these models \cite{Harvey:1995fq,Harvey:1996gc}. Despite the efforts, however, the relationship between Borcherds-Kac-Moody algebras and $\N=4$ models in four dimensions has not been explained. 

Finally, string theories on K3 seem to be the most promising framework where the  mysterious Mathieu \cite{Eguchi:2010ej} and Umbral \cite{Cheng:2012tq,Cheng:2013wca} moonshine phenomena could be understood. In particular, there are various proposals relating symmetries of these string theories to the Mathieu and the other Umbral groups appearing in the moonshine conjectures.

A great deal of information in these string theory models is encoded in the geometry of the compact K3 surface or, more generally, in the nonlinear sigma model (NLSM) describing the type IIA string worldsheet in the perturbative limit. In recent years, some considerable progress has been made in the study of the finite symmetry groups of these models and  their action on the BPS states. This is interesting for a variety of reasons. First of all, to each such symmetry $g$ that commutes with the spacetime supersymmetries, we can associate a 4d $\N=4$ string compactification, called a CHL model \cite{Chaudhuri:1995fk,Chaudhuri:1995dj,Sen:1995ff,Chaudhuri:1995bf}, by taking an appropriate orbifold of the K3$\times T^2$ compactification described above. These generalize the unorbifolded case (which is the CHL model associated to the identity element of the symmetry group), with which they share many features. In particular, arguments similar to those alluded to above allow one to compute the $1/2$-BPS state counting functions. They also allow one to relate the $1/4$-BPS counting functions to refined supersymmetric indices, called twining genera, of the K3 NLSM that take into account the action of $g$ on states in a short representation of the worldsheet superconformal algebra \cite{DVV,DMVV,Jatkar:2005bh,David:2006ji,DS, DJS1, DJS2, dabholkar2011counting, Dabholkar:2006xa, Dabholkar:2007vk, gaiotto2005re, strominger2006recounting, shih2006exact}. Returning to string theory on K3$\times T^2$, the twining genus is sufficient to determine the action of the corresponding symmetry on the sector of 1/4 BPS states.   Finally, an explicit knowledge of all twining genera is expected to provide strong evidence for (or to disprove) the conjectural relations between K3 sigma models and Umbral moonshine \cite{Cheng:2014zpa,slowpaper}. Unfortunately, fundamental difficulties have, to date, prevented the determination of many twining genera, as we now describe.

There exists a serious obstacle in the study of K3 NLSMs and their symmetries: very few of these models are known explicitly. Indeed, no explicit metric for a K3 surface has ever been determined, so all K3 NLSMs that have been studied to date have been specified by data other than a K3 metric and B-field, and then shown to be equivalent to a K3 NLSM. The spectrum (or, equivalently, the partition function) is known  only for some torus orbifolds or Gepner models. On the other hand, we have a detailed understanding of the $80$-dimensional moduli space and its duality group \cite{Aspinwall:1996mn,Aspinwall:1994rg,Nahm:1999ps}. In \cite{Gaberdiel:2011fg}, these general properties of K3 sigma models were used to classify all finite groups of symmetries commuting with the $\N=(4,4)$ algebra at all points in the moduli space. Unfortunately, this classification only provides a description of the symmetries as abstract groups, but not their action on the states of the theory. A precise description of the symmetry action on the full spectrum of the NLSM  at all points in the moduli space seems completely out of reach, since for most of these models we don't even know the partition function! Nevertheless, as we demonstrate in this paper, by computing all K3 NLSM twining genera, it is possible to understand the action on a subsector of states.

An important step toward the completion of this program was made in \cite{slowpaper}, where it was proved that distinct twining genera -- and, therefore, distinct CHL $1/4$-BPS state counting functions -- correspond to different conjugacy classes in the NLSM duality group. As a consequence, there are only a finite number -- at most $81$ -- of such genera.\footnote{We note two minor caveats in this result. First, there was one case that could not be completely determined: there were either one or two conjugacy classes. So, the number of conjugacy classes may actually be 82. If there are indeed two classes, they have identical twining genera, so the results of \cite{slowpaper} do, indeed, serve to classify all possible twining genera. Second, a number of distinct classes have coincident twining genera, so the total number of twining genera is actually fewer than 81.}  In the case where $g$ is the identity, the twining genus is the elliptic genus, which has been known for decades\cite{EOTY}. However, the twining genera associated to many other symmetries have yet to be determined. This might seem surprising, since -- like the elliptic genus -- twining genera do not depend on the moduli of the K3 NLSM, as long as we remain at a point in moduli space where $g$ is a symmetry. However, for about half of the 81 cases mentioned above, explicit descriptions of NLSMs at these points in moduli space have not been found.

In this paper, we propose a general approach to compute \emph{all} twining genera just using general properties of K3 string compactifications, and without the need of any explicit description of the NLSM. More precisely, for each point in the moduli space of K3 models and for each symmetry $g$ of the corresponding sigma model, we either determine the corresponding twining genus precisely or, in the worst case, we limit the possible genus to only two explicit possibilities. 

Our strategy is to start from the results of \cite{slowpaper} and, for each of the relevant classes of symmetries, to derive constraints on the twining genus both from the general properties of conformal field theory and from the full string theory on K3. General CFT arguments imply that the twining genus has suitable modular transformations (it is a weak Jacobi form of weight $0$ and index $1$) under certain subgroups of $SL(2,\ZZ)$, which were determined in \cite{slowpaper}. The space of such Jacobi forms is always finite dimensional, so that the precise form of a twining genus can in principle be determined once a sufficient number of Fourier coefficients is known. The lowest of these Fourier coefficients gets contributions only from the Ramond-Ramond ground states of the theory, and  the action of the symmetry $g$ on such states is known. Thus, the first Fourier coefficient can always be computed, and this is sufficient to determine the twining genus in a few cases. This technique was already exploited in \cite{slowpaper}. A straightforward generalization of this technique is the following. One considers the functions (the twisted-twining genera) obtained by taking generic $SL(2,\ZZ)$ transformations of a twining genus. The physical interpretation of these functions is as $g^i$-twined traces over the $g^j$-twisted sector of the theory, for some $i,j\in \ZZ$. Formally, the Fourier expansions of these $SL(2,\ZZ)$ transforms correspond to expansions of the original twining genus at different `cusps' (i.e., points at the boundary) of the upper half-plane. As for the standard twining genus, one can determine the action of $g^i$ on the $g^j$-twisted ground states and therefore compute the lowest Fourier coefficient of the corresponding $g^j$-twisted $g^i$-twining genus. This information is sufficient to determine the twining genus whenever the modular group $\hat\Gamma_g$ has genus zero, i.e. the quotient $\hat\HH/\hat\Gamma_g$ of the upper half-plane by $\hat\Gamma_g$ is a sphere.

These conformal field theory techniques fail whenever the modular group $\hat\Gamma_g$ admits a cusp form, i.e. when the space of Jacobi forms contains an element whose first Fourier coefficient vanishes at all cusps. In these cases, in order to compute the twining genus, one should know the action of $g$ on some states of higher conformal weight, but this cannot be determined without an explicit description of the NLSM. Fortunately, further constraints come from considering the full string theory rather than just the sigma model. As mentioned above, the twining genus for a symmetry $g$ is related to the generating function for 1/4 BPS dyons in the CHL orbifold  corresponding to $g$. The latter enjoys a phenomenon known as wall crossing. As one moves around the Siegel upper half space parametrizing the arguments of this function, the function's Fourier expansion (whose coefficients are the $1/4$ BPS degeneracies) is typically unchanging. However, as one crosses certain real codimension $1$ submanifolds (walls) in moduli space, one must employ a different Fourier expansion: the previous expansion diverges due to poles of the function located along the walls \cite{DVV,Cheng:2007ch,CV,sen2007walls}. The connection between moduli and arguments of the function is provided by a contour prescription \cite{Cheng:2007ch} (see section \ref{sec:chl} for details), which describes how to extract dyon degeneracies from contour integration. The degeneracies will jump as the contour crosses poles of its integrand and picks up a contribution from the corresponding residue. Physically, this means that while $1/4$ BPS degeneracies are locally constant as a function of moduli, they jump discretely at walls. These jumps occur because some of the 1/4 BPS states are given by bound states of pairs of 1/2 BPS states, which can become unstable and disappear from the spectrum as one varies the moduli. This physical interpretation of wall crossing allows us to precisely locate the walls, as they are the boundaries of the regions of stability for these bound states. This yields a very strong constraint on twining genera, since general choices of coefficients in a twining genus would yield unphysical poles in the corresponding CHL model's $1/4$-BPS counting function.

Before proceeding with our analysis, we highlight the study of these twining genera in the context of moonshines associated to K3 surfaces. Independently, twining genera associated to various K3 SCFT symmetries $g$ have been proposed in the context of the Mathieu \cite{Eguchi:2010ej}, Umbral \cite{Cheng:2012tq, Cheng:2013wca}, and Conway \cite{duncan2007super, Duncan:2014eha} moonshines. One way to view the classification of $\mathcal{N}=(4, 4)$-preserving symmetries of K3 SCFTs is as certain four-plane preserving subgroups of $Co_0$, the finite `Conway-0'  group which governs the automorphisms of the famous Leech lattice. Putative twining genera associated to the corresponding $Co_0$ conjugacy classes (which by abuse of notation we also label by some representative $g$) have been derived from a certain $c=12$ CFT which enjoys a global $Co_0$ symmetry \cite{Duncan:2015xoa}. We refer to these twining genera as arising from `Conway moonshine'. For the former two moonshines (of which Mathieu may be viewed as a special case of Umbral), twining genera associated to appropriate 4-plane preserving subgroups of $Co_0$ that are moreover subgroups of the Umbral groups have also been proposed in \cite{Cheng:2014zpa} (with  twining genera in the $M_{24}$ case first computed in \cite{Cheng:2010pq, Gaberdiel:2010ch, Gaberdiel:2010ca,Eguchi:2010fg}). These Umbral genera coincide, in most  cases, with those of \cite{Duncan:2015xoa} for compatible 4-plane preserving conjugacy classes; see \cite{slowpaper} for details. Remarkably, our study singles out the proposed twining genera associated to the Mathieu, Umbral, and Conway moonshines from an infinite family of candidate Jacobi forms. This result perfectly agrees with a recent conjecture in \cite{slowpaper}. 

The structure of the paper is as follows. We begin with sections reviewing CHL compactifications and their BPS state counts, and K3 NLSMs and their symmetries, emphasizing the relations between these subjects. In particular, the former introduces $1/4$-BPS counting functions, while the `twisted-twining' genera of K3 NLSMs are described in the latter section. We then explain, in section \ref{sec:twining}, how modular properties of the twining genera allow us to strongly constrain (and even completely determine, in many cases) these functions by studying the actions of symmetries on Ramond-Ramond ground states in various twisted sectors. Section \ref{sec:secquant} introduces the other constraints we will need, which arise from considerations in the full CHL string compactification. We combine these constraints in section \ref{sec:determining} and explain how we determine all twining genera. We conclude with a discussion of our results and ideas for future work.

\section{BPS state counts in 4d $\N=4$ theories}\label{sec:chl}


In this section we review some salient facts about CHL orbifolds, which are $\mathcal{N}=4$ compactifications of the heterotic string to four dimensions on $T^6/\mathbb{Z}_N$, or equivalently type II on $(K3 \times T^2)/\mathbb{Z}_N$. We largely follow the discussions in \cite{sen2007walls, Cheng:2008kt, PV}; see \cite{DS, DJS1, DJS2, dabholkar2011counting, Banerjee:2008yu, Dabholkar:2006bj, Dabholkar:2006xa, Dabholkar:2007vk, gaiotto2005re, strominger2006recounting, shih2006exact, Govindarajan:2008vi, Govindarajan:2009qt, Govindarajan:2010fu, Govindarajan:2011mp, David:2006ji, banerjee2008partition, Banerjee:2008pv,Bossard:2017wum} for further results on CHL orbifolds. Our primary reason for interest in these CHL models is that they provide a class of string compactifications in which the spectrum of $1/4$-BPS dyons can be computed exactly. In particular, the moduli dependence of this spectrum, which arises from decays of $1/4$-BPS states into two $1/2$-BPS states at so-called walls of marginal stability, is well understood.

\subsection{Construction of CHL models}\label{sec:chlBasics}


The prototypical example of four dimensional string theory with half-maximal supersymmetry ($16$ supercharges) is given by heterotic strings compactified on $T^6$. The resulting four dimensional $\N=4$ theory has a gauge group $U(1)^{28}$ (at generic points in the moduli space), corresponding to $22$ vector multiplets in addition to the six graviphotons. Now and henceforth, we assume that we are not at a point of enhanced gauge symmetry.  The moduli are given by the axio-dilaton $S $, the metric and $B$-field along $T^6$, and Wilson lines for the 16 gauge fields of the 10-dimensional heterotic theory.  They parametrize the usual heterotic moduli space,
\be\label{modspace} O(\Gamma^{6,22})\backslash O(6,22)/(O(6)\times O(22))\times (SL(2,\ZZ)\backslash SL(2,\RR)/U(1))\ ,\ee
the product of a Narain moduli space and the axio-dilaton moduli space. The discrete groups acting on the left -- $O(\Gamma^{6, 22})$ and $SL(2, \mathbb{Z})$ -- are the T-duality and S-duality groups, respectively. Here, $\Gamma^{6,22}$ denotes the usual Narain lattice of winding-momentum for  heterotic strings on $T^6$; it is the unique (up to isomorphism) even unimodular lattice  of signature $(6,22)$.
This model admits dual type IIA and type IIB descriptions. More precisely, upon choosing a splitting $T^6=T^4\times S^1 \times \hat S^1$, this heterotic compactification is related via string-string duality to type IIA on K3 $ \times S^1 \times \hat S^1$, and via T-duality along one of the circles (say, $\hat S^1$) to type IIB on K3 $ \times S^1\times \tilde S^1$.
The heterotic axio-dilaton $S$ has dual descriptions as either the complex structure modulus of the $S^1\times \tilde S^1$ torus in type IIB or the complexified K\"ahler modulus of the $S^1\times \hat S^1$ torus in type IIA. In the type IIA picture, the Narain lattice $\Gamma^{6,22}$ can be interpreted as the direct sum 
\be\label{latticesplit}\Gamma^{6,22}=\Gamma^{4,20}\oplus\Gamma^{2,2}\ee 
of the lattice $H^*(K3,\ZZ)\cong \Gamma^{4,20}$ of integral cohomology for K3 and the lattice $\Gamma^{2,2}$ of winding-momentum along $S^1\times \hat S^1$.

Starting from this compactification, one can obtain a whole class of four dimensional $\N=4$ models (CHL models, \cite{Chaudhuri:1995fk,Chaudhuri:1995dj,Sen:1995ff,Chaudhuri:1995bf}) by taking the orbifold by a cyclic symmetry group $\ZZ_N$  commuting with the $\N=4$ supersymmetry.  In the type IIA frame, the generator $\hat g$ of $\ZZ_N$ acts on the nonlinear sigma model (NLSM) on K3 as an order $N$ symmetry $g$, and on the $T^2=S^1\times\hat S^1$ via a shift $\delta$ around $S^1$ by $1/N$ times its circumference. We write $\hat g=(\delta,g)$. The simplest and most studied examples to keep in mind are the ones where $g$ is a geometric symmetry of the K3 target space that preserves the holomorphic 2-form  (in order to preserve the $SU(2)$ holonomy that yields $\N=4$ supersymmetry), a.k.a. a symplectic automorphism.  
The orbifold procedure projects some massless fields out of the spectrum. Furthermore, thanks to the shift $\delta$, the fields in the twisted sector are necessarily massive and therefore the low energy spectrum is different from the unorbifolded case.  
 Note that $\hat g$ does not act on $\hat S^1$, so T-dualizing this circle to translate between IIA and IIB is simple.

We can easily get a larger class of models if we allow $g$ to be a symmetry of the $\N=(4,4)$ K3 sigma model (which we describe in section \ref{sec:K3}) but not of the geometric K3 surface itself\cite{PV}. As we will explain in section \ref{sec:K3}, we are interested in symmetries $g$ of the sigma model that fix the spectral flow generators and worldsheet superconformal algebras, since these are the conditions for $\hat g$ to commute with all spacetime supercharges.  Each such symmetry $g$ corresponds to a duality in the subgroup\footnote{The notation $O^+(4,20)$ denotes the subgroup of $O(4,20)$ whose maximal compact subgroup is $SO(4)\times O(20)$. The group $O^+(\Gamma^{4,20})$ is the group of automorphisms of the lattice $\Gamma^{4,20}$ that are contained in $O^+(4,20)$. } $O^+(\Gamma^{4,20})\subset O(\Gamma^{6,22})$, where $\Gamma^{4,20}$ is the K3 lattice in the splitting \eqref{latticesplit}. In particular, $g$ acts trivially on the torus $S^1\times \hat S^1$ and on the heterotic axio-dilaton. The condition that supersymmetry generators are preserved restricts us to the elements of $O^+(\Gamma^{4,20})$ that fix a positive definite 4-plane in $\Gamma^{4,20}\otimes \RR$.
Considering non-geometric symmetries introduces a new complication: if $g$ acts asymmetrically on left-movers and right-movers, then the orbifold of the sigma model by $g$ may be inconsistent, due to a failure of level matching in a twisted sector that destroys modular invariance. As will be explained in section \ref{sec:levelfail} and appendix \ref{a:mess}, even for such $g$'s the corresponding CHL model can be consistently defined by requiring the order of the shift $\delta$ to be a suitable multiple  $\hat N=N\lambda$ of the order $N$ of $g$  \cite{PV}. For simplicity, in this section we will mostly focus on the case $\lambda=1$, corresponding to the case where the level matching condition for $g$ is satisfied.

Inequivalent CHL models correspond to different $O(\Gamma^{6,22})$ conjugacy classes of pairs $(\delta,g)$. In \cite{PV} it was shown that such classes are labeled by the eigenvalues of $g\in O^+(\Gamma^{4,20})$ in the defining 24-dimensional representation. This set of eigenvalues is conveniently encoded into the (generalized) Frame shape of $g$, i.e. a formal product 
\be
\pi_g=\prod_{a|N} a^{m(a)}
\ee where $N$ is the order of $g$ and $m(a)$ are integers such that the characteristic polynomial of $g$ is\footnote{In words, this definition means than when $m(a)>0$ we add the $a$-th roots of unity $m(a)$ times to the list of eigenvalues of $g$, and when $m(a)<0$ we subtract the $a$-th roots of unity $|m(a)|$ times from the list of eigenvalues. For example, the eigenvalues corresponding to the Frame shape $1^{-8}2^{16}$ are eight $1$'s and sixteen $-1$'s.}
\be \det(t-g)=\prod_{a|N} (t^a-1)^{m(a)}\ .
\ee The Frame shape always exists because $g$ acts rationally in the $24$-dimensional representation. Furthermore, when $g$ acts by permutations, $\pi_g$ is simply the cycle shape. There are $42$ possible Frame shapes corresponding to symmetries of K3 sigma models \cite{Gaberdiel:2011fg,Gaberdiel:2012um}. Different Frame shapes obviously correspond to different $O^+(\Gamma^{4,20})$ classes, but the converse is not always true --  indeed, there are $81$ different $O^+(\Gamma^{4,20})$ classes corresponding to these Frame shapes\footnote{In the subsequent section (and see especially \cite{slowpaper}), we explain why the relevant duality group for our purposes is $O^+(\Gamma^{4, 20})$ rather than $O(\Gamma^{4, 20})$.}\cite{slowpaper}.

The moduli of the CHL model corresponding to the symmetry $\hat g=(g,\delta)$ are simply given by the $g$-invariant moduli of the parent theory -- all moduli are invariant under $\delta$.  As a result, the moduli space is given by a quotient of
\begin{equation}
\frac{O(6, r-6)}{O(6) \times O(r-6)} \times \frac{SL(2,\RR)}{U(1)}\label{eq:CHLModuli}
\end{equation}
by a discrete U-duality group. Here $r$,  with $8\le r\le 28$, is the number of gauge fields that survive the orbifold projection -- that is, the rank of the CHL model's gauge group. The rank of the gauge group corresponds to dimension of the $g$-fixed subspace in $\Gamma^{6,22}\otimes \RR$. This subspace has signature $(6,d+2)$ because by construction $g$ fixes the sublattice $\Gamma^{2,2}$ in \eqref{latticesplit} and a four-dimensional subspace in $\Gamma^{4,20}\otimes \RR$. 

As is typical in toroidal compactifications of heterotic string theory, one can conveniently encode the moduli parametrizing the first factor in \eqref{eq:CHLModuli} in an $r \times r$ matrix $M$ satisfying
\begin{equation}
MLM^T = M, \ M^T = M \ ,
\end{equation} 
where $L$ is an $O(6, r-6)$-invariant matrix with $6$ (+1)-eigenvalues and $(r-6)$ (-1)-eigenvalues; we define an inner product on $\RR^{6,r-6}$ using $L$: $v\cdot w=v^TLw$.
It is sometimes convenient to express $M$ in terms of a $(6\times r)$-dimensional vielbein $\mu$ as $M = \mu^T \mu$. We will also be interested in the $r$-dimensional vectors of electric charge, $Q$, and magnetic charge, $P$, that in particular characterize our dyonic states. We will combine these into a vector that lives in the lattice of electric-magnetic charges:
\begin{equation}
\begin{pmatrix} Q \\ P \end{pmatrix} \in \Lambda_{e}\oplus \Lambda_{m}.
\end{equation}
In the unorbifolded theory, the even unimodular lattice $\Gamma^{6, 22}$ is isomorphic to both the electric and magnetic charge lattices, $\Gamma^{6, 22} \simeq \Lambda_e \simeq \Lambda_m$. For a CHL model based on a symmetry $(\delta,g)$ of order $N$ (where $g$  satisfies the level matching condition), the lattice of electric charges is
\be \Lambda_e=\Gamma^{1,1}\oplus \Gamma^{1,1} (1/N)\oplus (\Gamma_{K3}^g)^*\ .
\ee
and the lattice of magnetic charges is its dual
\be \Lambda_m=\Gamma^{1,1}\oplus \Gamma^{1,1} (N)\oplus \Gamma_{K3}^g=\Lambda_e^*\ .
\ee Here, $\Gamma^{1,1}(N)$ and $\Gamma^{1,1}(1/N)$ denote the even unimodular lattice of signature $(1,1)$ with quadratic form rescaled by $N$ and $1/N$, respectively, while $\Gamma^g_{K3}$ is the $g$-fixed sublattice of the K3 lattice $\Gamma^{4,20}$. For $\lambda>1$ (i.e. when the level matching condition for $g$ is not satisfied), the lattice of electric-magnetic charges is more complicated \cite{PV} (see appendix \ref{a:mess}).


The full U-duality group is, in general, rather complicated \cite{PV} -- in particular, it is not simply a product of a T-duality group acting on the left factor and  an S-duality group acting on the right factor -- but we can nonetheless identify  subgroups that act in these ways, which we call T- and S-dualities. The S-duality group acts trivially on the moduli $M$ and as $\Gamma_1(\hat N)\subseteq SL(2,\ZZ)$ on the axio-dilaton
\begin{equation}
S' = \frac{a S + b}{c S + d}, \qquad \begin{pmatrix} Q' \\ P' \end{pmatrix} = \begin{pmatrix} a & b \\ c & d \end{pmatrix}\begin{pmatrix} Q \\ P \end{pmatrix}.
\end{equation} 
(See appendix \ref{sec:congruence} for the definition of $\Gamma_1(\hat N)$). This is easiest to understand in the type IIB picture: $SL(2,\ZZ)$ acts on the basis of $H^1(T^2;\ZZ)$, and $\Gamma_1(\hat N)$ is the subgroup that commutes with the $1/\hat N$ shift.\footnote{In addition, for each $\gamma\in\Gamma_0(\hat N)$, there is an element of the U-duality group that acts on the axio-dilaton as $\gamma$\cite{Dabholkar:2006bj,Jatkar:2005bh}. This fact -- rather surprising from the above geometric reasoning -- relies on the fact that when $a$ is coprime to $\hat N$, a CHL model obtained by orbifolding by $\hat g=(a\delta,g)$ is dual to a model with $g$  replaced by $g^a$. We then note that $\avg{(a\delta,g^a)}=\avg{(\delta,g)}$\cite{PV}.}
The T-duality group $O(\Lambda_e)$ leaves the heterotic axio-dilaton invariant but acts  on the moduli $M$ and on the charge vector as
\begin{equation} 
P'= (\Upsilon^T)^{-1}P, \ Q' = (\Upsilon^T)^{-1} Q 
\end{equation}
and
\begin{equation}
M' = \Upsilon M \Upsilon^T \ ,
\end{equation} where $\Upsilon \in O(\Lambda_e)$.


\subsection{BPS State Counts}\label{sec:BPScount}


Having introduced CHL models, we now describe the main focus of our paper: generating functions that count BPS states in these models. We begin with  1/2 BPS state counts, both as a warm-up for the more interesting 1/4 BPS case and because  1/2 BPS states play an important role in describing `mortal' 1/4 BPS states -- that is, states that exist only in parts of the CHL moduli space. We will then proceed to describe 1/4 BPS state counts.

We begin by describing the set of all 1/2 BPS states, although we will shortly specialize to a subset thereof. As above, we denote the electric-magnetic charge of  a state by $\column{Q}{P}$. One can then easily show that the 1/2 BPS condition implies that $Q$ and $P$ are parallel (when thought of as $r$-dimensional vectors in $\Lambda_e\otimes\RR$). For each such charge vector, one can consider an index counting the number of  `bosonic' minus `fermionic' 1/2 BPS supermultiplets with the given charges. Here, a supermultiplet is called bosonic or fermionic depending on the spin of its lowest component. We will loosely refer to these indices as `degeneracies'. Crucially, they are the same 
at all points in moduli space; this is demonstrated via the standard argument for moduli-independence of supersymmetric indices.

In the CHL model associated to the identity -- that is, heterotic on $T^6$ -- all $1/2$ BPS states can be mapped via S-duality to states carrying purely electric charges, i.e. with $P=0$. At a perturbative point in moduli space, 1/2 BPS states with these charges are given by perturbative heterotic states that have only left-moving excitations, a.k.a. Dabholkar-Harvey states \cite{Dabholkar:1990yf}. In particular, the level matching condition tells us that for these states the level of the left-moving oscillators is $n=1+Q^2/2$.  As a consequence, we can encode all 1/2 BPS state degeneracies $b(n)$ in the generating function
\be\label{halfBPS}\frac{1}{\Delta(\tau)}=\frac{1}{\eta^{24}(\tau)}=\sum_n b(n)q^{n-1}\ ,\quad q=e^{2\pi i\tau}\ .\ee
See appendix \ref{a:basics} for the definition of $\eta(\tau)$.

 Unfortunately, the story for more general CHL models is not as nice -- in particular, the  S-duality group  $\Gamma_1(\hat N)$ is not always sufficient to map every 1/2 BPS charge vector to a purely electric one.  Even if we restrict to purely electric states, the degeneracies depend not only on $Q^2$, but also on other discrete T-duality invariants \cite{Banerjee:2008pv}.\footnote{Note, however, that certain quantities, such as the asymptotic degeneracy of 1/2 BPS states in the limit of large charges, are expected to be invariant under the `classical'  duality group and therefore depend only on $Q^2$.} For the purpose of this paper, we will only need a formula for the degeneracy of a certain class of purely electric 1/2 BPS states that are obtained as Dabholkar-Harvey states in the $\hat g$-twisted sector of the CHL orbifold. For such a class of charges, the generating function is a simple generalization of \eqref{halfBPS} (see eq.\eqref{eq:halfTwisted}).
%
We refer the reader to refs. \cite{Sen:2005bh,Dabholkar:2005dt} for the more general results.


We can now proceed to describe $1/4$ BPS state counts. $1/4$ BPS states are characterized by a charge vector $\column{Q}{P}$ where $Q$ and $P$ are not parallel and are therefore dyonic in every duality frame.   The degeneracies of 1/4 BPS states (i.e., the number of bosonic minus fermionic quarter BPS supermultiplets carrying given charges $\column{Q}{P}$) are invariant under dualities. The only quadratic invariants under the `classical' duality group $O(6,r-6)$ are $Q^2$, $P^2$ and $P\cdot Q$, so that quantities that can be computed in the supergravity approximation, such as the macroscopic BHW entropy of a $1/4$ BPS black hole, only depend on them.
  Signed degeneracies of 1/4 BPS states are usually described as functions $(-1)^{P\cdot Q+1}d(Q^2/2,P^2/2,P\cdot Q)$ of these invariants.\footnote{In the following, with some abuse of language, we will ignore the sign $(-1)^{P\cdot Q+1}$ and simply refer to the functions $d(Q^2/2,P^2/2,P\cdot Q)$ as degeneracies. } However, at a microscopic (quantum) level, the relevant T-duality group is the discrete $O(\Lambda_e)\subset O(6,r-6)$. Vectors $\column{Q}{P}$ with the same invariants $Q^2$, $P^2$, $P\cdot Q$ can belong to different $O(\Lambda_e)$ orbits and have different degeneracies.   In the unorbifolded case ($g=e$), most results in the literature focus on the case where the charges $P,Q\in \Gamma^{6,22}$ span a primitive sublattice of rank $2$ in $\Gamma^{6,22}$ -- this condition ensures that there is a single T-duality orbit for each value of the invariants $Q^2$, $P^2$, $P\cdot Q$. We will impose the analogous condition also in the CHL models. However, in this case, this might not be sufficient to ensure that there is a unique T-duality orbit.  Rather than attempt a complete classification of the discrete invariants, we will consider only the T-duality orbits of a specific set of charges, for which the calculation of the degeneracies is particularly simple. 

First of all, we will consider states that are charged only under the four gauge fields given by the metric and B-field with one leg along the torus $S^1\times \hat S^1$ and one leg in the uncompactified directions.
 In the heterotic frame, these are states with winding $-\hat{w}$ and momentum $\hat{m}$ along $\hat{S}^1$, winding $-w'$ and momentum $m'$ along $S^1$, and Kaluza-Klein and H-monopole charges\footnote{From the ten-dimensional perspective, an H-monopole along one of the circles roughly corresponds to NS5-branes wrapping $T^4$ times the other circle, while a Kaluza-Klein monopole may be thought of as arising from a Taub-NUT space with the appropriately identified asymptotic circle.} $\hat{M}, -\hat{W}$ along $\hat{S}^1$ and $M', -W'$ along $S^1$. Focusing on the sublattice of the electric-magnetic charge lattice that contains these states, we can describe these dyons with the charge vectors
\begin{equation}\label{chargesQuant}
Q = \begin{pmatrix} \hat{m} \\ m' \\ \hat{w} \\ w' \end{pmatrix}, \ P = \begin{pmatrix} \hat{W} \\ W' \\ \hat{M} \\ M' \end{pmatrix},
\end{equation}
where $m'$ is quantized in units of $1/N$, $M'$ is quantized in units of $N$, and the other quantum numbers are integrally quantized.\footnote{We are using the conventions of \cite{sen2007walls} where before (after) orbifolding the circles $\hat{S}^1, S^1$ have radii $2\pi \sqrt{\alpha'}, 2 \pi N \sqrt{\alpha'}$ ($2\pi \sqrt{\alpha'}, 2 \pi \sqrt{\alpha'}$).} 

Then, we further restrict ourselves to a subclass of these dyons that has a perhaps more transparent description in the IIB frame (called frame 1 in \cite{sen2007walls}). The counting of $1/4$ BPS dyons in the unorbifolded theory was originally carried out in this picture\cite{DS, DJS1, DJS2}. One can get from IIB to the heterotic picture by first making an S-duality transformation, then a T-duality on $\tilde{S}^1$ to go to type IIA, and then finally using string-string duality to go to the heterotic string. Consider a single D5-brane on $K3 \times S^1$, a single Kaluza-Klein monopole on $\hat{S}^1$, momentum $-n/N$ on $S^1$, momentum $l$ along $\hat{S}^1$, and $m$ units of D1-brane charge on $S^1$. The D5-brane has an induced D1-charge coming from wrapping $K3$, shifting the total D1 charge by $-\chi(K3)/24 = -1$, which we have included in $m$. Going through the aforementioned chain of dualities, we see that this configuration maps to a configuration in the heterotic string with momentum $-n/N$ on $S^1$, a KK monopole on $\hat{S}^1$, $-m$ units of NS5-brane charge along $T^4 \times S^1$, $l$ NS5-brane charge wrapped along $T^4 \times \hat{S}^1$ and one unit of fundamental string charge along $S^1$. This gives the charge vectors
\begin{equation}\label{dyonCharges}
Q = \begin{pmatrix} 0 \\ -n/N \\ 0 \\ -1 \end{pmatrix}, \ P = \begin{pmatrix} m \\ -l \\ 1 \\ 0 \end{pmatrix}
\end{equation}
with T-duality invariants \be Q^2 = 2n/N, \ P^2 = 2 m, \ P\cdot Q = l\ .\ee In the simplest model where $\hat g$ is the identity, this set of charges contains a representative for each T-duality orbit of a  primitive charge vector $\column{Q}{P}$.  
\bigskip

Let us consider the generating function $1/\Phi_{g,e}$\footnote{The reason for the second subscript denoting the identity element is to emphasize the similarity with the twisted-twining genera of the next section. We consider elliptic genera with twisted boundary conditions along both cycles of the torus, i.e. $\phi_{g, h}$, and indeed more general second-quantized functions $\Phi_{g, h}$---counting $h$-twining dyons in the CHL model labeled by $g$--- can be defined.} for the degeneracies $d(Q^2/2,P^2/2,P\cdot Q)$ of these orbits of 1/4 BPS states in the CHL model corresponding to a symmetry $\hat{g}$, namely
\be\label{dyonGen} \frac{1}{\Phi_{g,e}(\Omega)}=\sum_{n,m,l} d(m,n,l) p^m q^{\frac{n}{N}}y^l\ ,
\ee where we have defined the symmetric matrix 
\be \Omega=\twoMatrix{\sigma}{z}{z}{\tau}.\ \ee  $\sigma,\tau,z\in \CC$ are complexified chemical potentials for $P^2/2$, $Q^2/2$ and $P\cdot Q$, and
\be q=e^{2\pi i\tau},y=e^{2\pi i z},p=e^{2\pi i\sigma}\ .
\ee
The series $1/\Phi_{g,e}$ can be computed in a weak coupling limit in the type IIB frame. This requires an explicit knowledge of the action of $g$ on the states of the nonlinear sigma model. For this reason,  we postpone the statement  and derivation of the precise expression to section \ref{sec:secquant}. In general, the inverse generating function $\Phi_{g,e}$ converges (in a suitable domain) to a meromorphic Siegel modular form of genus $2$ with respect to some discrete subgroup of $Sp(4,\RR)$.

\subsection{Contour prescription and wall crossing}\label{sec:contour}


The above dyon degeneracies were determined at a particular point in moduli space. By the standard argument for moduli-independence of supersymmetric indices, one would naively expect this function to count BPS states at all points in moduli space, similarly to the 1/2 BPS counting function. A more careful analysis shows that the degeneracies $d(Q^2/2,P^2/2,P\cdot Q)$, while locally constant on the moduli space, jump discontinuously at certain real codimension one subspaces, called walls, in moduli space. This is due to the fact that at these walls $1/4$ BPS states can decay into pairs of $1/2$ BPS states.\footnote{Other decay channels of $1/4$ BPS dyons (for example, into a pair of $1/4$ BPS dyons) are allowed at submanifolds in moduli space of higher codimension and as such those loci can be avoided as we continuously move around moduli space\cite{sen2007rare}. (Strictly speaking, this has been proven  only for charges satisfying the primitivity condition described above).} These decaying $1/4$ BPS states are bound states of $1/2$ BPS states, and as we approach a wall (from the side where the bound state exists) the constituent $1/2$ BPS states approach infinite separation.

Following \cite{sen2007walls}, let us classify all possible decay channels of a 1/4 BPS state into a pair of 1/2 BPS states.  Consider a 1/4 BPS dyon with charges $\column{Q}{P}$ as in \eqref{dyonCharges} and consider the splitting into a pair of 1/2 BPS charge vectors
\be\label{sumcharge} \column{Q}{P}\rightarrow \column{Q_1}{P_1}+\column{Q_2}{P_2}\ .
\ee Since the electric and magnetic charge vectors  of 1/2 BPS states are parallel, $\column{Q_1}{P_1}$ and $\column{Q_2}{P_2}$ can be written  $\column{d M_1}{-c M_1}$ and $\column{-b M_2}{a M_2}$, for some $a,b,c,d\in \RR$ and some vectors $M_1,M_2$. We can take $M_1$ and $M_2$ to take values in the real space $\RR^{2,2}$ spanned by charges of the form \eqref{chargesQuant}, and  normalize them so that $ad-bc=1$. Then, requiring these charges to sum to $(Q,P)$ as in \eqref{sumcharge} determines $M_1=aQ+bP$ and $M_2=cQ+dP$. Therefore, the charges of the 1/2 BPS states are encoded in the matrix $\left(\begin{smallmatrix}
a & b\\ c & d
\end{smallmatrix}\right)$ as
\be\label{generaldec} \column{Q_1}{P_1}=\twoMatrix{da}{db}{-ca}{-cb}\column{Q}{P}\ ,
 \qquad \column{Q_2}{P_2}=\twoMatrix {-bc}{-bd}{ ac}{ad}\column{Q}{P}\ .
\ee Given the expression \eqref{dyonCharges} for the charges $Q$ and $P$, the requirement that   $\column{Q_1}{P_1}$ and $\column{Q_2}{P_2}$ satisfy the quantization conditions described below equation \eqref{chargesQuant} is equivalent to
\be\label{discrete} bc,\ bd,\ ad\in \ZZ\qquad ac\in N\ZZ\ , \qquad \qquad \text{for }\lambda=1\ .
\ee More precisely, this is the constraint on $a,b,c,d$ in the simpler case when the level matching condition for $g$ is satisified (i.e. $\lambda=1$). For $\lambda>1$, the quantization conditions \eqref{chargesQuant} are modified and lead to more complicated constraints on $a,b,c,d$ (see appendix \ref{a:mess}).
Note that there are infinitely many matrices $\twoMatrix{a}{b}{c}{d}$ corresponding to the same splitting \eqref{sumcharge}. In particular, one can rescale $\twoMatrix{a}{b}{c}{d}\to \twoMatrix{xa}{xb}{x^{-1}c}{x^{-1}d}$ for any real non-zero $x$. Using this freedom, we can assume that $a$ and $b$ are integral and coprime (or equal to $0$ and $\pm 1$, in case one of the two vanishes); this fixes $x$ up to a sign. With this choice, \eqref{discrete} implies that $c$ and $d$ are integral as well, so that $\twoMatrix{a}{b}{c}{d}\in SL(2,\ZZ)$. Finally, the two matrices  $\twoMatrix{a}{b}{c}{d}$ and $\twoMatrix{-c}{-d}{a}{b}$ determine the same wall, just with $Q_1,P_1$ and $Q_2,P_2$ exchanged. We conclude that the distinct splittings \eqref{sumcharge} are in one to one correspondence with elements of $PSL(2,\ZZ)/\ZZ_2$, where the $\ZZ_2$ is generated by $S=\twoMatrix{0}{-1}{1}{0}$, with the additional constraint $ac\in N\ZZ$.

Let us now determine the location of the wall in the moduli space corresponding to a matrix $\twoMatrix{a}{b}{c}{d}$. It is useful to arrange the charges as \cite{CV}
\be \Lambda_{Q,P}=\twoMatrix{Q\cdot Q}{-Q\cdot P}{-Q\cdot P}{P\cdot P}=\twoMatrix{2n/N}{-l}{-l}{2m}
\ee
and define the `left-moving' charge vector
\begin{equation}
\Lambda_{Q_L, P_L} = \begin{pmatrix} Q_L \cdot Q_L & -Q_L \cdot P_L \\ -Q_L \cdot P_L & P_L \cdot P_L \end{pmatrix}
\end{equation}
in terms of the projections of the charge vectors onto the `left', positive-definite 6-dimensional space:
\begin{equation}
Q^a_L = \mu^a_I Q^I, \ P^a_L = \mu^a_I P^I,\qquad  a = 1, \ldots, 6, I = 1, \ldots , r.
\end{equation}
Notice that this projection introduces dependence on the T-moduli in $M$ (more precisely through the vielbein $\mu$). 
Introduce the norm $\|X\|^2=-2\det X$ on the space $\mathbb{R}^{2,1}$ of symmetric matrices.\footnote{The signature is easily determined by looking at the norms of the following basis elements:
$$\twoMatrix{0}{1}{1}{0},\quad\twoMatrix{-1}{0}{0}{1},\quad\twoMatrix{1}{0}{0}{1}.$$} 
The polarization identity then gives
$$(X,Y)=\frac{1}{4}(\|X+Y\|^2-\|X-Y\|^2)=-a_1 d_2-a_2 d_1+2b_1 b_2=-\det Y\Tr(XY^{-1}),$$
where $X=\twoMatrix{a_1}{b_1}{b_1}{d_1}$ and $Y=\twoMatrix{a_2}{b_2}{b_2}{d_2}.$ This scalar product is invariant under $SL(2,\RR)$ transformations
\be (\gamma X\gamma^T,\gamma Y\gamma^T)=(X,Y)\ .
\ee

Then, following \cite{CV}, we define a `central charge vector' $\mathcal{Z}$ in terms of $\Lambda_{Q_L, P_L}$ and the axio-dilaton $S = S_1 + i S_2$ as
\begin{equation}
\mathcal{Z}\equiv \Z\left(\smcolumn{Q_L}{P_L},S\right) = \frac{1}{\sqrt{\|\Lambda_{Q_L, P_L}\|}} \Lambda_{Q_L, P_L} + \frac{\sqrt{\|\Lambda_{Q_L, P_L}\|}}{S_2}\begin{pmatrix}
|S|^2 & S_1 \\ S_1 & 1
\end{pmatrix}\ .
\end{equation} From this definition, it is easy to see that $\Z$ transforms covariantly under $SL(2,\RR)$ transformations  
\be\Z\left(\twoMatrix{a}{b}{c}{d}\smcolumn{Q_L}{P_L},\frac{aS+b}{cS+d}\right)=\twoMatrix{a}{b}{c}{d} \mathcal{Z} \twoMatrix{a}{b}{c}{d}^T\ , \qquad \twoMatrix{a}{b}{c}{d}\in SL(2,\RR)\ .\ee Here, $SL(2,\RR)$ is  the classical S-duality group. As discussed in the sections above, quantization breaks this classical real group to a discrete subgroup. Nevertheless, formally $\Z$ transforms covariantly under the full $SL(2,\RR)$.

The central charge vector $\mathcal{\Z}$ has the property that its norm equals (up to a normalization) the mass of the corresponding BPS state
\be M\left(\smcolumn{Q_L}{P_L},S\right)^2=-\frac{1}{2}\|\mathcal{Z}(\smcolumn{Q_L}{P_L},S)\|^2=\frac{(Q_L+\bar S P_L)\cdot (Q_L+SP_L)}{S_2}+2\sqrt{Q_L^2P_L^2-(Q_L\cdot P_L)^2}\ .
\ee In particular, $M$ is formally invariant under $SL(2,\RR)$ transformations of its arguments
\be\label{Minvar} M\left(\twoMatrix{a}{b}{c}{d}\column{Q_L}{P_L},\frac{aS+b}{cS+d}\right)=
M\left(\smcolumn{Q_L}{P_L},S\right)\ .
\ee

The domain wall corresponding to a decomposition \eqref{sumcharge} is the submanifold of the moduli space characterized by the equation
\be M\left(\smcolumn{Q_L}{P_L},S\right)=M\left(\smcolumn{Q_{1L}}{P_{1L}},S\right)+M\left(\smcolumn{Q_{2L}}{P_{2L}},S\right)\ .
\ee It is useful to regard this as an equation in the unknown $S$ for fixed values of the charges $P,Q$ and the moduli $\mu$. For the simplest decomposition
\be \column{Q}{P}\to \column{Q}{0}+\column{0}{P}\ ,
\ee corresponding to the matrix $\twoMatrix{a}{b}{c}{d}=\twoMatrix{1}{0}{0}{1}$, it is easy to check that the wall equation 
\be \label{simplewall} M\left(\smcolumn{Q_L}{P_L},S\right)=M\left(\smcolumn{Q_{L}}{0},S\right)+M\left(\smcolumn{0}{P_{L}},S\right)
\ee
 is equivalent to
\be \left(\Z(\smcolumn{Q_L}{P_L},S),\alpha_0\right)=0\ ,\qquad \alpha_0=\twoMatrix{0}{-1}{-1}{0}\ .
\ee The most general decomposition \eqref{generaldec} can be written as
\be \column{Q}{P}\to \gamma^{-1}\twoMatrix{1}{0}{0}{0}\gamma\column{Q}{P}+\gamma^{-1}\twoMatrix{0}{0}{0}{1}\gamma\column{Q}{P}
\ee where $\gamma=\twoMatrix{a}{b}{c}{d}\in SL(2,\RR)$ and $a,b,c,d$ satisfy \eqref{discrete}. Using the $SL(2,\RR)$ invariance \eqref{Minvar} of the mass formula, the wall equation
\be M\left(\smcolumn{Q_L}{P_L},S\right)=M\left(\gamma^{-1}\smtwoMatrix{1}{0}{0}{0}\gamma\smcolumn{Q_L}{P_L},S\right)+M\left(\gamma^{-1}\smtwoMatrix{0}{0}{0}{1}\gamma\smcolumn{Q_L}{P_L},S\right)
\ee is equivalent to
\be M\left(\gamma\smcolumn{Q_L}{P_L},\gamma\cdot S\right)=M\left(\smtwoMatrix{1}{0}{0}{0}\gamma\smcolumn{Q_L}{P_L},\gamma\cdot S\right)+M\left(\smtwoMatrix{0}{0}{0}{1}\gamma\smcolumn{Q_L}{P_L},\gamma \cdot S\right)\ .
\ee This is of the same form as \eqref{simplewall}, so its solutions are
\be \left(\Z(\gamma\smcolumn{Q_L}{P_L},\gamma\cdot S),\alpha_0\right)=0\ .
\ee  Finally, using covariance of $\Z$ and invariance of the scalar product under $SL(2,\RR)$ transformations, this equation is equivalent to
\be \left(\Z(\smcolumn{Q_L}{P_L},S),\alpha_\gamma\right)=0\ , 
\ee where, explicitly
\be \label{eq:alphaMat} \alpha_\gamma:=\gamma^{-1}\alpha_0\gamma^{-T}=\twoMatrix{2bd}{-ad-bc}{-ad-bc}{2ac}\ .
\ee

 Formally, the moduli dependence of the coefficients $d(m, n, \ell)$ is directly related to the fact that different Fourier expansions of $1/\Phi_{g,e}$ converge for different values of the arguments $\sigma,z,\tau$.\footnote{This raises the question of why $1/\Phi_{g, e}$ should be the correct counting function at all points in moduli space. \cite{sen2007walls, Cheng:2007ch} show that the jumps at walls are exactly as we would expect from the description of walls as arising from decays of $1/4$ BPS states into pairs of $1/2$ BPS states. \cite{gaiotto2005re, Dabholkar:2006bj} explain the appearance of the genus 2 Siegel form $1/\Phi_{g,e}$ via a string web construction.}
For a CHL orbifold with orbifold group $\mathbb{Z}_N$, we can extract the signed degeneracy of dyons with charges $(Q, P)$ -- a Fourier coefficient of $1/\Phi_{g, e}$ -- at a point in the moduli space via
\begin{equation}\label{eq:partfun}
d(Q^2/2, P^2/2, P \cdot Q) = \frac{1}{N}\oint_{\mathcal{C}} d\Omega \frac{e^{\pi i (\Lambda_{Q, P}, \Omega)}}{\Phi_{g, e}(\Omega)}\ ,
\end{equation}
where $\C$ is a contour over the periods of the real parts $\Re\sigma,\Re\tau,\Re z$ of the chemical potentials.

At first sight, eq. \eqref{eq:partfun} is independent of the moduli, in contradiction with our discussion of dyon decay. Even more confusingly, extracting the degeneracy seems ambiguous since the Fourier expansion of $1/\Phi_{g, e}$ converges only in a certain proper subdomain of the Siegel upper half-space; in a different subdomain, there can be another Fourier expansion with different coefficients. These confusions can be resolved simultaneously by first noticing that $1/\Phi_{g, e}$ has poles, so that the integral in \eqref{eq:partfun} is not invariant under deformations of the contour $\C$: namely, when the contour $\C$ crosses one of the poles of $1/\Phi_{g, e}$, the integral picks up the corresponding residue. Therefore, in order to make sense of \eqref{eq:partfun}, it is necessary to give a precise prescription for (the imaginary part of) the contour $\C$. It has been proposed in \cite{Cheng:2007ch} that the contour depends on both the charges and the moduli through the  central charge vector $\Z$:
%
\begin{equation}\label{contour}
\mathcal{C}:= \mathcal{C}(Q, P)\vert_{\mu, S} = \left\lbrace \Im\Omega = \epsilon^{-1}\mathcal{Z}, 0 \leq \Re\sigma,  {1 \over N}\Re \tau, \Re z < 1 \right\rbrace\ ,
\end{equation}
where $\epsilon \ll 1$. This prescription was proposed in \cite{Cheng:2007ch} for the unorbifolded case ($g=e$), based on the observation that it provides the right Fourier coefficients at all points in the moduli space. The fundamental reason for this is that, with such a definition, the contour $\C$ crosses a pole of $1/\Phi_{e, e}$ when there is wall crossing and nowhere else. Furthermore, the difference between the two contour integrals on the two sides of the wall matches exactly the degeneracy of the pair of 1/2 BPS states disappearing from the spectrum. Subsequently, this contour prescription was given an independent interpretation in terms of 1/4 BPS string networks \cite{gaiotto2005re,Dabholkar:2006bj,Banerjee:2008pv}  and as a saddle point in a 1/4 BPS instanton contribution to a 3D effective coupling \cite{Bossard:2016zdx}. It is therefore natural to assume that even in CHL orbifolds this contour prescription provides the exact counting of 1/4 BPS states everywhere in the moduli space. This assumption holds in all cases where the generating function $1/\Phi_{g, e}$ is known.

\section{K3 nonlinear sigma models and their symmetries}\label{sec:K3}

Having explained the important role played by K3 nonlinear sigma models (NLSMs) and their symmetries in CHL models, we now review a number of important results pertaining to them. 

NLMS on K3 are two-dimensional $\N=(4,4)$ superconformal field theories at central charge $c=\bar c=6$. They arise as the worldsheet description of perturbative type IIA string theory on a K3 surface. The 80-dimensional moduli space of K3 NLSMs  is given by
\be \M_{K3} = O^+(\Gamma^{4,20})\backslash O^+(4,20)/(SO(4)\times O(20))\ ,\label{eq:moduli}\ee
and parametrizes the metric and the B-field on the K3 surface, which are part of the $132$-dimensional moduli space \eqref{modspace} 
 parametrizing compactification of type IIA on K3$\times T^2$. The space $\M_{K3}$ is the quotient of the Grassmannian of positive definite oriented four-planes $\Pi\subset \Gamma^{4,20}\otimes\RR$ by the duality group $O^+(\Gamma^{4,20})$; this implies that it is connected. The spacetime (string theory on K3) interpretation of $\Gamma^{4,20}$ is that it is the lattice of charges of D-branes wrapping cycles of the K3. The NLSM is believed  to become a singular (inconsistent) CFT in the limit where when $\Pi$ becomes orthogonal to a root, i.e. a vector $v\in 
\Gamma^{4,20}$ with $v^2=-2$. At these points in the moduli space, the D-branes corresponding to the cycle $v$ become massless and the compactification develops an enhanced gauge symmetry. Since we are only interested in compactifications with a generic abelian gauge group, we can henceforth assume our NLSMs are non-singular.

 At such a point in $\M_{K3}$, we can define a supersymmetric index, called the elliptic genus, as the following trace over the Ramond-Ramond sector of the theory:
\be \phi(\tau,z)=\Tr_{RR}\parens{(-1)^{2(J_0+\bar J_0)}q^{L_0-c/24}\bar q^{\bar L_0-\bar c/24}y^{2J_0}}\qquad q:=e^{2\pi i\tau},y:=e^{2\pi i z}\ . \label{eq:ellipticGenus}\ee
Here, $c=\bar c=6$ are the holomorphic and anti-holomorphic central charges, $L_0$ and $\bar L_0$ are Virasoro generators, and $J_0$ and $\bar J_0$ are the Cartan generators of the left- and right- moving $SU(2)$ R-symmetries of the $\N=4$ superconformal algebra.  Although the right hand side appears to depend on $\bar q$, this is misleading: the elliptic genus gets non-vanishing contributions only from states that are BPS with respect to the right-moving superconformal algebra -- that is, states with $\bar L_0=1/4$. As a consequence, $\phi(\tau,z)$ is actually holomorphic both in $\tau$ and in $z$.  The elliptic genus satisfies modular and elliptic properties that are the characteristic features of a weak Jacobi form of weight $0$ and index $1$ (see \cite{eichler_zagier}):
\be\label{modular} \phi\left(\frac{a\tau+b}{c\tau+d},\frac{z}{c\tau+d}\right) = e^{\frac{2\pi i cz^2}{c\tau+d}}\phi(\tau,z)\ ,\qquad \begin{pmatrix}
a &  b\\ c & d
\end{pmatrix}\in SL(2,\ZZ)
\ee
\be\label{elliptic} \phi(\tau,z+\ell\tau+\ell')=e^{-2\pi i(\ell^2\tau+2\ell z)}\phi(\tau,z)\ ,\qquad
\ell, \ell'\in \ZZ\ ,
\ee and its Fourier expansion
\be \phi(\tau,z)=\sum_{n=0}^\infty\sum_{\ell\in\ZZ} c(n,\ell)q^ny^\ell
\ee has only non-negative powers of $q$. Furthermore, it is invariant under (supersymmetry preserving) exactly marginal deformations, and therefore it is the same function 
\be \phi(\tau,z)=2y+2y^{-1}+20+O(q)\ ,
\ee
for all NLSMs on K3, since $\M_{K3}$ is connected. This function may be easily determined by considering a K3 that is realized as an orbifold $T^4/\ZZ_2$. Setting $y=1$ in \eqref{eq:ellipticGenus} yields the Witten index, i.e. the Euler characteristic of the target space, which for K3 NLSMs is 24.  
\bigskip

It is believed that NLSMs on K3 and on $T^4$ are the only examples of $\N=(4,4)$ superconformal field theory at $c=\bar c=6$  giving rise to string models with spacetime supersymmetry (we will implicitly assume that this is the case in the following). Both the elliptic genus and the Witten index of NLSMs on $T^4$ vanish identically. Such torus models can be also be characterized by the presence of R-R ground states with $L_0=\bar L_0=\frac{1}{4}$ and $(-1)^{2(J_0+\bar J_0)}=-1$.  Such states, which would contribute $-y^{\pm 1}$ to the elliptic genus, are absent in NLSMs on K3  \cite{Nahm:1999ps}. We will use this fact in the following sections.

\subsection{Symmetries of NLSMs on K3}

Our interests in the  CHL models described in section \ref{sec:chl} motivate us to study the groups $G$ of discrete symmetries of NLSMs on K3 that preserve all $16$ spacetime supersymmetries of type IIA strings compactified on K3. From a worldsheet viewpoint, spacetime supersymmetries are a consequence of the worldsheet $\N=(4,4)$ superconformal algebra and of the independent left- and right-moving half-integral spectral flow symmetries of NLSMs on K3. Half-integral spectral flow exchanges the Ramond and Neveu-Schwarz sectors and transforms $L_0$ and $J_0$ as follows (in the case of left-moving spectral flow)
\be  L_0\mapsto L_0\pm J_0+1/4,\qquad J_0\mapsto J_0\pm 1/2\label{eq:spectralFlow}\ .\ee Analogous formulae hold for $\bar L_0$ and $\bar J_0$ for right-moving spectral flow. In particular, the action of this transformation on the Ramond sector ground states, labeled by their $(L_0,J_0)$ eigenvalues, reads $(1/4,1/2)\leftrightarrow (0,0)$, $(1/4,0)\leftrightarrow (1/2,-1/2)$, $(1/4,-1/2)\leftrightarrow (1,-1)$.
Therefore, the condition that the symmetries $g\in G$ preserve all spacetime supersymmetries translates, from the worldsheet perspective, to the condition that these symmetries  commute with the worldsheet $\N=(4,4)$ superconformal algebra and with the half-integral spectral flows. From now on, whenever we talk about symmetries of K3 NLSMs, we will implicitly assume that these properties hold.

Such symmetries turn out to have a simple mathematical characterization \cite{Gaberdiel:2011fg}. To see this, note that the spacetime picture of $\Gamma^{4,20}$ as a lattice of D-brane charges translates to a sigma model interpretation of $\Gamma^{4,20}\otimes\RR$ as the $24$-dimensional space of Ramond-Ramond ground states contributing to the $q^0$ term in the elliptic genus. The four plane $\Pi\subset  \Gamma^{4,20}\otimes\RR$, which is parametrized by the moduli space $\M_{K3}$, can be identified with the four states contributing $2y+2y^{-1}$ to the elliptic genus. The latter are very special states: the corresponding Ramond-Ramond vertex operators are the generators of simultaneous left- and right-moving half-integral spectral flow \cite{EOTY}. By the arguments above, symmetries $g$ of the NLSM preserving all spacetime supersymmetries must fix these four states, and each such $g$ can be identified with an element in the duality group $O^+(\Gamma^{4,20})$ fixing the four-plane $\Pi$ pointwise \cite{Gaberdiel:2011fg}.

\bigskip

 Let us consider a NLSM on K3 with a group of symmetries $G\subset O^+(\Gamma^{4,20})$ preserving the $\N=(4,4)$ superconformal algebra and the four R-R spectral flow generators. For each $g\in G$, we define the twining genus
\be \phi_g(\tau,z):=\Tr_{RR}(g\,(-1)^{2(J_0+\bar J_0)}q^{L_0-\frac{c}{24}}\bar q^{\bar L_0-\frac{\bar c}{24}} y^{2J_0})\ .\label{eq:twiningGenus}
\ee 
These functions share a number of properties with the elliptic genus. In particular, they receive contributions only from BPS states, so they are holomorphic in both $\tau$ and $z$. Furthermore, they are invariant under deformations of the NLSM that preserve the symmetry $g$, i.e. that are generated by $g$-invariant exactly marginal operators.
Since the $24$ R-R ground states transform in the defining $24$-dimensional representation of $G\subset O(\Gamma^{4,20})$, and in particular the spectral flow generators are fixed by $G$, the twining genus has the form
\be\label{eq:twiningPert} \phi_g(\tau,z)=2y+2y^{-1}+(\Tr_{\bf 24}(g)-4)+O(q)\ .
\ee

The twining genus is invariant under conjugations by the duality group $O^+(\Gamma^{4,20})$, i.e.
\be \phi_g(\tau,z)=\phi_{hgh^{-1}}(\tau,z) \qquad h\in O^+(\Gamma^{4,20})\ .\label{eq:twiningDuality}
\ee
As discussed in \cite{Nahm:1999ps}, elements $h\in O(\Gamma^{4,20})$ that are not in $O^+(\Gamma^{4,20})$ correspond to dualities flipping the parity of the worldsheet. In particular, such $h$ do not commute with the generators of the $\N=(4,4)$ superconformal algebra: conjugation by $h$  exchanges the holomorphic and anti-holomorphic $\N=4$  algebras.  Since \eqref{eq:twiningGenus} is manifestly left-right asymmetric,  the identity \eqref{eq:twiningDuality} does not hold, in general, for  $h\in O(\Gamma^{4,20})\setminus O^+(\Gamma^{4,20})$ \cite{slowpaper}.  

The family $\F_g^{ns}$ of non-singular K3 NLSMs which share the symmetry $g\in O^+(\Gamma^{4,20})$ may be shown to form a connected subset of $\M_{K3}$ \cite{slowpaper}.  Therefore, physically independent twining genera, i.e. twining genera that are not related by dualities or continuous deformations, correspond to $O^+(\Gamma^{4,20})$ conjugacy classes fixing a subspace of $\Gamma^{4,20}\otimes \RR$ of signature $(4,d)$, $d\ge 0$, that is orthogonal to no roots. 
(The latter condition ensures that the family $\F_g^{ns}$ is non-empty, so that $\phi_g$ is well-defined). These $O^+(\Gamma^{4,20})$ conjugacy classes have been classified in \cite{slowpaper}.

\subsection{Orbifolds and quantum symmetries}

We conclude this section with a discussion of some general aspects of orbifolds, as well as of some issues that interest us because of the particular orbifolds that arise in the construction of CHL models.

Consider a $N=(4,4)$ SCFT $\C$ (a NLSM on K3 or $T^4$) with $c=\bar c=6$ and a symmetry $g$ of order $N$ preserving the superconformal algebra and the spectral flow. The orbifold $\C/\langle g\rangle$ of $\C$ by the cyclic group $\langle g\rangle$ is obtained by projecting on the $g$-invariant subspace of all the twisted and untwisted sectors $\Hh_{g^r}$, $r\in \ZZ/N\ZZ$.
It is often useful to split each $\Hh_{g^r}$ into its $g$-eigenspaces
\be \Hh_{r,s}:=\{v\in \Hh_{g^r}\mid g(v)=e^{\frac{2\pi i s}{N}}v\}
\ee so that the spectrum of $\C/\langle g\rangle$ is given by
\be\label{orbspectrum} \Hh_{\C/\langle g\rangle}= \oplus_{r=1}^N \Hh_{r,0}\ .
\ee The spaces $\Hh_{r,s}$ are also useful for describing the spectrum of the CHL model. Recall that this model is obtained by taking the orbifold of an NLSM on K3$\times S^1$ by a symmetry $(\delta,g)$, where $\delta$ is $1/N$-th of a period along $S^1$. The spectrum of this orbifold is given by tensoring states in $\Hh_{r,s}$ with states of the circle CFT with winding $r\mod N$ and momentum $s/N \mod 1$. 

The orbifold $\C/\langle g\rangle$ is a consistent CFT only if the level matching condition
\be L_0-\bar L_0\in \frac{1}{N}\ZZ\qquad \text{on } \Hh_g\ ,
\ee is satisfied. If this is the case, then $\C/\langle g\rangle$ is again an $\N=(4,4)$ SCFT at the same central charge and therefore an NLSM on K3 or $T^4$.  In general, the level matching condition might fail; in this case, one can still define the twisted sectors $\Hh_{g^r}$, but one cannot construct a consistent CFT with a local OPE  that includes $g$-twisted vertex operators. This general case will be considered in section \ref{sec:levelfail}.

The action of any $h\in G$ commuting with $g$  on the untwisted fields induces an action of $h$ on all twisted sectors. In particular, one has  $g^r=e^{2\pi i(L_0-\bar L_0)}$ on the $\Hh_{g^r}$ twisted sector. This leads to an obvious generalization of the twining genera $\phi_g$. For each commuting pair $g,h\in G$, we define the twisted-twining genus
\be \phi_{g,h}(\tau,z):=\Tr_{\Hh_g}(h\,(-1)^{2(J_0+\bar J_0)}q^{L_0-\frac{c}{24}}\bar q^{\bar L_0-\frac{\bar c}{24}} y^{2J_0})\ ,
\ee where the trace is taken over the $g$-twisted Ramond-Ramond sector $\Hh_g$.  The twining genera $\phi_h$ correspond to the special case where the `twist' $g$ is the identity: $\phi_h\equiv \phi_{e,h}$.

We will be mostly interested in the twisted-twining genera of the form $\phi_{g^i,g^j}$, i.e. when the commuting pair generate a cyclic subgroup of $G$. 
These twisted-twining genera are related to the characters for the spaces $\Hh_{r,s}$
\be\label{hatphidef} \hat \phi_{r,s}(\tau,z)=\Tr_{\Hh_{r,s}}((-1)^{2(J_0+\bar J_0)}q^{L_0-\frac{c}{24}}\bar q^{\bar L_0-\frac{\bar c}{24}} y^{2J_0})\ ,
\ee by a discrete Fourier transform
\be\label{discFour} \hat\phi_{r,s}(\tau,z):=\frac{1}{N}\sum_{k=1}^N e^{-\frac{2\pi i s k}{N}} \phi_{g^r,g^k}(\tau, z)\ .
\ee
By definition, all Fourier coefficients of $\hat\phi_{r,s}$
\be \hat\phi_{r,s}(\tau,z)=\sum_{n\in \QQ}\sum_{l\in \ZZ} \hat c^g_{r,s}(n,l)q^ny^l\ ,\label{eq:cHat}
\ee are (possibly negative) integers $\hat c^g_{r,s}(n,l)\in\ZZ$. For later convenience, we also define coefficients $\hat c_{r,s,t}(D)$, $r,s\in \ZZ/N\ZZ$, $t\in \ZZ/2\ZZ$, by 
\be \hat c^g_{r,s}(n,l)=\hat c^g_{r,s,l\,\text{(mod 2)}}(4n-l^2)\ .\label{eq:cHatPolarity}
\ee Here, we used the fact that, for weak Jacobi forms of index $1$, $\hat c^g_{r,s}(n,l)$ depends on $n,l$ only through the discriminant $D=4n-l^2$ and $l\modu 2$.
From \eqref{orbspectrum}, \eqref{discFour} and \eqref{eq:cHat}, it is also easy to derive the elliptic genus of the orbifold theory $\C/\langle g\rangle$
\be \phi^{\C/\langle g\rangle}(\tau,z)=\sum_{r\in \ZZ/N\ZZ} \hat \phi_{r,0}(\tau,z)=\frac{1}{N} \sum_{r,k\in \ZZ/N\ZZ} \phi_{g^r,g^k}(\tau, z)\ .
\ee

Mathematically, the twisted-twining genera are weak Jacobi forms of weight $0$ and index $1$ and are the components of a vector-valued representation of $SL(2,\ZZ)$
\be\label{SL2Ztransf} \phi_{g,h}(\tau,z)\left(\frac{a\tau+b}{c\tau+d},\frac{z}{c\tau+d}\right) = e^{\frac{2\pi i cz^2}{c\tau+d}}\phi_{g^ah^c,g^bh^d}(\tau,z)\ ,\qquad \begin{pmatrix}
a &  b\\ c & d
\end{pmatrix}\in SL(2,\ZZ)\ .
\ee
Note that this transformation law holds only when the level-matching condition is satisfied. In the $\lambda>1$ case, additional phases might appear on the right-hand side of this formula (see \eqref{eq:jacobiTwining}).

If $\C'=\C/\langle g\rangle$ is a consistent orbifold by a cyclic group $\langle g\rangle$, then $\C'$ has a symmetry $g'$ (often called the quantum symmetry) of the same order $N$ that acts by $e^{\frac{2\pi i r}{N}}$ on the $g^r$-twisted sector. By taking the orbifold of $\C'$ by $g'$, one recovers the original theory $\C$. More generally, the space $\Hh_{r,s}$, i.e. the $g=e^{\frac{2\pi i s}{N}}$ eigenspace in the $g^r$-twisted sector of $\C$, can be identified with $\Hh'_{s,r}$, i.e. the $g'=e^{\frac{2\pi i r}{N}}$ eigenspace in the $g'^s$-twisted sector of $\C'$.

\subsection{The $\lambda>1$ case}\label{sec:levelfail}

Many of the constructions of the previous section generalize to the case where the states in the $g$-twisted sector do not satisfy the level matching condition, i.e. when 
\be\label{levelfail} (L_0-\bar L_0)_{\rvert \Hh_g}\in \frac{\E'}{N\lambda}+\frac{1}{N}\ZZ\ .
\ee Here, $N$ is the order of $g$, $\lambda$ is a divisor of $N$ and $\E'\in \ZZ/\lambda\ZZ$ is coprime to $\lambda$, i.e. $\gcd(\E',\lambda)=1$. In this case, the orbifold $\C/\langle g\rangle$ is not a consistent CFT. However, one can still consider the twisted sectors as vector spaces or as modules over the untwisted sector (that is, the action of an untwisted vertex operator on a twisted state is well defined). One can still define an action $\rho_{g^i}(h)$ on the twisted sectors $\Hh_{g^i}$ of any symmetry $h$ commuting with $g$, in a way compatible with the action on the untwisted sector. However, this definition is ambiguous, since one can multiply $\rho_{g^i}(h)$ by a phase. This implies that the map $\rho_{g^i}:\langle g\rangle \to GL(\Hh_{g^i})$ is only a projective representation of $\langle g\rangle$; equivalently, $\rho_{g^i}$ can be thought of as a representation of a central extension of $\langle g\rangle$, which can be chosen to have order $N\lambda$. As a conseqeunce, the very definition of the spaces $\Hh_{r,s}$ and of the functions $\phi_{g^j,g^k}$ and $\hat \phi_{r,s}$ is also ambiguous. 

Assuming that a choice has been made for $\rho_{g^i}$, one can tentatively define the twisted-twining genera as
\be\label{tentdef} \phi_{g^i,g^j}(\tau,z):=\Tr_{\Hh_{g^i}}(\rho_{g^i}(g^j)\,(-1)^{2(J_0+\bar J_0)}q^{L_0-\frac{c}{24}}\bar q^{\bar L_0-\frac{\bar c}{24}} y^{2J_0})\ .
\ee  Eq.\eqref{SL2Ztransf} then generalizes to 
\be \phi_{g,h}(\frac{a\tau+b}{c\tau+d},\frac{z}{c\tau+d}) = \epsilon_{g,h}\begin{pmatrix}
a &  b\\ c & d
\end{pmatrix}e^{\frac{2\pi i cz^2}{c\tau+d}}\phi_{g^ah^c,g^bh^d}(\tau,z)\ ,\qquad \begin{pmatrix}
a &  b\\ c & d
\end{pmatrix}\in SL(2,\ZZ)\ ,\label{eq:jacobiTwining}
\ee where $\epsilon_{g,h}:SL(2,\ZZ)\to U(1)$ is a phase depending on the choice of the representations on the twisted sectors.

Note that, in the case $\lambda=1$, this ambiguity is fixed by setting $\rho_g(g)=e^{2\pi i(L_0-\bar L_0)}$ and requiring the fusion
\be \Hh_{r,s}\boxtimes\Hh_{r',s'}\to \Hh_{r+r',s+s'}\ , \qquad r,s,r',s'\in \ZZ/N\ZZ\ .
\ee These conditions cannot hold when $\lambda>1$. One can always require that $\rho_g(g)=e^{2\pi i(L_0-\bar L_0)}$. However, by \eqref{levelfail}, $\rho_g(g)^N=e^{2\pi i\frac{\E'}{\lambda}}\neq 1$  is only proportional to the identity, up to a non-trivial phase, so that the $\rho_g(g)$ eigenvalues are not, in general, $N$-th roots of unity. As a consequence, the very definition of $\Hh_{r,s}$ is more subtle in this case. A simple way to circumvent these issues is to think of $g$ as generating a central extension $\ZZ_{N\lambda}$ of $\ZZ_{N}$, with the central element $g^N$ acting trivially on the untwisted sector and by a phase on each twisted sector. From this viewpoint, it is natural to define $(N\lambda)^2$ spaces $\Hh_{r,s}$, $r,s\in \ZZ/N\lambda\ZZ$ such that
\be \Hh_{r,s}:=\{v\in \Hh_{g^r}\mid \rho_{r}(g)(v)=e^{\frac{2\pi i s}{N\lambda}}v\}\qquad r,s\in \ZZ/N\lambda\ZZ\ ,
\ee where the maps $\rho_{r}$, $r\in \ZZ/N\lambda\ZZ$ are such that $\rho_{r}(g^r)=e^{2\pi i(L_0-\bar L_0)}$ and the fusion rules
\be \label{fusion} \Hh_{r,s}\boxtimes\Hh_{r',s'}\to \Hh_{r+r',s+s'}\ , \qquad r,s,r',s'\in \ZZ/N\lambda\ZZ
\ee hold. The spaces $\Hh_{r,s}$ satisfy 
\be\label{emptyspace} s-r\E_g'\not\equiv 0\modu \lambda\qquad \Rightarrow\qquad \Hh_{r,s}=0\ .
\ee 
Furthermore, there are isomorphisms (as $\Hh_{0,0}$ modules)
\be\label{isomodule} \Hh_{r,s}\cong \Hh_{r+N,s-\E'N}\ ,
\ee so that there are only $N^2$ independent non-trivial irreducible  $\Hh_{0,0}$-modules, as expected for a symmetry of order $N$. 

In the same  spirit and  with a certain abuse of notation, we will conventionally define the twisted-twining genera $\phi_{g^i,g^j}$ as $(N\lambda)^2$ independent functions 
\be\label{newdef} \phi_{g^i,g^j}(\tau,z)=\Tr_{\Hh_{g^i}}(\rho_i(g)^j\,(-1)^{2(J_0+\bar J_0)}q^{L_0-\frac{c}{24}}\bar q^{\bar L_0-\frac{\bar c}{24}} y^{2J_0})\ , \qquad i,j\in \ZZ/N\lambda\ZZ\ ,
\ee
where indices $i$ labeling the representations $\rho_i$ take values in $i\in \ZZ/N\lambda\ZZ$ rather than $\ZZ/N\ZZ$, and $\rho_i(g)$ has (in general) order $N\lambda$. With this definition, all the factors $\epsilon_{g,h}$ in \eqref{eq:jacobiTwining} get absorbed into the definition of $\phi_{g^i,g^j}$ and eq.\eqref{SL2Ztransf} formally holds also for $\lambda>1$. Of course, the functions $\phi_{g^i,g^j}$ and $\phi_{g^k,g^l}$ are not really independent for $i\equiv k,j\equiv l\modu N$: consistently with equation \eqref{tentdef}, they just correspond to different choices of the representation $\rho_{g^i}$ on the $g^i$-twisted sector, so that  they just differ by an overall phase (see section \ref{sec:twining} for more details).

Similarly, one can define the $(N\lambda)^2$ characters $\hat\phi_{r,s}$ of the spaces $\Hh_{r,s}$, $r,s\in \ZZ/N\lambda\ZZ$, that are related to $\phi_{g^i,g^j}$ by a discrete Fourier transform
\be  \hat \phi_{r,s}(\tau,z)=\Tr_{\Hh_{r,s}}((-1)^{2(J_0+\bar J_0)}q^{L_0-\frac{c}{24}}\bar q^{\bar L_0-\frac{\bar c}{24}} y^{2J_0})=\frac{1}{N\lambda}\sum_{k=1}^{N\lambda} e^{-\frac{2\pi i s k}{N\lambda}} \phi_{g^r,g^k}(\tau, z)\ .
\ee The Fourier coefficients of $\hat\phi_{r,s}$, defined as in see \eqref{eq:cHat} and \eqref{eq:cHatPolarity}, are still denoted by  $\hat c^g_{r,s}(n,l)$ or $\hat c^g_{r,s,l}(4n-l^2)$ , but the subscripts $r,s$ now run in $\ZZ/N\lambda\ZZ$ rather than $\ZZ/N\ZZ$.

As stressed in section \ref{sec:chl}, the CHL orbifold can be defined also when $\lambda>1$, simply by taking the shift $\delta$ to have order $N\lambda$ rather than $N$. Indeed, the spectrum of the CHL model can be described exactly in the same way as for $\lambda=1$, i.e. it is obtained by tensoring states in $\Hh_{r,s}$ with states in the circle CFT with winding $r\mod N\lambda$ and momentum $\frac{s}{N\lambda}\mod 1$. Eq.\eqref{fusion} ensures that the OPE is well-defined. Notice however that, because of \eqref{emptyspace}, certain values of winding and momentum do not correspond to any state in the theory. As a consequence, the lattice of electric-magnetic charges is more complicated for $\lambda>1$. We refer the reader to appendix \ref{a:mess} and to \cite{PV} for more details.

\section{Modular properties of twining genera}\label{sec:twining} 
%
%
We next develop some modular machinery which will provide us with many constraints on twisted-twining genera in K3 NSLMs. Our logic builds upon ideas in \cite{slowpaper}, which employed modularity arguments to determine the twining genera associated to the Frame shapes $3^8$ and $4^6$. In the sequel, we will combine these considerations with constraints from wall crossing.

\subsection{The modular groups of twining genera}\label{s:modulargroup}


Let $g$ be a symmetry of order $N$ of a NLSM on K3. 
The transformation law \eqref{SL2Ztransf} implies that $\phi_{e,g}$ transforms into itself under a group $\hat\Gamma_g$, which we call the fixing group of $\phi_{e,g}$.
More precisely, we have
\be \phi_{e,g}(\frac{a\tau+b}{c\tau+d},\frac{z}{c\tau+d}) = e^{\frac{2\pi i cz^2}{c\tau+d}}\phi_{e,g}(\tau,z)\ ,\qquad \begin{pmatrix}
a &  b\\ c & d
\end{pmatrix}\in \hat\Gamma_g\ .\label{eq:jacobiFixingGroupTwining}
\ee
As we discuss in the next section, this transformation law strongly constrains $\phi_{e,g}$. 
When $\lambda>1$, rather than describing $\hat \Gamma_g$ directly, it is convenient to first consider a larger group $\Gamma_g$, which we call the eigengroup of $g$, such that $\phi_{e,g}$ transforms into itself up to a phase, i.e.
\be \phi_{e,g}(\frac{a\tau+b}{c\tau+d},\frac{z}{c\tau+d}) = \xi_{e,g}\begin{pmatrix}
a &  b\\ c & d
\end{pmatrix}
e^{\frac{2\pi i cz^2}{c\tau+d}}\phi_{e,g}(\tau,z)\ ,\qquad \begin{pmatrix}
a &  b\\ c & d
\end{pmatrix}\in \Gamma_g\ ,\label{eq:jacobiEigengroupTwining}
\ee
where  $\xi_{e,g}:\Gamma_g\to U(1)$  (the multiplier of $\phi_{e,g}$) is a suitable homomorphism. Invariance of $\phi_{e,g}$ under charge conjugation and conjugation of $g$ in $O^+(\Gamma^{4,20})$ implies\footnote{Strictly speaking, we cannot exclude that the eigengroup is larger just by accidental coincidences. For simplicity, we will ignore this possibility.} \cite{slowpaper}
\be \Gamma_g:=\{ \left(\begin{smallmatrix}
a & b\\ c & d
\end{smallmatrix}\right)\in SL(2,\ZZ)\mid c\equiv 0\modu{N},\ \exists h\in O^+(\Gamma^{4,20}) \text{ s.t. } g^d =hgh^{-1}\text{ or } g^d =hg^{-1}h^{-1}\}\ .
\ee
A non-trivial multiplier $\xi_{e,g}$ can only arise when $\lambda>1$, and in this case it has order $\lambda$, i.e. $\xi_{e,g}^\lambda=1$. In particular, if one adopts the definition where there are $N^2$ functions $\phi_{g^i,g^j}$ labeled by $i,j\in \ZZ/N\ZZ$, then by eq.\eqref{eq:jacobiEigengroupTwining} the multiplier $\xi_{e,g}$ is simply the restriction of $\epsilon_{e,g}$ to $\Gamma_g$. Equivalently, as described in section \ref{sec:levelfail}, one could eliminate the multiplier on the right hand side of \eqref{eq:jacobiEigengroupTwining} at the cost of considering $\phi_{g^i,g^j}$ as $(N\lambda)^2$ distinct functions labeled by $i,j\in \ZZ/N\lambda\ZZ$.  With this convention (that we adopt henceforth), $\xi_{e,g}$ determines the relative phases between $\phi_{g^i,g^j}$, $i\equiv 0 \modu N$, $j\equiv 1\modu N$, and $\phi_{e,g}$, i.e.
\be \phi_{g^c,g^d}(\tau,z)=\xi_{e,g} \begin{pmatrix}
 a & b \\ c& d
\end{pmatrix} \phi_{e,g}(\tau,z)\ , \qquad \begin{pmatrix}
 a & b \\ c& d
\end{pmatrix}\in \Gamma_1(N)\subset \Gamma_g\ .
\ee
Unlike $\epsilon_{e,g}$, $\xi_{e,g}$ is independent of the choice of the representations $\rho_{g^i}$; this is clear from the fact that $g$ has an unambiguous action on the untwisted sector. When $\lambda=1$, the fixing group and eigengroup coincide (i.e. $\hat\Gamma_g=\Gamma_g$); in general, the fixing group is the kernel of $\xi_{e,g}$.  

The groups $\Gamma_g$ and the order $\lambda$ of the multiplier depend only on the Frame shape of $g$ and have been determined in \cite{slowpaper}. 
When the multiplier is trivial, $\Gamma_g$ is either
\be \Gamma_0(N):=\{\left(\begin{smallmatrix}
a & b\\ c & d
\end{smallmatrix}\right)\in SL(2,\ZZ)\mid c\equiv 0\modu N\}\ ,
\ee or 
\be \Gamma_{\langle-1\rangle}(N):=\{\left(\begin{smallmatrix}
a & b\\ c & d
\end{smallmatrix}\right)\in SL(2,\ZZ)\mid c\equiv 0,\ a\equiv\pm1\modu N \}\ .
\ee When the multiplier has order $\lambda>1$, then the eigengroup is always $\Gamma_0(N)$.

%

It turns out that the possible orders of a non-trivial multiplier are $2,3,4$ and $6$ \cite{slowpaper}. The multiplier is always of the form
\be \xi_{e,g}\left(\begin{smallmatrix}
a & b\\ c & d
\end{smallmatrix}\right)=e^{-2\pi i \frac{\E'}{\lambda} \frac{cd}{N}}\ ,\qquad \left(\begin{smallmatrix}
a & b\\ c & d
\end{smallmatrix}\right)\in \Gamma_0(N)\ .
\ee Here, $\E'\in \ZZ/\lambda \ZZ$, $\gcd(\E',\lambda)=1$, parametrizes the possible different multipliers of order $\lambda$, and depends on the given $g$. In fact, it is easy to check that it is the same $\E'$ determining the spectrum of $L_0-\bar L_0$ of the $g$-twisted sector (see eq.\eqref{levelfail}). Indeed, by \eqref{levelfail}, the $g$-twisted genus $\phi_{g,e}$, which is the S-transform of $\phi_{e,g}$, has a Fourier expansion of the form
\be \phi_{g,e}(\tau,z)=\sum_{n\in \frac{\E'}{N\lambda}+\frac{1}{N}\ZZ}\sum_{l\in \ZZ}c_{g,e}(n,l)q^ny^l\ ,\label{eq:multiplierSTransform}
\ee so that $\phi_{e,g}$ gets multiplied by a phase $e^{\frac{2\pi i\E'}{\lambda}}$ under the transformation $ST^NS^{-1}=\left(\begin{smallmatrix} 1 & 0\\ -N & 1
\end{smallmatrix}\right)$. 
 For $\lambda=2$, there is only one possible non-trivial multiplier ($\E'\equiv1\modu 2$). For each $\lambda>2$,  there are two possible multipliers  ($\E'\equiv \pm1\modu\lambda$), that are complex conjugate to each other. Using worldsheet parity, one can show that both multipliers must appear for the twining genera of a given Frame shape \cite{slowpaper}.

Finally, we provide the fixing groups, $\hat\Gamma_g$. Of course, if the multiplier is trivial, one has $\hat\Gamma_g=\Gamma_g$. One can show by a case by case analysis that whenever $\lambda>1$, the fixing group is equal to $\Gamma_0(N\lambda)$. (For all of our multipliers, $\hat\Gamma_g \subseteq \Gamma_0(N\lambda)$ follows from $\E'\equiv \pm 1\modu\lambda$ and the fact that $\lambda$ divides $N$. To see this, suppose that $\twoMatrix{a}{b}{c}{d}\in\hat\Gamma_g\subset\Gamma_0(N)$; in particular, this implies that $(c,d)=1$ and $N|c$. The condition $\E'\equiv\pm 1\modu\lambda$ implies $N\lambda|cd$; since $N|c$ and $\lambda|N$, we must have $\lambda|c$. Recalling that $c$ and $d$ are coprime, we find $N\lambda|c$). 

\subsection{Constraints from modularity}\label{sec:modularConstraints}

The ring of weak Jacobi forms under a subgroup of $SL(2,\ZZ)$ is generated by the standard forms $\chi_{0,1}$ and $\chi_{-2,1}$ and by a suitable set of modular forms (see appendix \ref{a:basics}). In particular, any twining genus can be written as
\be \phi_{e,g}(\tau,z)=\frac{\Tr_{\bf{24}}(g)}{12}\chi_{0,1}(\tau,z)+F_{e,g}(\tau)\chi_{-2,1}(\tau,z)\ ,\label{eq:phiConstraint}
\ee where  \be F_{e,g}(\tau)=2-\frac{\Tr_{\bf{24}}(g)}{12} +O(q)\ee is a modular form of weight $2$ with multiplier $\xi_{e,g}$ for the eigengroup $\Gamma_g$ described in the previous section. In particular, $F_{e,g}$ is a modular form of weight $2$ with trivial multiplier for $\hat\Gamma_g$. Since $\chi_{0,1}$ has no multiplier, \eqref{eq:phiConstraint} makes it clear that the multiplier is necessarily trivial when $\phi_{e,g}(\tau,0)\equiv \Tr_{\bf{24}}(g)\neq 0$. 

See appendix \ref{sec:congruence} for background on modular forms for congruence subgroups; we summarize here the most important definitions and results. Let us denote by $M_2(\Gamma)$ the space of modular forms of weight $2$ with trivial multiplier for a group $\Gamma\subseteq SL(2,\ZZ)$. Modular forms $F(\tau)$ of weight $2$ under $\Gamma$ correspond to meromorphic $1$-differentials $F(\tau)d\tau$ on $\hat \HH/\Gamma$, with at most single poles at the cusps and which are holomorphic elsewhere. Indeed, in a neighborhood of a cusp $\tau\to \cc$ of width $w_\cc$, a good coordinate is given by $q_{\cc}= e^{\frac{2\pi i\tau_\cc}{w_\cc}}$, where $\tau_\cc=\gamma(\tau)$ and $\gamma\in SL(2,\ZZ)$ is such that $\gamma(\cc)=i\infty$. By the latter statement, we mean that if we write
\be F(\tau)d\tau=F_\cc(\tau_\cc)d\tau_\cc\ ,\label{eq:Fc}\ee
then we have an expansion
\be F_\cc(\tau_\cc)=a_0(\cc)+a_1(\cc)q_{\cc}+a_2(\cc)q_\cc^2+\ldots\ee
about $\tau\to\cc$ (or $\tau_\cc\to i\infty$) in integral powers of $q_\cc$. (Note that there is nothing stopping us from replacing $\tau_\cc$ by $\tau'_\cc=\tau_\cc+1$ even though, when $w_\cc\not=1$, this yields a different expansion. Thus, implicit in the notation $\tau_\cc$ is our choice of $\gamma$). Making another change of coordinates,
\be F(\tau)d\tau=\tilde F_\cc(q_\cc)dq_\cc \ ,\ee
and using $d\tau_\cc=\frac{w}{2\pi i}\frac{dq_\cc}{q_\cc}$, we obtain
\be \tilde F_\cc(q_\cc)=a_0(\cc)\frac{w_\cc}{2\pi i} q_\cc^{-1}+\ldots
\ee  Therefore, the residue of $F(\tau)d\tau$ at the cusp $\cc$ is determined by the $q_\cc^0$ Fourier coefficient $a_0(\cc)$ in the expansion about $\cc$. If $\hat \HH/\Gamma$ has genus $0$ and $n$ cusps, then the dimension of $M_2(\Gamma)$ is the number of independent residues $a_0(\cc)$, i.e. $\dim M_2(\Gamma)=n-1$ (the $-1$ is due to the fact that the sum over all residues must be zero). In general, one has
\be \dim M_2(\Gamma)=\text{genus}(\Gamma)+n-1\ .
\ee The fixing groups relevant for the twining genera are all genus $0$ or $1$.

When the group $\hat\Gamma_g$ is genus zero and has $n$ cusps, the space of weak Jacobi forms of index $1$ and weight $0$ has dimension $n$ ($1$ parameter from the constant in front of $\chi_{0,1}$ and $n-1$ parameters for $F_{e,g}(\tau)$). Therefore, $\phi_{e,g}$ is completely determined by the leading ($q_\cc^0$) term in the expansion of $\phi_{e,g}$ around each cusp (actually, it is sufficient to know the expansion around $n-1$ cusps, since at $\infty$ we know both the coefficient of $y$ and the constant term of $\phi_{e,g}$). The expansion at a given cusp corresponds to the expansion at $\infty$ for a twisted-twining genus $\phi_{g^r,g^s}$ and the leading coefficient is determined by the action of $g^s$ on the \emph{ground states} of the $g^r$ twisted sector. More explicitly, say that $\gamma=\twoMatrix{s}{a}{-r}{b}\in SL(2,\ZZ)$ maps the cusp $\cc=b/r$ to $\infty$. Then, the expansion of $\phi_{e,g}$ about $\cc$ corresponds to the expansion of $\phi_{g^r,g^s}$ about $\infty$. The latter takes the form 
\begin{align}
\phi_{g^r,g^s}(\tau_\cc,z)&=\frac{\Tr_{\bf 24}(g)}{12}\chi_{0,1}(\tau_\cc,z)+F_{g^r,g^s}(\tau_\cc)\chi_{-2,1}(\tau_\cc,z)\label{eq:grgs}\\
&=b_1(\cc)(y+y^{-1})+b_2(\cc)+O(q_\cc),
\end{align}
where $b_1(\cc)$ and $b_2(\cc)$ satisfy
\be 2b_1(\cc)+b_2(\cc)=\Tr_{\bf 24}(g)\ .
\ee For a general $\phi_{g^r,g^s}$, one would have $2b_1+b_2=\Tr_{\bf 24}(g^{\gcd(r,s,N)})$, but we are focusing on those $\phi_{g^r,g^s}$ obtained by an $SL(2,\ZZ)$ transformation of $\phi_{e,g}$, and in this case one has $\gcd(r,s,N)=1$. In particular, \eqref{eq:twiningPert} shows that when $r=0$, $b_1(\cc)=2$. In \eqref{eq:grgs}, we have introduced $F_{g^r,g^s}(\tau)$, which is given by
\be F_{e,g}(\tau)d\tau= F_{g^r,g^s}(\tau_\cc)d\tau_\cc. \ee
Rearranging \eqref{eq:grgs} yields the residue of $F_{e,g}(\tau)d\tau$ at $\cc$:
\be \label{eq:bToA} a_0(\cc)=b_1(\cc)-\frac{\Tr_{\bf 24}(g)}{12}\ .\ee
With the values $a_0(\cc)$ at each cusp in hand, we may expand $F_{e,g}$ in a basis for $M_2(\hat\Gamma_g)$. In fact, frequently a smaller basis suffices, due to the fact that $F_{e,g}$ lies in the space $M_2^{\xi_{e,g}}(\Gamma_g)\subset M_2(\hat\Gamma_g)$ of modular forms which are modular for $\Gamma_g$ with multiplier $\xi_{e,g}$. Appendix \ref{sec:modularCusp} describes how knowledge of the values $a_0(\cc)$ at all cusps of $\hat\HH/\hat\Gamma_g$ allows us to expand $F_{e,g}$ in a basis for $M_2^{\xi_{e,g}}(\Gamma_g)$.

%

When $\hat\Gamma_g$ has genus $1$, the leading Fourier coefficients at the cusps are not sufficient to determine $\phi_{e,g}$. Indeed, in this case $M_2(\hat\Gamma_g)$ contains a cusp form (a form with vanishing residues at all cusps) $f$, and one is free to add $\alpha f(\tau)\chi_{-2,1}(\tau,z)$, for any $\alpha\in \CC$, without affecting the leading coefficients at the different cusps. There is a physically motivated restriction on $\alpha$: the discrete Fourier transforms
$ \hat\phi_{r,s}$ defined in \eqref{discFour}  are interpreted as $\ZZ_2$-graded dimensions of the spaces $\Hh_{r,s}$, i.e. the $g=e^{\frac{2\pi is}{N}}$  eigenspaces in the $g^r$-twisted sector. As such, the Fourier coefficients must be (possibly negative) integers. Therefore, for a suitable normalization of $f$, we are only allowed to add to the twining genus a term $\alpha f(\tau)\chi_{-2,1}(\tau,z)$ for $\alpha$ \emph{integral} rather than complex. 

Even with this restriction, there are still infinitely many possibilities for the twining genus. Therefore, to determine $\phi_{e,g}$ in these cases, one should know its action on the massive BPS states, which is a priori difficult. We will shortly show how additional constraints can be derived from string theory arguments. However, we first demonstrate, via an example, the reasoning that allows us to compute many twining genera for which $\hat\Gamma_g$ has genus 0.

\subsection{An example: Frame shape $2^36^3$}\label{sec:g0Ex}

We compute the twining genus associated to the Frame shape $2^3 6^3$. The eigengroup is $\Gamma_0(6)$; however, there is a multiplier with $\lambda=2$, so we are really interested in the genus 0 fixing group $\hat\Gamma_g=\Gamma_0(12)$. This has cusps at $\infty$, $0$, $1/2$, $1/3$, $1/4$, and $1/6$.
The expansion of $\phi_{e,g}$ at each of these cusps corresponds to the expansion at $\tau\to \infty$ of
\be \phi_{e,g}\ ,\qquad \phi_{g,e}\ ,\qquad \phi_{g^2,g}\ , \qquad \phi_{g^3, g}\ , \qquad \phi_{g^4, g^5}\ , \quad\mbox{and}\quad \phi_{g^6,g}\ ,
\ee respectively. If we remember to account for multipliers, $\Gamma_0(6)$ transformations let us replace these by
\be \phi_{e,g}\ ,\qquad \phi_{g,e}\ ,\qquad \phi_{g^2,g}\ , \qquad \phi_{g^3, g}\ , \qquad -\phi_{g^2, g}\ , \quad\mbox{and}\quad -\phi_{e,g}\ .\label{eq:multiplierTricks1}
\ee
From \eqref{eq:twiningPert}, we have
\be a_0(\infty)=2\ .
\ee The orbifolds by $g$ and $g^3$ are inconsistent, so the twisted sectors associated to these symmetries do not satisfy the level matching condition. By \eqref{levelfail}, this implies that $\phi_{g,e}$ and $\phi_{g^3,g}$ have no term of order $q^0$ and
\be a_0(0)=a_0(1/3)=0\ .
\ee 
Next, we argue that because $g^2$ is a K3 orbifold quantum symmetry, $b_1(1/2)=0$ and
\be a_0(1/2)=0\ .
\ee
We do so by arguing that there are no states in the CFT that could contribute to the coefficient of $q^0y$ in $\phi_{g^2,g}$. Suppose otherwise, towards a contradiction. This coefficient receives contributions from the states in the R-R $g^2$-twisted sector with $L_0=\bar L_0=1/4$, $J_0 =1/2$ and $(-1)^{F_L+F_R}=\pm 1$, i.e. $(-1)^{F_R}=\mp 1$. These states are necessarily $g^2$-invariant, since in the $g^2$-twisted sector the $g^2$ eigenvalue is always the eigenvalue of $e^{2\pi i (L_0-\bar L_0)}$. This means that these states are contained in the orbifold of the K3 model by $g^2$.  By spectral flow to the NS-NS sector, the states with $(-1)^{F_R}=-1$ flow to states with $L_0=\bar L_0=0$, while states with $(-1)^{F_R}=+1$ flow to $(L_0,\bar L_0)=(0,1/2)$. The first case is impossible, because states in the twisted sector cannot have zero conformal weight, by uniqueness of the vacuum. On the other hand, if the orbifold theory contains states with weights $(0,1/2)$, then it must necessarily be an NLSM on $T^4$. But we know that the orbifold by $g^2$ is a K3 model, since we can compute its Witten index. We conclude that there cannot be any states contributing to $b_1(1/2)$. 
Finally, we note that \eqref{eq:multiplierTricks1} gives us $a_0(1/4)=0$ and $a_0(1/6)=-2$ for free. As a check on our work, we note that the sum of residues
\be \sum_{\cc} w_{\cc} a_0(\cc)=2-2=0 \ee
vanishes, as expected. We find a unique twining genus:
\be F_{e,g}(\tau)=-\frac{1}{4}\E_2-\frac{1}{4}\E_3+\frac{1}{6}\E_4+\frac{3}{4}\E_6-\frac{1}{2}\E_{12}\ .\ee
The functions on the right hand side are defined in appendix \ref{sec:modularCusp}.

A K3 NLSM with a symmetry whose Frame shape is $2^36^3$ can be explicitly constructed  as an orbifold of $T^4$ \cite{slowpaper}; as expected, the twining genus one obtains agrees with our above result. Remarkably, this function also agrees with the weight 0, index 1 weak Jacobi form that Umbral ($A_2^{12}$, $A_3^8$, and $A_6^4$) and Conway moonshines associate to this Frame shape (see section \ref{sec:moonshine}).

\section{Second quantized twining genera and 4d physics}\label{sec:secquant}

 In this section, we will derive additional constraints on the twining genera $\phi_{e,g}$ coming from the properties of CHL models. This will enable us to fix $\phi_{e, g}$ in the case where $\Gamma_g$ is genus 1, i.e. fix the coefficient of the nontrivial cusp form of $\Gamma_g$. In particular, with these additional considerations we can fix all $\phi_{e, g}$ associated to 4-plane preserving symmetries of K3 NSLMs.
 
 The line of reasoning is the following. As discussed in section \ref{sec:chl}, given a symmetry $g$ of a K3 NLSM one can consider the corresponding four dimensional CHL model and the generating function $1/\Phi_{g,e}$ of 1/4 BPS degeneracies in this model. As we will review in the next subsection, $\Phi_{g,e}$ is determined in terms of the twining genus $\phi_{e,g}$ (or rather from $\hat\phi_{r,s}$, which also depend on $\phi_{e, g^n}$ for higher powers of $g$) via a multiplicative lift
\be \{\hat\phi_{r,s}\}\qquad \rightarrow \qquad \Phi_{g,e}\ ,
\ee mapping a (vector valued) weak Jacobi form for $SL(2,\ZZ)$ to a Siegel modular form under some subgroup of $Sp(4,\ZZ)$. As described in section \ref{sec:chl}, the 1/4 BPS  degeneracies are computed by taking suitable contour integrals of $1/\Phi_{g,e}$. In particular,  the poles of $1/\Phi_{g,e}$ are related via a precise contour prescription to the decay of 1/4 BPS dyons into a pair of 1/2 BPS particles. 

Now---due to the multiplicative lift, which we will write explicitly below---the aforementioned singular divisor of $\Phi_{g,e}$ is determined by certain Fourier coefficients of $\hat \phi_{r,s}$  (the so-called polar coefficients). We will show that all expected poles, i.e. the ones related to physically meaningful wall-crossing, are completely taken into account by the polar coefficients of $\hat \phi_{0,0}$. The requirement that there are no additional unphysical poles puts strong constraints on the Fourier coefficients of $\hat \phi_{r,s}$. 

We will show that the constraints from wall-crossing, together with the ones from modularity discussed in the previous section, are sufficient to single out a finite set of weak Jacobi forms.


\subsection{BPS degeneracies from twining genera}

In this section, we describe the relation between the generating functions $1/\Phi_{g,e}$ of 1/4 BPS states in CHL models and the twining genera $\phi_{e, g}$ of the corresponding NLSMs on K3 in more detail.

The dyon partition function for the unorbifolded theory was postulated in \cite{DVV} and analogous formulas for the CHL orbifolds were  computed  in \cite{DS, DJS1, DJS2}. Although these references consider only geometric symmetries of the K3 sigma model, the extension to more general $g$ presents only minor technical modifications. This subsection is essentially a reformulation of these results in order to include this general case.
 
 We are interested in counting the microstates for the set of charges described in section \ref{sec:chl} which correspond,  in the type IIB frame, to a D1-D5 system with momentum  in a KK-monopole background. At weak coupling in the type IIB frame, the function $\frac{1}{\Phi_{g,e}}$ is the product of three contributions:  
\be \frac{1}{\Phi_{g, e}(\Omega)}=Z_{D1}(p,q,y)Z_{KK}(q)Z_{CM}(q,y),\quad \Omega=\twoMatrix{\sigma}{z}{z}{\tau}.\ee
$Z_{D1}$ counts the states associated with the worldvolume of the D1-D5 bound state, $Z_{KK}$ is the contribution associated with a KK monopole with momentum, and $Z_{CM}$ counts states associated with the center of mass of the D1-D5 system in the Taub-NUT background. We can evaluate $Z_{D1}$ by noting that, in the limit where the volume of the K3 is small compared to the radius of $S^1$, the effective worldvolume theory describing the bound state of a D5-brane and  $m+1$ D1-branes is (a deformation of) the symmetric product $\text{Sym}^{m+1} K3$ obtained by orbifolding the $(m+1)$-fold product of the K3 sigma model by the symmetric group. Any symmetry $g$ of the original NLSM induces a symmetry of the $n$-th symmetric product, so that one can define the $g$-twining and $g$-twisted genera $\phi_{e,g}^{\Sym^n K3}$ and $\phi_{g,e}^{\Sym^n K3}$ in each of these CFTs.\footnote{More generally, one can defined twisted-twining genera $\phi_{g,h}^{\Sym^n K3}$ for any commuting pair of symmetries $g,h$. We will only focus on the cases $(g,e)$ and $(e,g)$.} These functions are Jacobi forms of index $n$ under suitable congruence subgroups of $SL(2,\ZZ)$. The contribution $Z_{D1}$ is essentially the generating function for all twisted genera $\phi_{g,e}^{\Sym^n K3}$, namely
\be Z_{D1}(p,q,y)=p^{-1}\Psi_{g,e}(\begin{smallmatrix}
\sigma & z \\ z & \tau
\end{smallmatrix})=\sum_{m=-1}^\infty p^{m} \phi_{g,e}^{\text{Sym}^{m+1} K3}(\tau,z)\ .\ee
The function $\Psi_{g,e}$ is known as the second quantized elliptic genus \cite{DMVV}. This may seem like it will be unwieldy to deal with, as it involves computations in an infinite tower of CFTs, but fortunately a remarkable identity allows us to compute this function by only studying the original K3 sigma model \cite{DMVV}. Specifically, we have
\be \Psi_{g,e}(\begin{smallmatrix}
\sigma & z \\ z & \tau
\end{smallmatrix})=\prod_{m=1}^{\infty}\prod_{n=0}^\infty\prod_{l\in \ZZ}(1-p^m q^{\frac{n}{N\lambda}}y^l)^{-\hat c^g_{m,n}(\frac{mn}{N\lambda},l)}\ ,
\ee
where $\hat c^g_{m, n}$ are the Fourier coefficients of the functions $\hat \phi_{m, n}$ (see eq.\eqref{eq:cHatPolarity}). The next factor, $Z_{KK}$, is easier to deal with: a chain of dualities relates BPS KK monopoles with momentum in type IIB to perturbative heterotic left-movers -- that is, the 1/2 BPS states we discussed in section \ref{sec:BPScount}. In particular, one KK monopole along $\hat S^1$ and $-n/N\lambda$ units of momentum along $S^1$ in type IIB get mapped to a fundamental heterotic string with winding $1$ and with momentum $-n/N\lambda$ along $S^1$. The generating function for the multiplicity of these states is the partition function for $24$ bosonic oscillators in the $g$-twisted sector -- recall that, in CHL models, states with winding number $w\modu{N\lambda}$ along $S^1$ belong to the $\hat g^w$-twisted sector. Taking into account the ground level of the twisted sector, the partition function is \cite{Sen:2005bh,Dabholkar:2005dt,Govindarajan:2010fu}
\begin{align}\label{eq:halfTwisted}
Z_{KK}=&q^{-\frac{1}{24}\sum_{a|N}  \frac{m(a)}{a}}\prod_{i=1}^{24}\prod_{n=1}^\infty(1-q^{r_i+n})^{-1}
=q^{-\frac{A}{24N\lambda }}\prod_{n=1}^\infty(1-q^{\frac{n}{N\lambda}})^{-\sum_{l\in\ZZ}\hat c^g_{0,n}(0,l)}  \ .
\end{align} 
 Here, $\prod_{a|N} a^{m(a)}$ is the Frame shape of $g$, $r_1,\ldots,r_{24}$ are rational numbers with  $0\le r_i<1$  such that $e^{2\pi i r_i}$ are the eigenvalues of $g$ in the $24$-dimensional representation, and $\sum_{\ell\in\ZZ} \hat c^g_{0,n}(0,\ell) $ is the multiplicity of the eigenvalue $e^{\frac{2\pi i n}{N\lambda}}$ (which, by \eqref{emptyspace}, vanishes unless $n\equiv 0\modu \lambda$). Furthermore, the constant
\be A= \sum_{a|N} m(a) \frac{N\lambda}{a}= \sum_{m=0}^{N\lambda-1}\sum_{\ell\in\ZZ} \hat c^g_{m,0}(0,\ell)=\begin{cases}24 & \text{if  $g^\lambda$-orbifold is K3 NLSM}\\
0 & \text{if $g^\lambda$-orbifold is $T^4$ NLSM}\end{cases},
\ee  is the Witten index of the $g^\lambda$-orbifold (see \cite{PV} for a proof of these identities).
The computation of $Z_{CM}$ is slightly more complicated, so since the derivation of \cite{DS} applies directly to the case of a general $g$, we  simply state the result\footnote{cf. also section 4.2 of \cite{Persson:2013xpa}.}:
\be Z_{CM}(q,y) = {1 \over \chi_{-2, 1}(\tau, z)}=\frac{\prod_{n=1}^\infty (1-q^n)^4}{y(1-y^{-1})^2\prod_{n=1}^\infty (1-q^ny)^2 (1-q^ny^{-1})^2}\ .\ee
$\chi_{-2, 1}$ is a standard weak Jacobi form of weight $-2$ and index $1$ defined in appendix \ref{a:basics} and in particular is $g$-independent.\footnote{The argument $\hat\rho$ in \cite{DS} is related to  $\tau$ in our conventions by a rescaling $\hat\rho=N\tau$; this introduces a dependence on the order $N$ of $g$ in \cite{DS}. }  For consistency with the automorphic forms literature, it is convenient to repackage  the factors $Z_{CM}$ and $Z_{KK}$ into a $g$-dependent Jacobi form called $\psi_{g, e}$, defined as
\be \psi_{g,e}(\tau,z) \equiv {1 \over Z_{KK} Z_{CM}} =q^{\frac{A}{24N\lambda}}y\prod_{l\in \ZZ_{<0}} (1-y^l)^{\hat c^g_{0,0}(0,l)} \prod_{n=1}^\infty \prod_{l\in \ZZ}(1-q^{\frac{n}{N\lambda}}y^l)^{\hat c^g_{0,n}(0,l)}\ ,\ee 
where we used  $\sum_{l\in \ZZ_{<0}}\hat c^g_{0,0}(0,l)=\hat c^g_{0,0}(0,-1)=2$. 

To summarize, we find that, up to the automorphic correction $\psi_{g,e}$, the $1/4$-BPS counting function is essentially equal to the second quantized elliptic genus:
\be \frac{1}{\Phi_{g,e}(\Omega)}=\frac{\Psi_{g,e}(\Omega)}{p\psi_{g,e}(\tau,z)}\ .\ee
The factor $p\psi_{g,e}$ is known as the `automorphic correction', so named because it restores the $p \leftrightarrow q$ exchange symmetry characteristic of Siegel modular forms.
More explicitly, we have
\be\label{Phige} \Phi_{g,e}(\begin{smallmatrix}
\sigma & z \\ z & \tau
\end{smallmatrix})= pq^{\frac{A}{24N\lambda}}y\prod_{(m,n,l)>0}  (1-p^m q^{\frac{n}{N\lambda}}y^l)^{\hat c^g_{m,n}(\frac{nm}{N\lambda},l)}\ ,
\ee where  $(m,n, l)>0$ means
\be m,n\in \ZZ_{\ge 0}\qquad \text{and}\qquad \begin{cases} l\in \ZZ_{<0} & \text{if }m=n=0\\
l\in \ZZ & \text{otherwise}\end{cases}\ .
\ee 
In mathematics, infinite products of the form \eqref{Phige} are known as multiplicative lifts and were studied in  \cite{Borcherds1995,Borcherds1998,Gritsenko1998I,Gritsenko1998II}. In general, they are automorphic forms for some congruence subgroup of $Sp(4,\ZZ)$. When $g$ is the identity, we obtain the famous Igusa cusp form $\Phi_{e,e}=\Phi_{10}$ of weight 10 under $Sp(4,\ZZ)$.

\subsection{Wall crossing and poles}\label{sec:wallsAndPoles}

As discussed in \S\ref{sec:chl}, the Fourier coefficients defining the 1/4 BPS multiplicities jump whenever the contour of integration crosses a pole of $\frac{1}{\Phi_{g,e}}$. In this section, we  study the locations of the poles of ${1 \over \Phi_{g,e}}$ that contribute to wall crossing. In particular, we show that only a subset of the potential poles correspond to locations of physical wall crossing. We thus constrain $\frac{1}{\Phi_{g,e}}$ by demanding the nonexistence of any additional poles, which we conjecture are unphysical. As in \cite{CV}, we  restrict our attention to the poles that `intersect the cusp at infinity', i.e. that intersect the region in the Siegel upper half-space where $\Im\Omega$ has very large eigenvalues, so that the product formula \eqref{Phige} converges. As \eqref{contour} demonstrates, this is the region that is relevant for extracting 1/4 BPS state degeneracies. For simplicity, in this section we only consider the case $\lambda=1$, i.e. we assume that the orbifold of the K3 NLSM by $g$ satisfies level-matching; the case $\lambda>1$ is described in appendix \ref{a:mess}.

The divisors that intersect the cusp at infinity for $\Phi_{g,e}$ are clear from the product formulas  \eqref{Phige} (see \cite{Borcherds1998} for a rigorous proof). The possible zeroes or poles of $\Phi_{g, e}$ are given by
\be\label{polegen} m\sigma+ n\frac{\tau}{N}+lz =k\ ,
\ee for $m,n,l,k\in \ZZ$ with $4\frac{mn}{N}-l^2<0$ (this is a necessary and sufficient condition for \eqref{polegen} to have a solution in the Siegel upper half-space), and  the multiplicity is $\hat c^g_{m,n}(\frac{mn}{N},l)\equiv \hat c^g_{m,n,l}(4\frac{mn}{N}-l^2)$. Of course, whether the divisor is a zero or a pole depends on the sign of $\hat c^g_{m,n,l}(4\frac{mn}{N}-l^2)$.  

A special subset of poles is the one given by $m\equiv 0\modu N$. In this case, the only non-vanishing polar coefficient (i.e. with negative discriminant) is $\hat c^g_{0,0,1}(-1)=2$, so that $n$ must also be a multiple of $N$. Therefore, $\Phi_{g,e}$ has double zeroes (hence, ${1 \over \Phi_{g, e}}$ has double poles) at
\be\label{polespec} Nr\sigma+s\tau+lz =k\ ,
\ee for $r,s,l,k\in \ZZ$ with $4Nrs-l^2=-1$. Notice that we have set $m=Nr$ and $n=Ns$. This subset of poles exists for any $g$ of order $N$.

Besides the poles of the form \eqref{polespec}, we  have additional potential poles for ${1 \over \Phi_{g,e}}$  if some $\hat c^g_{m,n,l}(4\frac{mn}{N}-l^2)>0$ for some $4\frac{mn}{N}-l^2<0$ with $m\neq 0\modu N$. In particular, if  for $\phi_{g,e}$ the Fourier coefficient relative to $q^{s/N}y$ is positive, for some $s/N<1/4$, then there is a pole of order $\hat c^g_{1,s,1}(s/N-1/4)=\hat c_{1,s}(s/N,1)>0$ corresponding to $m=1$, $n=s$, $l=1$ with equation
\be\label{eq:badpoles} \sigma+s\frac{\tau}{N}+z =k\ .
\ee for $s,k\in \ZZ$, $s/N<1/4$. This is \emph{not} of  the form \eqref{polespec}.\footnote{Naively, rescaling \eqref{eq:badpoles} by $N$ gives an equation of the form \eqref{polespec}. However, it is easy to see that the coefficients of the resulting equation do not satisfy the condition $4Nrs-l^2=-1$.}

As described in section \ref{sec:chl}, the degeneracy \eqref{eq:partfun} of 1/4 BPS dyons `jumps' whenever the integration contour \eqref{contour} crosses one of the poles of $1/\Phi_{g,e}$. The contour comprises a full period of the real part $\Re\Omega$ of the arguments, at a fixed value of their imaginary part $\Im\Omega=\epsilon^{-1} \Z$. Therefore, a necessary and sufficient condition for the contour to cross the pole of $1/\Phi_{g,e}$ is that $\Z$ satisfy the imaginary part of \eqref{polegen}, i.e.
\be (\Z,\left(\begin{matrix}
2n/N & -l\\ -l & 2m
\end{matrix}\right))=0\ .
\ee
On the other hand, as discussed in section \ref{sec:chl}  (see also \cite{sen2007walls, Cheng:2008kt}), `physical' wall-crossing is only expected for those values of the moduli where 1/4 BPS dyons can decay into a pair of 1/2 BPS states, namely for 
\be (\Z,\alpha)=0\ , 
\ee for the matrices 
\be\label{alphaagain} \alpha=\twoMatrix{2bd}{-(ad+bc) }{-(ad+bc) }{ 2ac}, \qquad \text{for }a,b,c,d\in \ZZ,\ ad-bc=1,\ ac\in N\ZZ\ ,
\ee
 given in eq.\eqref{eq:alphaMat}.


\bigskip


We will now show that the walls corresponding to the subset of poles \eqref{polespec}, i.e.
\be\label{poleeq} (\Z,\left(\begin{matrix}
2s & -l\\ -l & 2Nr
\end{matrix}\right))=0,\qquad  \text{for } r,s,l\in \ZZ,\ 4Nrs-l^2=-1\ ,
\ee 
 are in \textit{one to one} correspondence with the locations of the `physical' domain walls labeled by \eqref{alphaagain}. This implies that all the other potential poles, and in particular \eqref{eq:badpoles}, never arise in a  function $1/\Phi_{g,e}$ counting 1/4 BPS dyons in a CHL model. This argument puts strong constaints on the Fourier coefficients of $\hat \phi_{r,s}$ that will be discussed in the next section.


Let us consider a wall located at $(\Z,\alpha)=0$, where $\alpha$ is as in \eqref{alphaagain}.
For convenience, we can define
$$s=bd,\quad l=ad+bc,\quad Nr=ac\ ,$$
so that the equation of the wall becomes of the form \eqref{poleeq}, with
$$4Nrs-l^2=4acbd-(ad+bc)^2=4abcd-(ad)^2-(bc)^2-2abcd=-(ad-bc)^2=-1.$$
This shows one of the required directions of our argument: that every wall is a solution to \eqref{poleeq}.

Next, we show the reverse direction. Consider a pole of the form \eqref{polespec}, and the corresponding wall with equation \eqref{poleeq}, labeled by some $r,s,l\in \ZZ$ satisfying $4Nrs-l^2=-1$. We can trivially rewrite the latter equation as $4Nrs=l^2-1=(l+1)(l-1).$ Since $l$ is odd, both $l+1$ and $l-1$ are multiples of two. Thus, $Nrs=\frac{l+1}{2}\cdot\frac{l-1}{2}$ is a factorization of $Nrs$ as a product of consecutive integers (in particular, these factors are coprime). Make the following definitions for convenience:
\be\label{abcd} t=\frac{l-1}{2},\quad a=\text{ gcd }(Nr,t+1),\quad c=\text{ gcd }(Nr,t),\quad b=\text{ gcd }(s,t),\quad d=\text{ gcd }(s,t+1).\ee
Thinking for a moment about factors (and remembering that $\text{ gcd }(t,t+1)=1$) demonstrates the following facts:
$$ad=t+1,\quad bc=t\Rightarrow ad-bc=1.$$
$$ac=Nr,\quad bd=s\quad ad+bc=2t+1=l.$$
Thus, any pole \eqref{polespec} corresponds to a physical domain wall $(\alpha,\Z)=0$, with $\alpha$ as in \eqref{alphaagain} and $a,b,c,d$ as in \eqref{abcd}.

In summary, we indeed have a one-to-one correspondence between poles \eqref{poleeq} and walls corresponding to physically meaningful decay channels. Analogous results hold for the case where $\lambda>1$ (see appendix \ref{a:mess}): the physical domain walls are in one to one correspondence with a special set of poles corresponding to the $g^m$-twisted sector for $m\equiv 0\modu N$. Therefore, a physically consistent $1/\Phi_{g,e}$ cannot have any other pole related to the $g^m$-twisted sectors for $m\neq 0\modu N$.  In the next subsection, we will put constraints on (the sign of) $ \hat{c}^g_{m,n,l}(4\frac{mn}{N\lambda}-l^2)$ to eliminate the unphysical poles. 
 
A remark: notice that we could have run the same analysis for $\Phi_{e, g}$ or more general $\Phi_{g, h}$. In fact, one can easily show that the walls corresponding to the poles of ${1 \over \Phi_{e, g}}$ are precisely those of ${1 \over \Phi_{e, e}}$, as expected since ${1 \over \Phi_{e, g}}$ counts ($g$-equivariant) dyons in the unorbifolded model. Similarly, we expect that the constraints one may derive from studying the poles of ${1 \over \Phi_{g, h}}$ correspond to restricting to the set of physical walls obtained already for ${1 \over \Phi_{g, e}}$.

\subsection{Constraints on twining genera from wall crossing}

From the previous subsection, we learned that the physically meaningful walls of marginal stability are in one-to-one correspondence with the poles of $1/\Phi_{g,e}$ associated with the Fourier coefficient $\hat c^g_{0,0,1}(-1)$ of $\hat\phi_{0,0}$ (or, more generally, with the Fourier coefficient $\hat c^g_{Nr,-\E'Nr,1}(-1)$ of $\hat\phi_{Nr,-\E'Nr}$). The other potential poles, associated with the coefficients $\hat c^g_{m,n,l}(4\frac{mn}{N\lambda}-l^2)$ with $4\frac{mn}{N\lambda}-l^2<0$ and $m\neq 0\modu N$, are unphysical, in the sense that they do not correspond to any instability of 1/4 BPS dyons. 

We recall that the correspondence between walls of marginal stability and poles of $\Phi_{g,e}$ is based on the assumption that the degeneracy of 1/4 BPS states in CHL models is always recovered by a contour integral of $\Phi_{g,e}$ where the contour is given by the standard prescription. We refer to this as the \emph{standard contour assumption}. Thus, we have

\begin{claim} Let $g$ be a symmetry of a NLSM on K3. Under the standard contour assumption, the Fourier transformed twisted-twining genera $\hat \phi_{j,k}=\sum_{n,l} \hat c^g_{j,k}(n,l) q^ny^l$ for $j\neq 0\modu N$ have no positive polar Fourier coefficients, i.e.
\be \hat c^g_{j,k}(n,l)\le 0 \qquad \forall j,k,n,l\text{ with } j\neq 0\modu N,\ 4n-l^2<0\ .
\ee
\label{claim:strong}
\end{claim} Notice that $\hat c^g_{j,k}(n,l)$ is non-zero only when $n\equiv \frac{jk}{N\lambda}\modu \ZZ$. This follows from the fact that $g^j$ coincides $e^{2\pi i(L_0-\bar L_0)}$ in the $g^j$-twisted sector. On the other hand, for the states contributing to $\hat \phi_{j,k}$ one has $g=e^{2\pi i \frac{k}{N\lambda}}$ by definition, so that $e^{2\pi i(L_0-\bar L_0)}=g^j=e^{2\pi i \frac{jk}{N\lambda}}$.

The following (strictly weaker) corollary is often easier to utilize and will suffice for our purposes:
\begin{corollary}\label{cor:useful} Under the standard contour assumption, the twisted genus $\phi_{g,e}=\sum_{n,l} c_{g,e}(n,l) q^ny^l$ has no positive Fourier coefficients with $n<1/4$ and $l=\pm 1$, i.e.
\be c_{g,e}(n,\pm 1)\le 0\qquad \forall n<1/4\ .
\ee
\end{corollary}
This follows immediately by noticing that for $n=\frac{s}{N\lambda}$, one has 
\be c_{g,e}(n,\pm 1)=\hat c^g_{1,s,1}(4\frac{s}{N\lambda}-1)\ ,\ee so if $c_{g,e}(n,\pm 1)>0$ for some $n=\frac{s}{N\lambda}<\frac{1}{4}$, then $\hat \phi_{1,s}$ has a positive polar coefficient.

\bigskip

In the following section we will loosely refer to $\phi_{g, e}$ as the S transform of $\phi_{e, g}$, for succinctness.

\subsection{An example: Frame shape $1^2 11^2$}\label{sec:g1Ex}

As an example, let us consider a symmetry $g$ with Frame shape $1^211^2$. The fixing group $\Gamma_g=\hat\Gamma_g=\Gamma_0(11)$ has two cusps (at $\infty$ and $0$) and genus $1$. The expansion of $\phi_{e,g}$ at each of these cusps corresponds to the expansion at $\tau\to \infty$ of
\be \phi_{e,g} \quad\mbox{and}\quad \phi_{g,e}\ ,
\ee respectively.  From \eqref{eq:twiningPert}, we have
\be a_0(\infty)=2-\frac{2}{12}=\frac{11}{6}\ .
\ee Next, we argue as in \S \ref{sec:g0Ex} that because $g$ is a K3 orbifold quantum symmetry, $b_1(0)=0$ and
\be a_0(0)=-\frac{1}{6}\ .
\ee
As a check, we note that the sum of residues
\be \sum_{\cc} w_{\cc} a_0(\cc)=\frac{11}{6}+11\cdot\parens{-\frac{1}{6}}=0 \ee
vanishes, as expected.

The techniques of appendix \ref{sec:modularCusp} determine the twining genus, up to the addition of a cusp form proportional to $\eta[1^211^2]$:
\be F_{e,g}(\tau)=-\frac{11}{60}\E_{11} + const\times  \eta[1^2 11^2]\ . \ee
Writing the unknown constant as $11(\alpha-2/5)$ and $S$ transforming yields
\be \phi_{g,e}(\tau,z)=2+\alpha(-y-y^{-1}+2)q^{1/11}+2(\alpha-1)(y+y^{-1}-2)q^{2/11}+O(q^{3/11})  \ .\ee
(The coefficient of the cusp form was chosen so that the above $q$-expansion had integral coefficients when $\alpha$ was integral. We know that $\phi_{g,e}$ has integral coefficients because it is an untwined trace). Requiring all polar coefficients (that is, $y^{\pm 1}q^n$ coefficients with $n<1/4$) to be nonpositive gives
\be 0\le \alpha\le 1\ .\ee
Since $\alpha$ must be integral, this gives two twining genera. The $\alpha=0$ case yields the weak Jacobi form associated with $M_{24}$ moonshine, while the $\alpha=1$ case is associated with $2.M_{12}$ moonshine (cf. the Introduction and section \ref{sec:moonshine}) \cite{Cheng:2014zpa}. Amusingly, the $2.M_{12}$ function was found in an explicit K3 NLSM (more precisely, a UV Landau-Ginzburg orbifold description) in \cite{cheng2015landau}.

\section{Determining the genera}\label{sec:determining}

We now explain how to use the constraints explained above on twining genera in order to determine all possible twining genera of K3 NLSM symmetries. We begin with two simpler cases; we then proceed to the general case, which uses many ideas from the first two cases. We conclude this section with two tables: one outlines the calculation of all twining genera, and the other presents the complete set of possible twining genera.

\subsection{Pure K3 symmetries}

Let $g$ be a symmetry of a nonlinear sigma model on K3 and suppose that the orbifold of the NLSM by any power of $g$ is either inconsistent or a K3 model. We call such a $g$ a `pure K3 symmetry'. A case by case analysis shows that this case occurs exactly when the symmetry acts as a permutation on the $24$ dimensional representation, i.e. when the Frame shape $\prod_{a|N} a^{m(a)}$ contains only non-negative powers $m(a)\ge 0$.

The reasoning exemplified in sections \ref{sec:g0Ex} and \ref{sec:g1Ex} suffices to compute the twining genera of all pure K3 symmetries. Twisted twining genera $\phi_{g^i,g^j}$ with $i\not\equiv0 \modu N$ have no $q^0y^{\pm 1}$ terms, since states counted by the coefficient of $q^0y$ would spectral flow to states that cannot exist in a $g^i$-twisted sector. All other twisted-twining genera are related by a multiplier to $\phi_{e,g^j}$, which we can easily deduce from the Frame shape of $g^j$ (see \eqref{eq:twiningPert}). This information suffices to deduce the leading terms, $a_0(\cc)$, in the $q$-expansions of $F_{e,g}$ about all cusps, $\cc$. There are either one or two sets of values $\{a_0(\cc)\}$, corresponding to the cases where there are one or two multipliers. If $\hat\Gamma_g$ has genus 0, then the function(s) $F_{e,g}$ is (are) determined; otherwise, we are allowed to add a cusp form, the options for which are determined as in \S \ref{sec:g1Ex}. However, we note one subtlety in the genus 1 case when there are two distinct multipliers: we are only allowed to add the cusp form when its multiplier agrees with the multiplier we have chosen. (We mention this issue here because it only happens to arise for pure K3 symmetries -- in particular, those with Frame shapes $4^28^2$ and $6^4$. In fact, $\hat\Gamma_g$ is only ever genus 1 when $g$ is a pure K3 symmetry).

\subsection{Quantum symmetries in toroidal orbifolds}\label{sec:quantTorus}

If a K3 NLSM is the orbifold of a NLSM on $T^4$ by a cyclic group, then it has a quantum symmetry $Q$ (see section \ref{sec:K3}). Twining genera of quantum symmetries of toroidal orbifolds can be computed using the following formula
\be\label{twinQ} \phi^{K3}_{e,Q}(\tau,z)=\frac{1}{N}\sum_{j,k=1}^N e^{\frac{2\pi i j}{N}} \phi^{T^4}_{g^j,g^k}(\tau,z)\ ,
\ee
where, generically,
\be\label{T4twin} \phi^{T^4}_{g^j,g^k}(\tau,z)=(\zeta_L^n+\zeta_L^{-n}-2)(\zeta_R^{n}+\zeta_R^{-n}-2)\frac{\vartheta_1(\tau,z+r_L(j\tau+k))\vartheta_1(\tau,z-r_L(j\tau+k))}{\vartheta_1(\tau,r_L(j\tau+k))\vartheta_1(\tau,-r_L(j\tau+k))}\ ,
\ee are the twisted twining genera of the corresponding $T^4$ model \cite{Volpato:2014zla}. 
Here, $n=\gcd(j,k,N)$,
\be \zeta_L=e^{2\pi i r_L},\qquad \zeta_R=e^{2\pi i r_R}\ ,
\ee and the possible values of $r_L,r_R\in \frac{1}{N}\ZZ/\ZZ$ are given in table \ref{t:Qtwin}. (Formula \eqref{T4twin} needs to be modified when $nr_L\in \ZZ$ and $nr_R \notin \ZZ$, see \cite{Volpato:2014zla} and \cite{slowpaper} for more details).
More generally, one has 
\be\label{twisttwinQ} \phi^{K3}_{Q^a,Q^b}(\tau,z)=\frac{1}{N}\sum_{j,k=1}^N e^{\frac{2\pi i bj}{N}} e^{-\frac{2\pi i ak}{N}}\phi^{T^4}_{g^j,g^k}(\tau,z)\ .
\ee
For the $q^0$ term, one has (even when $nr_L\in\ZZ$ and $n r_R\not\in\ZZ$) 
\be \phi^{T^4}_{g^j,g^k}(\tau,z)_{\rvert q^0}=\begin{cases} (\zeta_L^n+\zeta_L^{-n}-2)(\zeta_R^{n}+\zeta_R^{-n}-2) &\text{for }jr_L\not\in \ZZ\ ,\\
(\zeta_L^n+\zeta_L^{-n}-2)(\zeta_R^{n}+\zeta_R^{-n}-2)\frac{\zeta_L^{k}+\zeta_L^{-k}-(y+y^{-1})}{\zeta_L^{k}+\zeta_L^{-k}-2} &\text{for }jr_L\in \ZZ, nr_L\not\in\ZZ\ ,\\
-\frac{1}{2}(\zeta_R^{n}+\zeta_R^{-n}-2)(2-y-y^{-1}) &\text{for }(nr_L,nr_R)=(0,\frac{1}{2})\modu \ZZ,  N\nmid j\ ,\\
-\frac{1}{3}(\zeta_R^{n}+\zeta_R^{-n}-2)(2-y-y^{-1}) &\text{for }(nr_L,nr_R)=(0,\frac{1}{3})\modu \ZZ,  N\nmid j\ ,\\
(\zeta_R^{n}+\zeta_R^{-n}-2)(2-y-y^{-1}) & \text{otherwise.}
\end{cases}
\ee Plugging this into \eqref{twisttwinQ} then yields the leading behavior of $F_{e,g}$ at each cusp; this allows us to expand $F_{e,g}$ in the $M_2(\hat\Gamma_g)$ basis described in appendix \ref{sec:modularCusp}. (Whenever $g$ is the quantum symmetry of a torus orbifold, $\hat\Gamma_g$ has genus 0 and the multiplier is trivial).
\begin{table}[h]
\centering
\begin{tabular}{|c|c|c|c|}
\hline
$r_L$ & $r_R$ & $\pi_Q$ & w-s parity\\
\hline
 $1/2$ & $1/2$ & $1^{-8}2^{16}$ & $\circ$\\
\hline
 $1/3$ & $1/3$ & $1^{-3}3^{9}$& $\circ$\\
\hline
 $1/4$ & $1/4$ & $1^{-4}2^{6}4^4$& $\circ$\\
\hline
 $1/6$ & $1/6$& $1^{-4}2^{5}3^46^1$& $\circ$\\
\hline
 $1/5$ & $2/5$& \multirow{2}{*}{$1^{-1}5^{5}$}& \multirow{2}{*}{$\updownarrow$}\\
 $2/5$ & $1/5$&  & \\
\hline
 $1/4$ & $1/2$ &  \multirow{2}{*}{$2^{-4}4^{8}$}& \multirow{2}{*}{$\updownarrow$}\\
 $1/2$ & $1/4$ & &\\
\hline
 $1/6$ & $1/2$ &  \multirow{2}{*}{ $1^{-2}2^{4}3^{-2}6^4$}& \multirow{2}{*}{$\updownarrow$}\\
 $1/2$ & $1/6$ & &\\
\hline
 $1/6$ & $1/3$ & \multirow{2}{*}{ $1^{-1}2^{-1}3^36^3$}& \multirow{2}{*}{$\updownarrow$}\\
 $1/3$ & $1/6$ & &\\
\hline
 $1/8$ & $5/8$ & \multirow{2}{*}{ $1^{-2}2^{3}4^18^2$}& \multirow{2}{*}{$\updownarrow$}\\
 $5/8$ & $1/8$ & &\\
\hline
 $1/10$ & $3/10$ & \multirow{2}{*}{ $1^{-2}2^{3}5^210^1$}& \multirow{2}{*}{$\updownarrow$}\\
 $3/10$ & $1/10$ & &\\
\hline
 $1/12$ & $5/12$ & \multirow{2}{*}{ $1^{-2}2^{2}3^24^112^1$}& \multirow{2}{*}{$\updownarrow$}\\
 $5/12$ & $1/12$ & &\\
\hline
\end{tabular}
\caption{Frame shapes corresponding to quantum symmetries of torus orbifolds. The twining genera can be obtained by applying formulae \eqref{twinQ} and \eqref{T4twin}. The last column reports whether world sheet parity fixes the twining genus for a quantum symmetry (symbol $\circ$) or if it relates two of them (symbol $\updownarrow$).}\label{t:Qtwin}
\end{table}

\subsection{General case}


We now explain how to compute the $q^0$ term of a general twisted-twining genus, $\phi_{g^i,g^j}$.
We distinguish between three cases (the reasoning in the first two of which is copied from sections \ref{sec:g0Ex} and \ref{sec:g1Ex}):
\begin{itemize}
\item Suppose $g$ has a non-trivial multiplier of order $\lambda>1$ and that $i$ is \emph{not} a multiple of $\lambda$. Then, the $g^i$-twisted sector does not satisfy the level-matching condition, so that $\phi_{g^i,g^j}$ has no term of order $q^0$ and $b_1(\cc)=b_2(\cc)=a_0(\cc)=0$ (note that $\lambda>1$ implies $\Tr_{\bf 24}(g)=0$).
\item Suppose $i$ is a multiple of $\lambda$ (but not of $N$), so that the orbifold by $g^i$ is consistent, and suppose that this orbifold is a K3 sigma model. Then, the $g^i$-twisted sector cannot contain any R-R states with $L_0=\bar L_0=\frac{1}{4}$ and $J_0=\pm 1/2$, because spectral flow to the NS-NS sector would lead either to an additional vacuum or to states with weights $(0,1/2)$. The latter are not contained in the orbifold K3 model (and the former are forbidden in the twisted sector of any orbifold). Since there are no such states, there cannot be any contribution to the $q^0y$ term in $\phi_{g^i,g^j}$, for any $j$. We conclude that $b_1(\cc)=0$ and $a_0(\cc)=-\Tr_{\bf 24}(g)/12$.
\end{itemize}
These first two bullet points may be summarized succinctly as follows: if the expansion of $\phi_{g^i,g^j}$ about $\infty$ corresponds to the expansion of $\phi_{e,g}$ about the cusp $\cc$, and if $g^i$ is not the quantum symmetry of a torus orbifold, then $b_1(\cc)=0$ and $a_0(\cc)=-\Tr_{\bf 24}(g)/12$.
\begin{itemize}
\item The remaining case is when $g^i$ is the quantum symmetry of a torus orbifold, so the orbifold by $g^i$ is a NLSM on $T^4$. (Consistency of this orbifold implies that $i$ is a multiple of $\lambda$). This is the most complicated case. It is convenient to first compute $\phi_{g^i,e}$ (using the formulae of the previous section) to learn how many right-moving ground states with $(L_0,J_0)=(1/4,1/2)$ are contained in the $g^i$-twisted R-R sector. Then, one should try to deduce the action of $g$ on these states.
\end{itemize}
The rest of this section is devoted to working through a few examples of the reasoning described in the last case.

\subsubsection{$1^82^{-8}4^8$}
We first work out the example of the Frame shape $1^82^{-8}4^8$, whose twining genus is unknown. The fixing group is the genus 0 group $\Gamma_g=\Gamma_0(4)$, which has cusps at $\infty$, $0$, and $1/2$. The expansion of $\phi_{e,g}$ at each of these cusps corresponds to the expansion at $\tau\to \infty$ of
\be \phi_{e,g}\ ,\qquad \phi_{g,e}\ , \quad\mbox{and}\quad \phi_{g^2,g}\ ,
\ee respectively. From \eqref{eq:twiningPert}, we have
\be a_0(\infty)=2-\frac{8}{12}=\frac{4}{3}\ .
\ee Since $g$ is a K3 orbifold quantum symmetry, $b_1(0)=0$ and
\be a_0(0)=-\frac{2}{3}\ .
\ee 
This is sufficient to fix $F_{e,g}(\tau)$ and the twining genus. Explicitly,
\be \phi_{e,g}=\frac{8}{12}\phi_{0,1}-\frac{4}{3}\phi_{-2,1}\E_2\ ,
\ee where $\E_2$ is defined in appendix \ref{sec:modularCusp}.
 However, as an exercise, let us consider also the expansion of $\phi_{g^2,g}$.
In order to calculate $b_1(1/2)$, we need to know the action of $g$ on the ground states of the $g^2$-twisted sector. This requires a bit of effort because the orbifold by $g^2$ (Frame shape $1^{-8}2^{16}$) is an NLSM on $T^4$.
The formulae of \S \ref{sec:quantTorus} yield
\be \phi_{g^2,e}=-2y-2y^{-1}-4+O(q^{1/2})\ .
\ee
The $q^0y$ coefficient tells us that the $g^2$-twisted sector has two R-R ground states that spectral flow to NS-NS fields with weight $(0,1/2)$. (Note that there are exactly two such states; that is, there are no states making positive contributions to the $q^0y$ coefficient, since they would spectral flow to twisted sector NS-NS vacua). These R-R states are $g^2$-invariant, since in the $g^2$-twisted sector the $g^2$ eigenvalue is always the eigenvalue of $e^{2\pi i (L_0-\bar L_0)}$. If these states were also $g$-invariant, they would be present in the orbifold of the model by $g$; since we know that this orbifold is a K3 model,  this cannot happen. We conclude that $g$ acts non-trivially on these two fields, which means by a minus sign, since $g^2$ acts trivially on them. Therefore,
\be \phi_{g^2,g}(\tau,z)=2y+2y^{-1}+4+O(q)\ ,
\ee so that $b_1(1/2)=2$ and $a_0(1/2)=2-\frac{2}{3}=\frac{4}{3}$. As a check, notice that
\be \sum_{\text{cusps }\cc} w_\cc a_0(\cc)=a_0(\infty)+4a_0(0)+a_0(1/2)=\frac{4}{3}+4\cdot\parens{-\frac{2}{3}}+\frac{4}{3}=0\ ,
\ee so that the sum over the residues vanishes, as expected. 
The fact that $a_0(\infty)=a_0(1/2)$ implies that $\phi_{e,g}$ is actually modular under $\Gamma_0(2)$ rather than $\Gamma_0(4)$; this is an accident.

\subsubsection{$1^42^{-2}4^{-2}8^4$}
Next, we work out the example of the Frame shape $1^42^{-2}4^{-2}8^4$; this Frame shape is expected to have two twining genera which are related by worldsheet parity, but neither of them is known. The fixing group is the genus 0 group $\Gamma_g=\Gamma_0(8)$, which has cusps at $\infty$, $0$, $1/2$, and $1/4$. The expansion of $\phi_{e,g}$ at each of these cusps corresponds to the expansion at $\tau\to \infty$ of
\be \phi_{e,g}\ ,\qquad \phi_{g,e}\ ,\qquad \phi_{g^2,g}\ , \quad\mbox{and}\quad \phi_{g^4,g}\ ,
\ee respectively. From \eqref{eq:twiningPert}, we have
\be a_0(\infty)=2-\frac{4}{12}=\frac{5}{3}\ .
\ee Since $g$ is not a torus orbifold quantum symmetry, $b_1(0)=0$ and
\be a_0(0)=-\frac{1}{3}\ .
\ee The other two genera require a bit more effort, as $g^2$ (Frame shape $2^{-4}4^8$) and $g^4$ (Frame shape $1^{-8}2^{16}$) are quantum symmetries of torus orbifolds. (As usual, we do not actually need the last case to fix $\phi_{e,g}$, but we use it to check our work). We begin with $\phi_{g^2,g}$. The formulae of \S \ref{sec:quantTorus} yield two possibilities
\be \phi_{g^2,e}=0+O(q^{1/4})\qquad\mbox{ or } \qquad\phi_{g^2,e}=2-1/y-y+O(q^{1/4}),\ee related by worldsheet parity.
In the former case, there are no $g^2$-twisted R-R ground states with $J_0=1/2$, so $b_1(1/2)=0$ and $a_0(1/2)=-1/3$. In the latter case, we find such a state; it is $g^2$-invariant, but not $g$-invariant (since the orbifold by $g$ gives a K3 sigma model). Thus, $g$ acts on this state as $-1$, yielding $b_1(1/2)=1$ and $a_0(1/2)=2/3$. We now proceed to determine $\phi_{g^4,g}$. We begin with
\be \phi_{g^4,e}=-2y-2/y-4+O(q^{1/2}).\ee
This indicates the existence of two $g^4$-invariant R-R ground states with $J_0=1/2$ in the $g^4$-twisted sector that are $g$-variant. If these states are not $g^2$-invariant, then their $g$ eigenvalues are $\pm i$. A $\Gamma_0(8)$ transformation relates $\phi_{g^4,g}$ to $\phi_{g^4,g^3}$, and so $\Tr_{g^4,q^0y}g=\Tr_{g^4,q^0y}g^3$. This rules out the choices $+i,+i$ and $-i,-i$, leaving us only with $\pm i,\mp i$. Thus,
$$\phi_{g^4,g}=4+O(q),$$
and $a_0(1/4)=-1/3$. If these states are $g^2$-invariant, then $g$ acts on them with a minus sign and
$$\phi_{g^4,g}=2y+2/y+O(q);$$
we then have $a_0(1/4)=5/3$. The sum
\be \sum_{\cc} w_{\cc} a_0(\cc)=\frac{5}{3}+8\cdot\parens{-\frac{1}{3}}+2\cdot\column{-1/3}{2/3}+\column{5/3}{-1/3}=0 \ee
vanishes, as expected; in addition, it tells us how the two cases around the cusps $1/2$ and $1/4$ match up. The twining genera in these two cases are specified by
\be F_{e,g}(\tau)=\frac{1}{3}\E_2-\frac{2}{3}\E_4 \ , \ee
and
\be F_{e,g}(\tau)=-\frac{5}{6}\E_2+\frac{1}{2}\E_4-\frac{1}{3}\E_8 \ . \ee

\subsubsection{$2^44^{-4}8^4$}
Finally, we work out the example of the Frame shape $2^44^{-4}8^4$; this Frame shape is expected to have two twining genera which are not related by worldsheet parity. One is known (it is the function denoted by $\phi_{\T H_a}$ in eq.(3.17) of \cite{Gaberdiel:2011fg}), while the other is not. The eigengroup is the genus 0 group $\Gamma_g=\Gamma_0(8)$. However, there is a multiplier, $\lambda=2$, so the fixing group is $\hat\Gamma_g=\Gamma_0(16)$, which has cusps at $\infty$, $0$, $1/2$, $1/4$, $3/4$, and $1/8$. (This is the only non-pure K3 case with a multiplier). The expansion of $\phi_{e,g}$ at each of these cusps corresponds to the expansion at $\tau\to \infty$ of
\be \phi_{e,g}\ ,\qquad \phi_{g,e}\ ,\qquad \phi_{g^2,g}\ , \qquad  \phi_{g^4,g}\ ,\qquad \phi_{g^4,g^{11}}\ ,\quad\mbox{and}\quad \phi_{g^8,g}\ ,
\ee respectively. If we remember to account for multipliers, $\Gamma_0(8)$ transformations let us replace these by
\be \phi_{e,g}\ ,\qquad \phi_{g,e}\ ,\qquad \phi_{g^2,g}\ , \qquad  \phi_{g^4,g}\ ,\qquad -\phi_{g^4,g}\ ,\quad\mbox{and}\quad -\phi_{e,g}\ . \ee
From \eqref{eq:twiningPert}, we have $a_0(\infty)=2$.
Since $g$ is not a torus orbifold quantum symmetry, $a_1(0)=b_1(0)=0$. Similarly, $g^2$ is not a torus orbifold quantum symmetry, so $a_0(1/2)=0$. Finally, we determine $\phi_{g^4,g}$. $g^4$ has Frame shape $1^{-8}2^{16}$, which is a torus orbifold quantum symmetry. This is the same Frame shape as that of $g^4$ in the previous section; as a reminder, we have
\be \phi_{g^4,e}=-2y-2/y-4+O(q^{1/2}).\ee
As in the previous section, the $q^0y$ coefficient indicates the existence of two $g^4$-twisted R-R ground states with $J_0=1/2$ that are $g^4$-invariant; however, unlike the previous section, these cannot be $g^2$-invariant, since the orbifold by $g^2$ gives a consistent K3 sigma model. $g^2$ therefore acts with a minus sign on these states. The arguments that we employed in the previous section to eliminate certain choices of $g$ eigenvalues fail here: the multiplier enables the cases which were forbidden in the previous section. Therefore, we seem to have three options. If the eigenvalues are $\pm i,\mp i$ (with opposite sign), then we find
\be \phi_{g^4,g}=0+O(q),\ee
and $a_0(1/4)=0$. If eigenvalues are $\pm i,\pm i$ (with the same sign), then
\be \phi_{g^4,g}=\mp 2iy\mp 2i/y\pm 4i +O(q) \ , \ee
and $a_0(1/4)=\mp 2i$.
We get the final $a_0$ values for free: $a_0(3/4)=-a_0(1/4)$ and $a_0(1/8)=-a_0(\infty)$. As a check on our work, we note that the sum
\be \sum_{\cc} w_{\cc} a_0(\cc)=2+\column{0}{\mp 2i}+\column{0}{\pm 2i}-2=0 \ee
vanishes.

The twining genera resulting from these options are as follows. 
If we choose $a_0(1/4)=0$, then
\be F_{e,g}(\tau)=-(1/6)\E_4+(1/2)\E_8-(1/3)\E_{16}.\ee
We can rule out this case by considering the S-transform $\phi_{g,e}$ of $\phi_{e,g}$. For, $\phi_{g,e}$ is an untwined trace, so its $q$-expansion coefficients should be (real) integers; in this case, we get fractions.  If, instead, we choose $a_0(1/4)=\mp 2i$, then
\be F_{e,g}(\tau)=(-1/6)\E_4+(1/2)\E_8-(1/3)\E_{16}\pm 8\eta[2^44^{-4}8^4]\ ,\ee
and the S transforms $\phi_{g,e}$ are now perfectly consistent.  Thus, eliminating the first case, we find two twining genera, as expected.

\subsection{Results}

In this section we present the fruits of our labor in the form of two tables. Table \ref{tab:residues} contains a set of information for each fixing group $\hat\Gamma_g$, where $g$ runs over all supersymmetry-preserving symmetries that exist at any point in the moduli space of non-singular K3 NLSMs. For each such $\hat\Gamma_g$, we provide the genus of $\hat\HH/\hat\Gamma_g$, the set of cusps and the widths of these cusps, and the twisted-twining genera whose expansions about $\infty$ are related by \eqref{eq:jacobiTwining} to the expansions of the twining genus $\phi_{e,g}$ about these cusps. In addition, we present each Frame shape whose fixing group is $\hat\Gamma_g$ and give the residues $a_0(\cc)$ of the possible twining genera for these Frame shapes. (Some Frame shapes have multiple entries, since there are multiple possible twining genera with different sets of residues for these Frame shapes). We emphasize the non-zero residues which differ only by a multiplier by writing only one such residue explicitly and expressing the remainder in terms of this residue. When doing so, we use the shorthand $r_\cc$ in place of $a_0(\cc)$. We also define $\zeta_n=e^{2\pi i/n}$.

Table \ref{tab:big} presents the full set of twining genera which meet the criteria we have laid out in earlier sections. (In particular, we note that we computed the $S$ transform of each of these functions to make sure that its coefficients were (real) integers that satisfied the constraints of Corollary \ref{cor:useful}). In general, we have not shown that these functions are in fact the twining genera of symmetries of K3 NLSMs, and we merely claim that the set of all K3 NLSM twining genera is contained within our set. However, in most cases, we find only one possible function for each Frame shape with a given multiplier, so that the twining genus is uniquely identified. In the remaining cases, we are left with two possibilities and we cannot determine which case is actually realized.
As we discuss in more detail below, the functions that we find are \emph{precisely} those which arise in Conway and Umbral moonshine. That our physical constraints identify the same functions that arise in a completely different context can be taken as evidence that each of these functions is, in fact, the twining genus of a K3 NLSM symmetry.

Table \ref{tab:big} is organized as follows. The first column lists the Frame shapes of all K3 NLSM symmetries. The second column provides the associated eigengroups $\Gamma_g$ and -- in the cases where there are non-trivial multipliers -- the orders $\lambda$ of the multipliers. The third column summarizes the classification of $O^+(\Gamma^{4,20})$ classes determined by \cite{slowpaper}, where a $\circ$ indicates an $O^+(\Gamma^{4,20})$ class that is also an $O(\Gamma^{4,20})$ class, while a $\updownarrow$ represents two $O^+(\Gamma^{4,20})$ classes that merge into a single class in $O(\Gamma^{4,20})$. (To the Frame shape $1^{-4}2^53^46^1$ there may correspond either one or two $O^+(\Gamma^{4,20})$ classes; we denote this by writing $\circ,\circ^*$. Even if there are two classes, they  are inverses of each other, so they have the same twining genera). The fourth column lists the (candidate) twining genera that we have found. Assuming the conjectures of \cite{slowpaper} upon which we expound further below, we are able in many cases to match our functions with $O^+(\Gamma^{4,20})$ classes; when this is possible, we place corresponding classes and twining genera in the same line. We indicate those cases in which we can not provide such a correspondence by surrounding the $O^+(\Gamma^{4,20})$ classes with brackets. The fifth column indicates whether or not the twining genus has been found in an explicit K3 model: $\checkmark$ indicates that the genus has been realized in a K3 CFT, $LG$ indicates that the genus was found in a Landau-Ginzburg orbifold model which flows to a K3 CFT in the IR, and $\times$ indicates that the genus has not yet been found. See \cite{slowpaper} for a description of the methods that have been employed to obtain these K3 NLSMs. (Our results provide strong evidence that the twining genera computed in Landau-Ginzburg orbifolds in the UV do, in fact, yield K3 NLSM twining genera). The sixth column relates the twining genera to various moonshines, as is explained in section \ref{sec:moonshine}.\footnote{The $+$ and $-$ subscripts on $\Lambda$ correspond to the signs that appear in table 3 of \cite{Duncan:2015xoa}.} Finally, we note that when there are multiple entries in table \ref{tab:residues} for a given Frame shape, we generally order the functions in table \ref{tab:big} in order for these results to correspond. The only Frame shapes for which this is not possible is $4^28^2$ and $6^4$, as the corresponding fixing groups in these cases are genus 1, so for some sets of residues there are multiple twining genera which differ by a cusp form. In these cases, the first and second twining genera in table \ref{tab:big} correspond, respectively, to the first and second entries in table \ref{tab:residues}.

\newcolumntype{C}{>{$}c<{$}}

\begin{center}
\begin{tabular}{ccc}
\begin{tabular}[t]{C|CC}
\midrule\midrule
\multicolumn{3}{C}{SL(2,\ZZ)\quad \text{Genus 0}}\\
\midrule
\text{Cusp} & \infty \\
\text{Width} & 1\\
\phi_{g^i,g^j} & \phi_{e,g}\\
\midrule
\pi_g & \multicolumn{2}{C}{a_0(\cc)} \\\midrule
1^{24} & 0 \\
\midrule\midrule
\end{tabular}
&
\begin{tabular}[t]{C|CC}
\midrule\midrule
\multicolumn{3}{C}{\Gamma_0(p),p=2,3,5,7\quad \text{Genus 0}}\\
\midrule
\text{Cusp} & \infty & 0\\
\text{Width} & 1 & p\\
\phi_{g^i,g^j} & \phi_{e,g} & \phi_{g,e}\\
\midrule
\pi_g & \multicolumn{2}{C}{a_0(\cc)} \\\midrule
1^82^8 & 4/3 & -2/3\\
1^{-8}2^{16} & 8/3 & -4/3\\
1^63^6 & 3/2 & -1/2\\
1^{-3}3^9 & 9/4 & -3/4\\
1^45^4 & 5/3 & -1/3\\
1^37^3 & 7/4 & -1/4\\
\midrule\midrule
\end{tabular}
&
\begin{tabular}[t]{C|CCC}
\midrule\midrule
\multicolumn{3}{C}{\Gamma_0(4)\quad \text{Genus 0}}\\
\midrule
\text{Cusp} & \infty & 0 & 1/2\\
\text{Width} & 1 & 4 & 1\\
\phi_{g^i,g^j} & \phi_{e,g} & \phi_{g,e} & \phi_{g^2,g}\\
\midrule
\pi_g & \multicolumn{2}{C}{a_0(\cc)} \\\midrule
2^{12} & 2 & 0 & -r_\infty\\
1^42^24^4 & 5/3 & -1/3 & -1/3\\
1^82^{-8}4^8 & 4/3 & -2/3 & 4/3\\
1^{-4}2^6 4^4 & 7/3 & -2/3 & 1/3\\
2^{-4}4^8 & 2 & 0 & -2\\
& 2 & -1 & 2\\
\midrule\midrule
\end{tabular}
\end{tabular}

\begin{tabular}{cc}
\begin{tabular}[t]{C|CCCCCCCCCCCCCC}
\midrule\midrule
\multicolumn{3}{C}{\gmo{5}\quad \text{Genus 0}}\\
\midrule
\text{Cusp} & \infty & 0 & 1/2 & 2/5\\
\text{Width} & 1& 5 & 5 & 1\\
\phi_{g^i,g^j} & \phi_{e,g} & \phi_{g,e} & \phi_{g^2,g} & \phi_{e,g^3}\\
\midrule
\pi_g & \multicolumn{2}{C}{a_0(\cc)} \\\midrule
1^{-1}5^5 & 25/12 & 1/12 & -11/12 & 25/12\\
& 25/12 & -11/12 & 1/12 & 25/12\\
\midrule\midrule
\end{tabular}
&
\begin{tabular}[t]{C|CCCC}
\midrule\midrule
\multicolumn{3}{C}{\Gamma_0(6)\quad \text{Genus 0}}\\
\midrule
\text{Cusp} & \infty & 0 & 1/2 & 1/3 \\
\text{Width} & 1& 6 & 3 & 2 \\
\phi_{g^i,g^j} & \phi_{e,g} & \phi_{g,e} & \phi_{g^2,g} & \phi_{g^3,g} \\
\midrule
\pi_g & \multicolumn{2}{C}{a_0(\cc)} \\\midrule
1^22^23^26^2 & 11/6 & -1/6 & -1/6 & -1/6\\
1^42^13^{-4}6^5 & 5/3 & -1/3 & -1/3 & 2/3\\
1^52^{-4}3^16^4 & 19/12 & -5/12 & 7/12 & -5/12\\
1^{-2}2^43^{-2}6^4 & 13/6 & 1/6 & 1/6 & -11/6\\
& 13/6 & -5/6 & 1/6 & 7/6\\
1^{-1}2^{-1}3^36^3 & 25/12 & 1/12 & -11/12 & 1/12 \\
& 25/12 & -11/12 & 13/12 & 1/12\\
1^{-4}2^53^46^1 & 7/3 & -2/3 & 1/3 & 1/3\\
\midrule\midrule
\end{tabular}
\end{tabular}

\begin{tabular}{cc}
\begin{tabular}[t]{C|CCCC}
\midrule\midrule
\multicolumn{3}{C}{\Gamma_0(8)\quad \text{Genus 0}}\\
\midrule
\text{Cusp} & \infty & 0 & 1/2 & 1/4 \\
\text{Width} & 1& 8 & 2 & 1 \\
\phi_{g^i,g^j} & \phi_{e,g} & \phi_{g,e} & \phi_{g^2,g} & \phi_{g^4,g} \\
\midrule
\pi_g & \multicolumn{2}{C}{a_0(\cc)} \\\midrule
2^44^4 & 2 & 0 & 0 & -r_\infty\\
1^2 2^1 4^1 8^2 & 11/6 & -1/6 & -1/6 & -1/6 \\
1^4 2^{-2} 4^{-2} 8^4 & 5/3 & -1/3 & -1/3 &  5/3\\
& 5/3 & -1/3 & 2/3 & -1/3\\
\midrule\midrule
\end{tabular}
&
\begin{tabular}[t]{C|CCCCCCCCCCCCCC}
\midrule\midrule
\multicolumn{3}{C}{\gmo{8}\quad \text{Genus 0}}\\
\midrule
\text{Cusp} & \infty & 0 & 1/2 & 3/8 & 1/3 & 1/4 \\
\text{Width} & 1& 8 & 4 & 1 & 8 & 2 \\
\phi_{g^i,g^j} & \phi_{e,g} & \phi_{g,e} & \phi_{g^2,g} & \phi_{e,g^3} & \phi_{g^3,g} & \phi_{g^4,g} \\
\midrule
\pi_g & \multicolumn{2}{C}{a_0(\cc)} \\\midrule
1^{-2} 2^3 4^1 8^2 & 13/6 & 1/6 & 1/6 & 13/6 & -5/6 & 1/6\\
& 13/6 & -5/6 & 1/6 & 13/6 & 1/6 & 1/6\\
\midrule\midrule
\end{tabular}
\end{tabular}

\begin{tabularx}{\textwidth}{C|CCCC}
\endhead
\midrule\midrule
\multicolumn{3}{C}{\Gamma_0(9)\quad \text{Genus 0}}\\
\midrule
\text{Cusp} & \infty & 0 & 2/3 & 1/3 \\
\text{Width} & 1& 9 & 1 & 1 \\
\phi_{g^i,g^j} & \phi_{e,g} & \phi_{g,e} & \phi_{g^3,g^8} & \phi_{g^3,g} \\
\midrule
\pi_g & \multicolumn{2}{C}{a_0(\cc)} \\\midrule
3^8 & 2 & 0 & \zeta_3^{\pm 1}r_\infty & \zeta_3^{\mp 1} r_\infty\\
1^3 3^{-2} 9^3 & 7/4 & -1/4 & -\zeta_3^{\mp 1}-1/4 & -\zeta_3^{\pm 1}-1/4\\
\midrule\midrule
\end{tabularx}

\begin{tabularx}{\linewidth}{C|CCCCCCCCCCCCCC}
\midrule\midrule
\multicolumn{3}{C}{\gmo{10}\quad \text{Genus 0}}\\
\midrule
\text{Cusp} & \infty & 0 & 1/2 & 2/5 & 1/3 & 3/10 & 1/4 & 1/5 \\
\text{Width} & 1& 10 & 5 & 2 & 10 & 1 & 5 & 2 \\
\phi_{g^i,g^j} & \phi_{e,g} & \phi_{g,e} & \phi_{g^2,g} & \phi_{g^5,g^8} & \phi_{g^3,g} & \phi_{e,g^7} & \phi_{g^4,g} & \phi_{g^5,g} \\
\midrule
\pi_g & \multicolumn{2}{C}{a_0(\cc)} \\\midrule
1^22^15^{-2}10^3 & 11/6 & -1/6 & -1/6 & (2+3\sqrt{5})/6 & -1/6 & 11/6 & -1/6 & (2-3\sqrt{5})/6 \\
& 11/6 & -1/6 & -1/6 & (2-3\sqrt{5})/6 & -1/6 & 11/6 & -1/6 & (2+3\sqrt{5})/6\\
1^32^{-2}5^110^2 & 7/4 & -1/4 & -1/4 & -1/4 & -1/4 & 7/4 & 3/4 & -1/4\\
& 7/4 & -1/4 & 3/4 & -1/4 & -1/4 & 7/4 & -1/4 & -1/4\\
1^{-2}2^35^2 10^1 & 13/6 & 1/6 & 1/6 & 1/6 & -5/6 & 13/6 & 1/6 & 1/6\\
& 13/6 & -5/6 & 1/6 & 1/6 & 1/6 & 13/6 & 1/6 & 1/6\\
\midrule\midrule
\end{tabularx}

\begin{tabular}{cc}
\begin{tabular}[t]{C|CCCCCCCCCCCCCC}
\midrule\midrule
\multicolumn{3}{C}{\Gamma_0(11)\quad \text{Genus 1}}\\
\midrule
\text{Cusp} & \infty & 0 \\
\text{Width} & 1& 11\\
\phi_{g^i,g^j} & \phi_{e,g} & \phi_{g,e}  \\
\midrule
\pi_g & \multicolumn{2}{C}{a_0(\cc)} \\\midrule
1^211^2 & 11/6 & -1/6\\
\midrule\midrule
\end{tabular}
&
\begin{tabular}[t]{C|CCCCCCCCCCCCCC}
\midrule\midrule
\multicolumn{3}{C}{\Gamma_0(12)\quad \text{Genus 0}}\\
\midrule
\text{Cusp} & \infty & 0 & 1/2 & 1/3 & 1/4 & 1/6\\
\text{Width} & 1& 12 & 3 & 4 & 3 & 1\\
\phi_{g^i,g^j} & \phi_{e,g} & \phi_{g,e} & \phi_{g^2,g} & \phi_{g^3,g} & \phi_{g^4,g} & \phi_{g^6,g} \\
\midrule
\pi_g & \multicolumn{2}{C}{a_0(\cc)} \\\midrule
2^36^3 & 2 & 0 & 0 & 0 & 0 & -r_\infty\\
1^22^{-2}3^24^26^{-2}12^2 & 11/6 & -1/6 & -1/6 & -1/6 & -1/6 & 11/6\\
& 11/6 & -1/6 & 5/6 & -1/6 & -1/6 & -7/6\\
\midrule\midrule
\end{tabular}
\end{tabular}

\begin{tabularx}{\linewidth}{C|CCCCCCCCCCCCCC}
\midrule\midrule
\multicolumn{3}{C}{\gmo{12}\quad \text{Genus 0}}\\
\midrule
\text{Cusp} & \infty & 0 & 3/4 & 2/3 & 1/2 & 5/12 & 1/3 & 1/4 & 1/5 & 1/6 \\
\text{Width} & 1& 12 & 3 & 4 & 6 & 1 & 4 & 3 & 12 & 2\\
\phi_{g^i,g^j} & \phi_{e,g} & \phi_{g,e} & \phi_{g^4,g^{11}} & \phi_{g^3,g^{11}} & \phi_{g^2,g} & \phi_{e,g^5} & \phi_{g^3,g} & \phi_{g^4,g} & \phi_{g^5,g} & \phi_{g^6,g} \\
\midrule
\pi_g & \multicolumn{2}{C}{a_0(\cc)} \\\midrule
1^12^23^14^{-2}12^2 & 23/12 & -1/12 & \frac{-1+12i}{12} & -1/12 & -1/12 & 23/12 & -1/12 & \frac{-1-12i}{12} & -1/12 & -1/12\\
& 23/12 & -1/12 & \frac{-1-12i}{12} & -1/12 & -1/12 & 23/12 & -1/12 & \frac{-1+12i}{12} & -1/12 & -1/12\\
1^23^{-2}4^16^212^1 & 11/6 & -1/6 & -1/6 & \frac{-6\zeta_3^{\mp 1}-1}{6} & -1/6 & 11/6 & \frac{-6\zeta_3^{\pm 1}-1}{6} & -1/6 & -1/6 & -1/6\\
1^{-2}2^23^24^112^1 & 13/6 & 1/6 & 1/6 & 1/6 & 1/6 & 13/6 & 1/6 & 1/6 & -5/6 & 1/6\\
& 13/6 & -5/6 & 1/6 & 1/6 & 1/6 & 13/6 & 1/6 & 1/6 & 1/6 & 1/6\\
\midrule\midrule
\end{tabularx}

\begin{tabular}{cc}
\begin{tabular}[t]{C|CCCCCCCCCCCCCC}
\midrule\midrule
\multicolumn{3}{C}{\Gamma_0(14)\quad \text{Genus 1}}\\
\midrule
\text{Cusp} & \infty & 0 & 1/2 & 1/7 \\
\text{Width} & 1& 14 & 7 & 2\\
\phi_{g^i,g^j} & \phi_{e,g} & \phi_{g,e} & \phi_{g^2,g} & \phi_{g^7,g} \\
\midrule
\pi_g & \multicolumn{2}{C}{a_0(\cc)} \\\midrule
1^12^17^114^1 & 23/12 & -1/12 & -1/12 & -1/12 \\
\midrule\midrule
\end{tabular}
&
\begin{tabular}[t]{C|CCCCCCCCCCCCCC}
\midrule\midrule
\multicolumn{3}{C}{\Gamma_0(15)\quad \text{Genus 1}}\\
\midrule
\text{Cusp} & \infty & 0 & 1/3 & 1/5 \\
\text{Width} & 1& 15 & 5 & 3 \\
\phi_{g^i,g^j} & \phi_{e,g} & \phi_{g,e} & \phi_{g^3,g} & \phi_{g^5,g} \\
\midrule
\pi_g & \multicolumn{2}{C}{a_0(\cc)} \\\midrule
1^1 3^1 5^1 15^1 & 23/12 & -1/12 & -1/12 & -1/12 \\
\midrule\midrule
\end{tabular}
\end{tabular}

\begin{tabular}{cc}
\begin{tabular}[t]{C|CCCCCCCCCCCCCC}
\midrule\midrule
\multicolumn{3}{C}{\Gamma_0(16)\quad \text{Genus 0}}\\
\midrule
\text{Cusp} & \infty & 0 & 3/4 & 1/2 & 1/4 & 1/8 \\
\text{Width} & 1& 16 & 1 & 4 & 1 & 1 \\
\phi_{g^i,g^j} & \phi_{e,g} & \phi_{g,e} & \phi_{g^4,g^{15}} & \phi_{g^2,g} & \phi_{g^4,g} & \phi_{g^8,g}\\
\midrule
\pi_g & \multicolumn{2}{C}{a_0(\cc)} \\\midrule
4^6 & 2 & 0 & \pm i r_\infty & 0 & \mp i r_\infty & -r_\infty\\
2^44^{-4}8^4 
& 2 & 0 & \pm 2i & 0 & -r_{3/4} & -r_\infty\\
\midrule\midrule
\end{tabular}
&
\begin{tabular}[t]{C|CCCCCCCCCCCCCC}
\midrule\midrule
\multicolumn{3}{C}{\Gamma_0(20)\quad \text{Genus 1}}\\
\midrule
\text{Cusp} & \infty & 0 & 1/2 & 1/4 & 1/5 & 1/10\\
\text{Width} & 1& 20 & 5 & 5 & 4 & 1 \\
\phi_{g^i,g^j} & \phi_{e,g} & \phi_{g,e} & \phi_{g^2,g} & \phi_{g^4,g} & \phi_{g^5,g} & \phi_{g^{10},g}\\
\midrule
\pi_g & \multicolumn{2}{C}{a_0(\cc)} \\\midrule
2^2 10^2 & 2 & 0 & 0 & 0 & 0 & -r_\infty\\
\midrule\midrule
\end{tabular}
\end{tabular}

\begin{tabularx}{\linewidth}{C|CCCCCCCCCCCCCC}
\endhead
\midrule\midrule
\multicolumn{3}{C}{\Gamma_0(24)\quad \text{Genus 1}}\\
\midrule
\text{Cusp} & \infty & 0 & 1/2 & 1/3 & 1/4 & 1/6 & 1/8 & 1/12\\
\text{Width} & 1& 24 & 6 & 8 & 3 & 2 & 3 & 1 \\
\phi_{g^i,g^j} & \phi_{e,g} & \phi_{g,e} & \phi_{g^2,g} & \phi_{g^3,g} & \phi_{g^4,g} & \phi_{g^6,g} & \phi_{g^8,g} & \phi_{g^{12},g}\\
\midrule
\pi_g & \multicolumn{2}{C}{a_0(\cc)} \\\midrule
2^14^16^112^1 & 2 & 0 & 0 & 0 & 0 & 0 & 0 & -r_\infty \\
\midrule\midrule
\end{tabularx}

\begin{tabularx}{\linewidth}{C|CCCCCCCCCCCCCC}
\endhead
\midrule\midrule
\multicolumn{3}{C}{\Gamma_0(32)\quad \text{Genus 1}}\\
\midrule
\text{Cusp} & \infty & 0 & 3/4 & 1/2 & 3/8 & 1/4 & 1/8 & 1/16\\
\text{Width} & 1& 32 & 2 & 8 & 1 & 2 & 1 & 1\\
\phi_{g^i,g^j} & \phi_{e,g} & \phi_{g,e} & \phi_{g^4,g^{31}} & \phi_{g^2,g} & \phi_{g^8,g^3} & \phi_{g^4,g} & \phi_{g^8,g} & \phi_{g^{16},g}\\
\midrule
\pi_g & \multicolumn{2}{C}{a_0(\cc)} \\\midrule
4^28^2 & 2 & 0 & 0 & 0 & \pm i r_\infty & 0 & \mp i r_\infty & - r_\infty\\
\midrule\midrule
\end{tabularx}

\addtocounter{table}{-5}

\begin{tabularx}{\linewidth}{C|CCCCCCCCCCCCCC}
\endhead
\midrule\midrule
\multicolumn{3}{C}{\Gamma_0(36)\quad \text{Genus 1}}\\
\midrule
\text{Cusp} & \infty & 0 & 1/2 & 1/3 & 2/3 & 1/4 & 1/6 & 5/6 & 1/9 & 1/12 & 5/12 & 1/18 \\
\text{Width} & 1& 36 & 9 & 4 & 4 & 9 & 1 & 1 & 4 & 1 & 1 & 1 \\
\phi_{g^i,g^j} & \phi_{e,g} & \phi_{g,e} & \phi_{g^2,g} & \phi_{g^3,g} & \phi_{g^3,g^{35}} & \phi_{g^4,g} & \phi_{g^6,g} & \phi_{g^6,g^{35}} & \phi_{g^9,g} & \phi_{g^{12},g} & \phi_{g^{12},g^5} & \phi_{g^{18},g} \\
\midrule
\pi_g & \multicolumn{2}{C}{a_0(\cc)} \\\midrule
6^4 & 2 & 0 & 0 & 0 & 0 & 0 & \zeta_6^{\pm 1}r_\infty & \zeta_6^{\mp 1}r_\infty & 0 & \zeta_6^{\pm 2}r_\infty & \zeta_6^{\mp 2}r_\infty & -r_\infty \\
\midrule\midrule
\bottomrule \caption{}\label{tab:residues}
\end{tabularx}
\end{center}

\begin{landscape}
\begin{tabularx}{\linewidth}{CCCCCC}
\pi_g & (\Gamma_g)_{|\lambda} &
\begin{matrix}
O^+(\Gamma^{4,20})\\ \text{ classes}
\end{matrix} 
&   F_{e,g}(\tau) & \begin{matrix}\text{Explicitly}\\\text{Computed}\\\text{in a CFT}\end{matrix} & \text{Niemeier}\\
\midrule\endhead
1^{24} & SL(2,\mathbb{Z}) 
&\begin{matrix}
\circ
\end{matrix} 
 & 0 & \checkmark & \text{All}
 \\
\rowcolor{gray!11}{} 1^82^8 & \Gamma_0(2) 
&\begin{matrix}
\circ 
\end{matrix} & -\frac{4}{3}\E_2
&\checkmark & \text{All except $A_6^4,A_{12}^2,D_{12}^2,A_{24},D_{16}E_8,D_{24}$}
\\
 1^{-8}2^{16} & \Gamma_0(2) &\begin{matrix}
\circ
\end{matrix} & -\frac{8}{3}\E_2
&\checkmark & \Lambda
\\
\rowcolor{gray!11}{} 2^{12} & \Gamma_0(2)_{|2} 
&\begin{matrix}
\circ
\end{matrix}& 2\E_2-\frac{4}{3}\E_4
&\checkmark & A_1^{24},A_2^{12},A_3^8,A_4^6,D_4^6,A_6^4,A_8^3,D_6^4,A_{12}^2,D_{12}^2,A_{24},\Lambda  
\\
 1^63^6 & \Gamma_0(3) & 
\begin{matrix}
\circ
\end{matrix} & -\frac{3}{4}\E_3
&\checkmark &  A_1^{24},A_2^{12},A_3^8,A_5^4D_4,D_4^6,A_6^4,D_6^4,E_6^4,\Lambda
\\
\rowcolor{gray!11}{} 1^{-3}3^{9}  & \Gamma_0(3) & 
 \begin{matrix}
\circ
\end{matrix}  & -\frac{9}{8}\E_3
&\checkmark & \Lambda
\\
 3^{8}    & \Gamma_0(3)_{|3}   &
  \begin{matrix} \updownarrow\end{matrix}
 & \frac{1}{2}\E_3-\frac{3}{8}\E_9\pm 9\eta[1^33^{-2}9^3]
 &\begin{matrix}\times\\ LG\end{matrix}
 & \begin{matrix}
A_2^{12}, D_4^6, A_8^3, E_8^3, \Lambda \\
A_1^{24}, A_4^6, D_8^3 
\end{matrix}
\\
\rowcolor{gray!11}{} 1^42^24^4 & \Gamma_0(4)
&\begin{matrix}
\circ
\end{matrix} & \frac{1}{3}\E_2-\frac{2}{3}\E_4
&\checkmark
&A_1^{24},A_2^{12},A_3^8,A_4^6,A_5^4D_4,D_4^6,A_9^2D_6,\Lambda 
\\
 1^82^{-8}4^8 & \Gamma_0(4)  &\begin{matrix}
\circ
\end{matrix}  &  -\frac{4}{3}\E_2
&\times & \Lambda
\\
\rowcolor{gray!11}{} 1^{-4}2^64^4 & \Gamma_0(4)   &\begin{matrix}
\circ
\end{matrix}  & -\frac{1}{3}\E_2-\frac{2}{3}\E_4
&\checkmark
& \Lambda
\\
 2^{-4}4^8  & \Gamma_0(4) 
 & \begin{matrix} \updownarrow\end{matrix}  & 
 \begin{matrix}
 2\E_2-\frac{4}{3}\E_4\\
 -2\E_2
 \end{matrix}
 &\begin{matrix}\checkmark\\\checkmark\end{matrix}
  & \begin{matrix}
 \Lambda_- \\ \Lambda_+
\end{matrix} 
\\
\rowcolor{gray!11}{} 2^{4}4^4  & \Gamma_0(4)_{|2} &\begin{matrix}
\circ
\end{matrix}  &   -\frac{1}{3}\E_2+\E_4-\frac{2}{3}\E_8
&\checkmark
& A_1^{24},A_2^{12},A_3^8,D_4^6,A_7^2D_5^2,E_6^4,\Lambda  
\\
 4^6 & \Gamma_0(4)_{|4}
 &   \begin{matrix} \updownarrow\end{matrix}  & -\frac{1}{6}\E_4+\frac{1}{2}\E_8-\frac{1}{3}\E_{16}\pm 8 \eta[2^44^{-4}8^4]
 &\begin{matrix}\times\\ LG\end{matrix}
 &\begin{matrix}
  A_3^8, A_4^6, A_{12}^2, \Lambda \\
  A_1^{24}, A_2^{12}, A_6^4, D_6^4
 \end{matrix}
 \\
\rowcolor{gray!11}{} 1^45^4 & \Gamma_0(5) 
&\begin{matrix}
\circ
\end{matrix} & -\frac{5}{12}\E_5
&\checkmark
&A_1^{24},A_2^{12},A_4^6,D_4^6,\Lambda
\\
 1^{-1}5^5 & \gmo{5} & \begin{matrix}\updownarrow\end{matrix}  & 
 -\frac{25}{48}\E_5\mp\frac{25\sqrt{5}}{2}\eta[1^{-1}5^5]
 &\begin{matrix}\checkmark\\\checkmark\end{matrix}
  &  \begin{matrix}
\Lambda_-\\ \Lambda_+
\end{matrix}
\\
\rowcolor{gray!11}{} 1^22^23^26^2 & \Gamma_0(6)
&\begin{matrix}
\circ
\end{matrix}  & \frac{1}{6}\E_2+\frac{1}{4}\E_3-\frac{1}{2}\E_6
&\checkmark
&A_1^{24},A_2^{12},A_3^8,D_4^6,E_6^4,\Lambda
\\
 1^{4}2^13^{-4}6^5 & \Gamma_0(6)  & \begin{matrix}
\circ
\end{matrix}   & \frac{1}{12}\E_2-\frac{1}{4}\E_3-\frac{1}{4}\E_6
&\checkmark
&\Lambda
\\
\rowcolor{gray!11}{} 1^{5}2^{-4}3^16^4 & \Gamma_0(6) & \begin{matrix}
\circ
\end{matrix}   & -\frac{7}{12}\E_2+\frac{1}{8}\E_3-\frac{1}{4}\E_6
&\checkmark
 & \Lambda
\\
 1^{-2}2^43^{-2}6^4 & \Gamma_0(6)
  & \begin{matrix} \updownarrow\end{matrix}  
 & \begin{matrix}
 \frac{1}{3}\E_2+\frac{5}{4}\E_3-\E_6\\
 -\frac{2}{3}\E_2-\frac{3}{4}\E_3
 \end{matrix}
 &\begin{matrix}\checkmark\\\checkmark\end{matrix}
 &  \begin{matrix}
\Lambda_-\\ \Lambda_+
\end{matrix}
\\
\rowcolor{gray!11}{} 1^{-1}2^{-1}3^36^3 & \Gamma_0(6)
 & \begin{matrix} {}\\\updownarrow\\ {}\end{matrix}
 &  \begin{matrix}
 \frac{11}{12}\E_2+\frac{3}{8}\E_3-\frac{3}{4}\E_6\\
 -\frac{4}{3}\E_2-\frac{3}{8}\E_3
 \end{matrix}
 &\begin{matrix}\checkmark\\\checkmark\end{matrix}
 & \begin{matrix}
\Lambda_-\\ \Lambda_+
\end{matrix} 
 \\
 1^{-4}2^53^46^1 & \Gamma_0(6) & \begin{matrix}
\circ,\circ^{*}
\end{matrix} 
 & -\frac{7}{12}\E_2-\frac{1}{4}\E_3-\frac{1}{4}\E_6
 &\checkmark  & \Lambda
 \end{tabularx}

\addtocounter{table}{-1}

\begin{tabularx}{\linewidth}{CCCCCC}
\pi_g & (\Gamma_g)_{|\lambda} &
\begin{matrix}
O^+(\Gamma^{4,20})\\ \text{ classes}
\end{matrix} 
&   F_{e,g}(\tau) & \begin{matrix}\text{Explicitly}\\\text{Computed}\\\text{in a CFT}\end{matrix} & \text{Niemeier}\\
\midrule\endhead
 2^36^3 & \Gamma_0(6)_{|2}  & \begin{matrix}
\circ
\end{matrix}   & -\frac{1}{4}\E_2-\frac{1}{4}\E_3+\frac{1}{6}\E_4+\frac{3}{4}\E_6-\frac{1}{2}\E_{12}
&\checkmark &  A_2^{12},A_3^8,A_6^4,\Lambda
\\ 
\rowcolor{gray!11}{} 6^{4} & \Gamma_0(6)_{|6} & \begin{matrix} \updownarrow\\\\\updownarrow\end{matrix}   & \begin{matrix}
 2\eta[1^22^23^26^{-2}]\\
 2\eta[1^52^{-1}3^16^{-1}]\\
 2\eta[1^22^23^26^{-2}]\\
 2\eta[1^52^{-1}3^16^{-1}]+36\eta[6^4]
  \end{matrix}
  &\begin{matrix}\times\\ LG\\\times\\\times\end{matrix}
  & \begin{matrix}
A_1^{24}, A_4^6 \\ A_2^{12}, D_4^6,\Lambda_- \\ A_1^{24}, A_4^6 \\ A_8^3, \Lambda_+
\end{matrix} 
\\
 1^37^3 & \Gamma_0(7) 
&\begin{matrix}
\circ
\end{matrix}   & -\frac{7}{24}\E_7
&\checkmark
&A_1^{24},A_3^8,\Lambda
\\
\rowcolor{gray!11}{} 1^22^14^18^2 & \Gamma_0(8) 
&\begin{matrix}
\circ
\end{matrix}   & \frac{1}{6}\E_4-\frac{1}{3}\E_8
&\checkmark
&A_1^{24},A_2^{12},A_5^4D_4,\Lambda
\\
 1^42^{-2}4^{-2}8^4 & \Gamma_0(8)  &  \begin{matrix} {}\\\updownarrow\\ {}\end{matrix}  
  &  \begin{matrix} \frac{1}{3}\E_2-\frac{2}{3}\E_4 \\ -\frac{5}{6}\E_2+\frac{1}{2}\E_4-\frac{1}{3}\E_8\end{matrix}
  &\begin{matrix}\times\\\times\end{matrix}
   &\begin{matrix}
\Lambda_-\\ \Lambda_+
\end{matrix} 
\\
\rowcolor{gray!11}{} 1^{-2}2^34^18^2 & \gmo{8} &  \begin{matrix} \updownarrow\end{matrix}  & 
 \begin{matrix}
 -\frac{1}{3}\E_2+\frac{1}{6}\E_4-\frac{1}{3}\E_8\mp 16\sqrt{2}\eta[1^{-2}2^34^18^2]
 \end{matrix}
 &\begin{matrix}\checkmark\\\checkmark\end{matrix}
  &\begin{matrix}
\Lambda_-\\ \Lambda_+
\end{matrix} 
\\
 2^44^{-4}8^4  & \Gamma_0(8)_{|2} 
  & \begin{matrix} \circ \\ \circ\end{matrix}  & -\frac{1}{6}\E_4+\frac{1}{2}\E_8-\frac{1}{3}\E_{16}\pm 
  8\eta[2^44^{-4}8^4]
  &\begin{matrix}\times\\\checkmark\end{matrix}
  & \begin{matrix}
\Lambda_+\\ \Lambda_-
\end{matrix}
\\
\rowcolor{gray!11}{} 4^28^2  & \Gamma_0(8)_{|4}  & \begin{matrix} \updownarrow\\\\\updownarrow\end{matrix}  & 
 \begin{matrix}
 16\eta[4^48^{-4}16^4]+2\eta[2^44^28^{-2}] - 8\eta[4^28^2]\\
 2\eta[2^44^28^{-2}]\\
 16\eta[4^48^{-4}16^4]+2\eta[2^44^28^{-2}] + 24\eta[4^28^2]\\
 2\eta[2^44^28^{-2}]
  \end{matrix}
  &\begin{matrix}LG\\\times\\\times\\\times\end{matrix}
  & \begin{matrix}
 A_3^8, \Lambda_- \\ A_2^{12} \\ E_6^4, \Lambda_+ \\A_2^{12}
\end{matrix}
\\
 1^33^{-2}9^3  & \Gamma_0(9)  & \begin{matrix} \circ \\ \circ \end{matrix}  & -\frac{1}{8}\E_3-\frac{3}{16}\E_9\pm\frac{9}{2}\eta[1^33^{-2}9^3]
 &\begin{matrix}\checkmark\\\checkmark\end{matrix}
  & \begin{matrix}
\Lambda_+\\ \Lambda_-
\end{matrix} 
\\
\rowcolor{gray!11}{} 1^{2}2^15^{-2}10^3   & \gmo{10} 
 & \begin{matrix} \updownarrow\end{matrix} & \frac{1}{24}\E_2-\frac{5}{24}\E_{10}\pm 2\sqrt{5}\eta[1^22^15^{-2}10^3]
 &\begin{matrix}\checkmark\\\checkmark\end{matrix}
  & \begin{matrix}
\Lambda_+\\ \Lambda_-
\end{matrix} 
\\
 1^{3}2^{-2}5^110^2  & \gmo{10} 
 & \begin{matrix} {}\\\updownarrow\\ {}\end{matrix} & -\frac{7}{24}\E_2+\frac{5}{48}\E_5-\frac{5}{24}\E_{10}\mp\frac{5\sqrt{5}}{2}\eta[1^32^{-2}5^110^2]
 &\begin{matrix}\checkmark\\\checkmark\end{matrix}
  & \begin{matrix}
\Lambda_-\\ \Lambda_+
\end{matrix} 
\\
\rowcolor{gray!11}{} 1^{-2}2^35^210^1  & \gmo{10} 
  & \begin{matrix} \updownarrow\end{matrix} 
 & -\frac{7}{24}\E_2-\frac{5}{24}\E_{10}\mp 10\sqrt{5}\eta[1^{-2}2^35^210^1]
 &\begin{matrix}\checkmark\\\checkmark\end{matrix}
 & \begin{matrix}
\Lambda_-\\ \Lambda_+
\end{matrix}
\\
 2^210^2 & \Gamma_0(10)_{|2}  
  & 
 \left[\;\begin{matrix} \updownarrow\\  \circ \\  \circ \\ \circ \end{matrix}\;\right] 
  &
  \begin{matrix}
  -\frac{1}{12}\E_2+\frac{1}{18}\E_4-\frac{5}{36}\E_5+\frac{5}{12}\E_{10}-\frac{5}{18}\E_{20}-\frac{20}{3}
  \eta[2^210^2]\\
  -\frac{1}{12}\E_2+\frac{1}{18}\E_4-\frac{5}{36}\E_5+\frac{5}{12}\E_{10}-\frac{5}{18}\E_{20}+\frac{40}{3}\eta[2^210^2]
\end{matrix}  
&\begin{matrix}LG\\\times\end{matrix}
 & \begin{matrix}
A_1^{24}, A_2^{12},\Lambda_- \\ A_4^6,\Lambda_+
\end{matrix}
\\
\rowcolor{gray!11}{} 1^211^2 & \Gamma_0(11)   & 
\left[\;\begin{matrix} \updownarrow\\  \circ \\  \circ \end{matrix}\;\right] 
 & \begin{matrix}
 -\frac{11}{60}\E_{11} -\frac{22}{5}\eta[1^211^2]\\
  -\frac{11}{60}\E_{11} + \frac{33}{5}\eta[1^211^2]
 \end{matrix}
 &\begin{matrix}\times\\ LG\end{matrix}
  & \begin{matrix}
 A_1^{24}, \Lambda_- \\ A_2^{12},\Lambda_+
\end{matrix} 
\\
 1^{2}2^{-2}3^24^{2}6^{-2}12^2  & \Gamma_0(12)  & \begin{matrix} {}\\\updownarrow\\ {}\end{matrix}  & \begin{matrix} \frac{1}{6}\E_2+\frac{1}{4}\E_3-\frac{1}{2}\E_6 \\ 
 -\frac{13}{12}\E_2-\frac{1}{4}\E_3+\frac{1}{2}\E_4+\frac{3}{4}\E_6-\frac{1}{2}\E_{12}\end{matrix}
 &\begin{matrix}\times\\\times\end{matrix}
  &  \begin{matrix}
\Lambda_-\\ \Lambda_+
\end{matrix}
\\
\rowcolor{gray!11}{} 1^{1}2^23^14^{-2}12^2  & \gmo{12} 
  & \begin{matrix}  \circ\\ \circ\end{matrix}  & -\frac{1}{24}\E_2+\frac{1}{12}\E_4+\frac{1}{8}\E_6-\frac{1}{4}\E_{12}\pm 3\sqrt{3}\eta[1^12^23^14^{-2}12^2]
  &\begin{matrix}\checkmark\\\checkmark\end{matrix}
    & \begin{matrix}
\Lambda_+\\ \Lambda_-
\end{matrix}
  \\
1^{2}3^{-2}4^16^212^1  & \gmo{12} 
& \begin{matrix}
\circ\\ \circ
\end{matrix}  & -\frac{1}{12}\E_2-\frac{1}{4}\E_3+\frac{1}{12}\E_4+\frac{1}{4}\E_6-\frac{1}{4}\E_{12}\pm 4\sqrt{3}\eta[1^23^{-2}4^16^212^1]
&\begin{matrix}\checkmark\\\checkmark\end{matrix}
& \begin{matrix}
\Lambda_+\\ \Lambda_-
\end{matrix} 
\\
\rowcolor{gray!11}{} 1^{-2}2^23^24^112^1  & \gmo{12} 
& \begin{matrix} \updownarrow\end{matrix}  & 
-\frac{5}{12}\E_2-\frac{1}{4}\E_3+\frac{1}{12}\E_4+\frac{1}{4}\E_6-\frac{1}{4}\E_{12}\mp 12\sqrt{3}\eta[1^{-2}2^23^24^112^1]
&\begin{matrix}\checkmark\\\checkmark\end{matrix}
& \begin{matrix}
\Lambda_-\\ \Lambda_+
\end{matrix} 
\\
 2^14^16^112^1 & \Gamma_0(12)_{|2}   
  & 
  \left[\;\begin{matrix} \updownarrow\\  \circ \\  \circ \end{matrix}\;\right] 
  & \begin{matrix}
  \frac{1}{24}\E_2-\frac{1}{8}\E_4-\frac{1}{8}\E_6+\frac{1}{12}\E_8+\frac{3}{8}\E_{12}-\frac{1}{4}\E_{24} - 6\eta[2^14^16^1 12^1]\\
    \frac{1}{24}\E_2-\frac{1}{8}\E_4-\frac{1}{8}\E_6+\frac{1}{12}\E_8+\frac{3}{8}\E_{12}-\frac{1}{4}\E_{24} +18\eta[2^14^16^1 12^1]
  \end{matrix}
  &\begin{matrix}\times\\\times\end{matrix}
  & \begin{matrix}
A_1^{24},\Lambda_- \\ D_4^6, \Lambda_+
\end{matrix} 
 \\
\rowcolor{gray!11}{} 1^12^17^114^1 & \Gamma_0(14) 
 & \left[\;\begin{matrix} \updownarrow\\  \circ \\  \circ \end{matrix}\;\right]  & 
 \begin{matrix}
 \frac{1}{36}\E_2+\frac{7}{72}\E_7-\frac{7}{36}\E_{14}-\frac{14}{3}\eta[1^12^17^114^1]\\
  \frac{1}{36}\E_2+\frac{7}{72}\E_7-\frac{7}{36}\E_{14}+\frac{28}{3}\eta[1^12^17^114^1]
 \end{matrix}
 &\begin{matrix}\times\\ LG\end{matrix}
 &  \begin{matrix} A_1^{24},\Lambda_- \\ A_3^8,\Lambda_+\end{matrix}
\\
 1^13^15^115^1  & \Gamma_0(15) & 
\left[\;\begin{matrix} \updownarrow\\  \circ \\  \circ \end{matrix}\;\right] 
 & 
 \begin{matrix}
 \frac{1}{32}\E_3+\frac{5}{96}\E_5-\frac{5}{32}\E_{15}-\frac{15}{4}\eta[1^13^15^115^1]\\
  \frac{1}{32}\E_3+\frac{5}{96}\E_5-\frac{5}{32}\E_{15}+\frac{45}{4}\eta[1^13^15^115^1]
 \end{matrix}
  &\begin{matrix}\times\\ LG\end{matrix}
  &  \begin{matrix}
 A_1^{24},\Lambda_-\\D_4^6, \Lambda_+
\end{matrix}
\\
\bottomrule \caption{}
\label{tab:big}
\end{tabularx}
\end{landscape}

\section{Implications for Moonshine}\label{sec:moonshine}

As discussed briefly in the introduction, Umbral and Conway moonshines associate -- via formal constructions that a priori have nothing to do with string theory on K3 -- weight 0, index 1 weak Jacobi forms to conjugacy classes of the appropriate Umbral or Conway groups, the construction of which we now explain \cite{Cheng:2014zpa,Duncan:2015xoa}. First, we define the Niemeier lattices\footnote{As we will stress momentarily, we use terminology such that `Niemeier lattices' includes the Leech lattice, whose root system is empty.} to be the 24 even unimodular 24-dimensional lattices of (in our conventions) negative definite signature, which are uniquely identified by their root systems. Each instance of Umbral moonshine is associated to one of the 23 such lattices with roots. The Umbral group $G_L$ associated to the lattice $L$ is defined by
\be\label{eq:umbralGroup} G_L := O(L)/W_L \ ,\ee
where $O(L)$ and $W_L$ are, respectively, the automorphism group of $L$ and the Weyl group of the root system of $L$. Although the weak Jacobi forms derived from the Conway ($Co_0$) moonshine module \cite{Duncan:2015xoa} are constructed in an entirely different manner from those of Umbral moonshine \cite{Cheng:2014zpa}, we may use similar notation if we define $\Lambda$ to be the Leech lattice -- the unique Niemeier lattice with no roots -- and define $W_\Lambda$ to be the trivial group. The corresponding $G_{\Lambda}$ is then $Co_0$, the automorphism group of the Leech lattice, and we call this case `Conway moonshine' for the purposes of this paper. We will sometimes collectively refer to the Umbral groups and $Co_0$ as the Niemeier groups.

There are two important differences between the Umbral and Conway constructions of Jacobi forms that we wish to highlight. First, unlike Umbral moonshine, Conway moonshine does not associate a weak Jacobi form to an arbitrary conjugacy class of $G_\Lambda$. Instead, the construction only works for conjugacy classes of elements $g\in O(\Lambda)$ that pointwise fix a 4-plane in $\Lambda\otimes\RR$. Second, Umbral moonshine associates a \emph{unique} weak Jacobi form to each conjugacy class.  In contrast, this is only the case for Conway moonshine for conjugacy classes that fix at least a 5-plane. When the subspace fixed by $[g]$ is precisely 4-dimensional, Conway moonshine associates two \emph{distinct} weak Jacobi forms to $[g]$.

The work \cite{slowpaper} has shown that the Niemeier lattices indeed play a role in the study of K3 nonlinear sigma models (NLSMs), building off of work associating the Niemeier lattices to K3 geometry advocated in \cite{nikulin2013kahlerian}. Consider a symmetry $g$ of perturbative type II string theory on K3, possibly at a singular point in the K3 CFT moduli space, that fixes a positive 4-plane in $\Gamma^{4,20}\otimes\RR$.\footnote{The following results have natural generalizations from sublattices fixed by cyclic groups of the form $\avg{g}$ to sublattices fixed by more general groups of symmetries. We restrict to the former for simplicity.} Denote by $\Xi^g$ the sublattice of $\Gamma^{4,20}$ that is pointwise fixed by $g$, and let $\Xi_g$ denote the orthogonal complement of $\Xi^g$. Then, there exists a (generally non-unique) Niemeier lattice $L$ such that $\Xi_g$ may be primitively embedded into $L$; if $\Xi_g$ has no roots (so that $g$ is a symmetry of a \emph{non-singular} K3 NLSM), then we can always choose $L$ to be the Leech lattice\cite{Gaberdiel:2011fg}. Denote this embedding by $i$ and the image of $\Xi_g$ under $i$ by $L_g$. Then, the group $\avg{g}$ generated by $g$ acts naturally on $L_g$; more precisely, $\tilde g=igi^{-1}$ generates a group that extends uniquely to a subgroup of $O(L)$ that fixes pointwise the orthogonal complement $L^g$ of $L_g$. When $\Xi_g$ contains no roots, $\avg{\tilde g}$ is isomorphic to a subgroup of $G_L$. Since $L^g$ is always at least 4-dimensional, we find that the Umbral symmetries that arise naturally in this setting are those that fix a 4-plane. 

The correspondence between groups of (supersymmetry-preserving) symmetries of non-singular K3 NLSMs and subgroups of the Niemeier groups $G_L$ is made particularly sharp by the following observation: for all such symmetries $g$ of non-singular K3 NLSMs for which the twining genus $\phi_{e,g}$ has been computed explicitly in the CFT, one of the Umbral or Conway weak Jacobi forms associated to $[\tilde g]$ has always been the same as the twining genus $\phi_{e,g}$. 
This  observation, among others, led the authors of \cite{slowpaper} to conjecture that every twining genus corresponding to a symmetry of a non-singular K3 NLSM equals a weak Jacobi form that Conway or Umbral moonshine associates to a 4-plane preserving conjugacy class, and conversely. Our results provide substantial further support for this conjecture, as physical considerations pertaining to string theory on K3 (and compactifications thereof) pick out exactly the set of weak Jacobi forms that Umbral and Conway moonshine associate to 4-plane fixing conjugacy classes, even when these weak Jacobi forms have yet to arise as the twining genera of K3 NLSMs. 

In fact, we obtain interesting results by proceeding formally and applying the method described in the preceding sections (as we describe in more detail below) to the Frame shapes of Umbral symmetries that do not fix a 4-plane. In particular, our findings suggest a possible broadening of the Conway moonshine construction; they also provide further evidence that the Niemeier groups capture symmetries of K3 string theory and their associated Jacobi forms determine spacetime BPS state counts. Of course, in these cases the physical motivation for employing our method does not apply, since these Frame shapes do not correspond to supersymmetry-preserving symmetries of non-singular K3 NLSMs. Nevertheless, as table \ref{tab:notFourFixing} indicates, we obtain a small list of functions, which contains -- for each Frame shape -- the Umbral moonshine weak Jacobi form(s) \cite{Cheng:2014zpa}.  It is remarkable that our constraints are able to identify such a limited set of functions, given that the relevant fixing groups have genera greater than 1 -- in some cases, much greater than 1. For instance, $\Gamma_0(144)$ has genus 13. We find our results for the Frame shape $2^122^1$ particularly surprising: we identify a unique weak Jacobi form (that of $L=A_2^{12}$ Umbral moonshine), even though the fixing group $\Gamma_0(44)$ has genus 4.      Since, in the case of Frame shapes corresponding to 4-plane fixing symmetries, our constraints identified precisely the Umbral and Conway moonshine weak Jacobi forms, one might hope that the extra (i.e. non-Umbral) functions we have obtained  play a role in an expanded version of Conway moonshine that encompasses all $Co_0$ conjugacy classes, and not only those that fix a 4-plane. Unfortunately, we have reason to believe this may not be the case for some of the functions we have found. The argument is the following. Suppose $g$ is a bona fide symmetry of order $N$ in a K3 NLSM, and that the multiplier of $\phi_{e,g}$ is determined by its order $\lambda$ and by $\E'\in \ZZ/\lambda\ZZ$, as described in section \ref{sec:levelfail}. Then, by standard CFT arguments, the multiplier of its higher powers $g^n$, $n|N$, is given by $\lambda_{g^n}=\frac{\lambda}{\gcd(\lambda,n)}$ and $\E'\in \ZZ/\lambda_{g^n}\ZZ$. The same property holds also for all twining genera of Umbral moonshine, including the ones that  have no interpretation as physical symmetries in a NLSM. It is natural to expect a similar behaviour for  the putative Conway twining genera associated to Frame shapes not fixing a 4-plane. Some of the functions reported in table \ref{tab:notFourFixing} are incompatible with the multipliers of known Conway twining genera; one cannot expect such functions to arise in any reasonable extension of Conway moonshine.


Even after imposing these constraints on the multiplier, it would not be too surprising if multiple Conway moonshine weak Jacobi forms were to exist for these new Frame shapes. In fact, we have already seen that, as the rank of the fixed lattice decreases from 5 (or higher) to exactly 4, the number of Conway moonshine weak Jacobi forms increases from 1 to 2. If there is indeed an extension of Conway moonshine into which we can incorporate these new Jacobi forms, we further speculate that applying our constraints to the other Conway Frame shapes -- that is, those that neither fix a 4-plane nor appear in the context of Umbral moonshine -- will yield additional weak Jacobi forms that will appear in this extended Conway moonshine.

Before proceeding, we wish to quickly detail precisely the method we employed in order to obtain the results of the last paragraph and of table \ref{tab:notFourFixing}. We assume that residues are still associated via \eqref{eq:bToA} to the coefficients of $q^0 y^{\pm 1}$ of appropriate twisted-twining genera, which are associated to cusps in the same way as above. Each of the Frame shapes under consideration essentially meets our criteria for being of pure K3 type, in the sense that all powers of these Frame shapes that are associated to symmetries of K3 NLSMs are the Frame shapes of genuine pure K3 symmetries. We therefore assume that twisted-twining genera $\phi_{g^i,g^j}$ with $i\not\equiv 0\modu N$ have no $q^0 y^{\pm 1}$ terms. The $q^0y^{\pm 1}$ terms of the remaining twisted-twining genera are related by a multiplier to those of $\phi_{e,g^j}$, which we assume are still determined by \eqref{eq:twiningPert}\footnote{We note that this last assumption, as well as the assumption that the twining genera must be weak Jacobi forms, were equivalent to the preservation of the supersymmetry and spectral flow generators in the 4-plane preserving case. In the cases where we do not preserve a 4-plane, it seems likely that these criteria hint that the appropriate physical setting to understand the forms is supersymmetry-preserving in some sense.}. These rules suffice to determine the twining genera, up to the addition of cusp forms, which we constrain with the results of Corollary \ref{cor:useful}. 

\subsection{Twining genera for the Frame shapes of Umbral moonshine that do not fix a 4-plane}

Table \ref{tab:notFourFixing} lists the functions that meet our criteria which are associated to Frame shapes of Umbral symmetries that do not fix a 4-plane. The first three columns are analogous to columns of table \ref{tab:big}. The fourth column states which functions appear in Umbral moonshine; in addition, it includes our speculations on which functions may appear in an extended Conway moonshine. (The rows without a $\Lambda$ correspond to those functions that are eliminated by the above multiplier considerations, if they are correct). The final column specifies the multipliers of these functions (which are easily determined from \eqref{eq:multiplierSTransform}). We note that our speculations yield at least one function for each multiplier. In particular, whenever a function has a complex multiplier, there is a function with the complex conjugate multiplier; this is not the case with only the Umbral functions. In the 4-plane fixing case, this is due to the fact that worldsheet parity takes each twining genus to a twining genus with the conjugate multiplier. Thus, this observation may be further evidence that our new functions play some role in string theory.

In order to make this table readable, we define the following (non-cusp) forms for $\Gamma_0(44)$, $\Gamma_0(63)$, $\Gamma_0(80)$, and $\Gamma_0(144)$:
\begin{align*}
\E^{(44)}&=-\frac{1}{60}\E_2+\frac{1}{90}\E_4-\frac{11}{180}\E_{11}+\frac{11}{60}\E_{22}-\frac{11}{90}\E_{44}\\
\E^{(63)}&=-\frac{1}{48}\E_3-\frac{7}{192}\E_7+\frac{1}{64}\E_9+\frac{7}{48}\E_{21}-\frac{7}{64}\E_{63}-\frac{7}{3}\eta[1^33^{-1}7^321^{-1}]+\frac{1}{3}\eta[1^63^{-2}]\\
\E^{(80)}&=\frac{1}{72}\E_4-\frac{1}{24}\E_8+\frac{1}{36}\E_{16}-\frac{5}{72}\E_{20}+\frac{5}{24}\E_{40}-\frac{5}{36}\E_{80}-2\eta[1^{-1}2^74^{-2}5^110^{-1}]\\
\E^{(144)}&= \frac{1}{6}\E_{12}-\frac{1}{4}\E_{144}-\frac{1}{12}\E_{16}-\frac{1}{2}\E_{24}-\frac{1}{8}\E_{36}-\frac{1}{24}\E_4+\frac{1}{3}\E_{48}+\frac{3}{8}\E_{72}+\frac{1}{8}\E_8\\
&-2\eta[1^42^{-1}4^16^112^{-1}]-2\eta[1^12^53^14^{-2}6^{-1}]-2\eta[1^22^23^{-2}6^412^{-2}]\ .
\end{align*}
We also define the following cusp forms for $\Gamma_0(144)$:
\begin{align*}
f^{(144a)}&=6f_{24}(q^2)+54f_{24}(q^6)+3f_{48}(q)-27f_{48}(q^3)-18f_{72}(q^2)+9f_{144,a}+18f_{144,b}\\
f^{(144b)}&=-12f_{24}(q^2)-108f_{24}(q^6)+120f_{36}(q^4)-36f_{72}(q^2)-36f_{144,b}\\
f^{(144c)}&=-6f_{24}(q^2)-54f_{24}(q^6)+48f_{36}(q^4)-3f_{48}(q)+27f_{48}(q^3)-18f_{72}(q^2)-9f_{144,a}-18f_{144,b}\\
f^{(144d)}&=6f_{24}(q^2)+54f_{24}(q^6)+3f_{48}(q)-27f_{48}(q^3)+18f_{72}(q^2)-9f_{144,a}-18f_{144,b}\\
f^{(144e)}&=108f_{24}(q^6)+12f_{24}(q^2)+6f_{48}(q)-54f_{48}(q^3)+36f_{72}(q^2)-18f_{144,a}-36f_{144,b}\ .
\end{align*}
As usual, all of the special modular forms appearing in these definitions, and in table \ref{tab:notFourFixing}, are defined in appendix \ref{sec:basis}.

\setlength\LTleft{-0.5in}
\begin{center}
\rowcolors{2}{gray!11}{}
\begin{tabularx}{\linewidth}{CCCCC}\label{tab:notFourFixing}
\pi_g & (\Gamma_g)_{|\lambda} &   F_{e,g}(\tau) & \text{Niemeier} & \E'\modu{\lambda}\\
\midrule\endhead
 12^2 & \Gamma_0(12)_{|12} & \begin{matrix}
 2\eta[1^42^{-1}4^16^112^{-1}]+f^{(144a)}\\
 2\eta[1^42^{-1}4^16^112^{-1}]\\
\E^{(144)}+f^{(144b)}\\
 \E^{(144)}+f^{(144c)}\\
\E^{(144)}-24f_{36}(q^4)-6f_{48}(q)+54f_{48}(q^3)-18f_{144,a}\\
 2\eta[1^12^53^14^{-2}6^{-1}]\\
 2\eta[1^12^53^14^{-2}6^{-1}]-72f_{36}(q^4)+18f_{144,a}\\
 2\eta[1^12^53^14^{-2}6^{-1}]+f^{(144d)}\\
 2\eta[1^12^53^14^{-2}6^{-1}]+f^{(144e)}\\
 2\eta[1^22^23^{-2}6^412^{-2}]
 \end{matrix} &
 \begin{matrix}
 \\
 A_1^{24}\\
 \Lambda\\
 \Lambda\\
 \Lambda\\
 A_2^{12}\\
 \\
 \\
 \\
 A_4^6
 \end{matrix}
 &\begin{matrix}
 11\\
 11\\
 1\\
 1\\
 1\\
 7\\
 7\\
 7\\
 7\\
 5
 \end{matrix}
\\
4^120^1 & \Gamma_0(20)_{|4} & \begin{matrix}
2\eta[1^{-1}2^74^{-2}5^110^{-1}]\\
2\eta[1^{-1}2^74^{-2}5^110^{-1}]-80f_{20}(q^4)-\frac{75}{16}f_{40}+\frac{395}{16}f_{80,a}+40f_{80,b}\\
2\eta[1^{-1}2^74^{-2}5^110^{-1}]-\frac{75}{16}f_{40}+20f_{40}(q^2)-\frac{85}{16}f_{80,a}\\
\E^{(80)}+\frac{80}{3}f_{20}(q^4)-20f_{40}(q^2)
\end{matrix} &
 \begin{matrix}
 A_2^{12}\\
 \\
 \\
 \Lambda
 \end{matrix}
 &\begin{matrix}
 3\\
 3\\
 3\\
 1
 \end{matrix}
 \\
3^121^1 & \Gamma_0(21)_{|3} &\begin{matrix}
\E^{(63)}-\frac{63}{8}f_{21}(q)+\frac{567}{8}f_{21}(q^3)+\frac{63}{4}f_{63,a}+21 f_{63,b}\\
\E^{(63)}-\frac{21}{8}f_{21}(q)+\frac{189}{8}f_{21}(q^3)-\frac{21}{2}f_{63,b}\\
\frac{7}{3}\eta[1^33^{-1}7^321^{-1}]-\frac{1}{3}\eta[1^63^{-2}]
\end{matrix} &
 \begin{matrix}
 \Lambda\\
 \Lambda\\
 A_1^{24}
 \end{matrix}
 &\begin{matrix}
 1\\1\\2
 \end{matrix}\\
2^122^1 & \Gamma_0(22)_{|2} &
\E^{(44)}+\frac{11}{5}\eta[1^211^2]+\frac{44}{5}\eta[2^222^2]+\frac{88}{5}\eta[4^244^2]-\frac{22}{3}f_{44}
&
 \begin{matrix}
 \Lambda, A_2^{12}
 \end{matrix} & \begin{matrix}1\end{matrix}\\
1^{1}23^1 & \Gamma_0(23) 
 & \begin{matrix}
 -\frac{23}{264}\E_{23}-\frac{69}{11}\eta[1^223^2]-\frac{23}{11}	f_{23}\\
 -\frac{23}{264}\E_{23}-\frac{69}{11}\eta[1^223^2]+\frac{230}{11}f_{23}
 \end{matrix} &
 \begin{matrix}
 \Lambda, A_1^{24}\\
 \Lambda
 \end{matrix}
 &\begin{matrix}\\\end{matrix}
 \\
\bottomrule \caption{}
\end{tabularx}
\end{center}

\section{Discussion}\label{sec:dis}
In this paper we have computed a set of candidate twining genera, or $g$-equivariant elliptic genera, $\phi_{e, g}$ in K3 SCFTs for all possible symmetries of K3 NSLMs that preserve the $\mathcal{N}=(4, 4)$ superconformal algebra and spectral flow generators. For most of the $81$  conjugacy classes of the duality group $O^+(\Gamma^{4, 20})$ we were able to determine the twining genus uniquely. For the remaining classes, we have found two possible candidates. 
K3 NLSM twining genera are closely related, via the so-called multiplicative lift, to the generating functions of 1/4 BPS dyons in the CHL model labeled by $g$. Our computations have therefore provided an interesting set of data to explore several outstanding questions in the study of supersymmetric vacua and properties of both worldsheet and spacetime BPS states in string theory. We briefly comment on several natural avenues for follow-up exploration.
\begin{itemize}
\item The twining genera we find in this work are exactly the Jacobi forms occurring in Umbral moonshine and Conway moonshine (in the 4-plane fixing cases), as explained in the introduction and in \S \ref{sec:moonshine}. This is a surprising and highly nontrivial finding  which demands an explanation. Is there a deeper connection between the way string theory singles out these Jacobi forms and the way independent number theoretic considerations from moonshine (e.g. genus zero properties--- cf. e.g. \cite{Duncan:2014vfa}, Rademacher summability \cite{Cheng:2011ay}, etc.) privilege these forms? In fact, a  host of worldsheet-based evidence led the authors of \cite{slowpaper}  to put forth several conjectures concerning the role of the Niemeier groups and twining genera in K3 NSLMs. Loosely speaking, they conjectured that for any symmetry of a  K3 NLSM, the corresponding twining genus would coincide with one of the Umbral or Conway moonshine functions and, conversely, that each of the Umbral/Conway functions appears as the twining genus of  some symmetry of a K3 NSLM. In our work, using very general spacetime considerations, we have essentially proved the first of these conjectures and provided strong evidence in favor of the second one. It is conceivable that our methods shed light on the physical role of the Jacobi forms (and even more speculatively, the mock modular forms) appearing in moonshine.

\item There has been interesting work connecting Borcherds-Kac-Moody (BKM) algebras to BPS states in the $K3 \times T^2$ compactification and in some of the simplest CHL models \cite{DVV,CV, Cheng:2008kt,Govindarajan:2008vi, Govindarajan:2009qt, Govindarajan:2010fu, Govindarajan:2011mp}. The generating functions themselves are essentially denominators of certain Borcherds-Kac-Moody algebras in favorable cases, and the Weyl group plays the role of a discrete analogue of the attractor flow, providing an algebraic interpretation of wall-crossing. However, for $N \geq 4$, it appears that there is no simple BKM interpretation: the connected components of the moduli space appear to be bounded by an infinite number of walls, which stymies the beautiful algebraic picture advocated in \cite{CV}. On the other hand, the poles of $1/\Phi_{g, e}$ should correspond to bosonic real roots, while the zeroes should correspond to fermionic real roots. Do fermionic real roots give additional generators of the Weyl group that ameliorate this problem? More generally, can we complete the dictionary between BKM data and BPS dyons established in \cite{CV, Cheng:2008kt}?

\item As indicated by the brackets in table \ref{tab:big}, we were unable to complete the association of twining genera to symmetries (or, more precisely, to $O^+(\Gamma^{4,20})$ classes). Perhaps considerations of the Borcherds-Kac-Moody symmetry enjoyed by the BPS states in many CHL compactifications will eliminate these persistent ambiguities.

\item Throughout our paper, we employ the classification of CHL models completed by \cite{PV}, which focuses on models that have a perturbative frame in which they are well-described by an orbifold of $K3 \times T^2$; this results in the Narain lattice splitting $\Gamma^{6, 22} = \Gamma^{4, 20} \oplus \Gamma^{2, 2}$. One could of course consider orbifolds of string theory with $16$ supercharges  by symmetries that are not symmetries of the perturbative K3. Can we still define the appropriate Siegel forms (either by a multiplicative lift or an alternative construction) and, if so, are they determined by our constraints? Such compactifications seem to be a natural place to look for symmetries that fix less than a four plane and, optimistically, to recover all Umbral twining functions in a \textit{physical} setting. 

\item One possible explanation of the relationship between Umbral/Conway moonshines and string theory on $K3 \times T^d$ for various $d$ has begun to emerge, in the setting of low-dimensional string compactifications \cite{kachru20163d,harrison2016heterotic}. In the 3d setting of  type II string theory on $K3\times T^3$, the Niemeier lattices appear at points in the moduli space where the lattice parameterizing (nonperturbative) points in the string moduli space, $\Gamma^{8,24}$, decomposes as $E_8\oplus L$ \cite{kachru20163d}. If we compactify even further to two dimensions, then there even exists a perturbative description of each of these points in moduli space, as the Narain lattice associated to the  heterotic string on $T^8$ (which is dual to type II on $K3\times T^4$) is $\Gamma^{8,24}$ \cite{harrison2016heterotic}. Referring back to the 3d picture, only 4-plane fixing symmetries will survive the decompactification to 6 dimensions (concomitantly taking the type II string coupling to zero so that the K3 sigma model description is good). Can we use this picture to identify which (Umbral or Conway) twining genera appear at a specified point in moduli space which allows for multiple Niemeier embeddings of $\Xi_g$ (in the notation of \S\ref{sec:moonshine})?

\item Building on the previous point---and as an alternative approach to the fourth item on our list---we may also hope to identify the non-4-plane preserving twining genera of Umbral and (proposed extended) Conway moonshine in the 2d or 3d pictures. In these theories, we evade the 4-plane preserving condition imposed upon us in the study of K3 NSLMs. For instance, although the 3d moduli space is nonperturbative, it is conceivable that the twining genera might appear in an appropriate physical quantity---e.g. a contribution to a term in the low-energy effective action---if we judiciously choose an appropriate duality frame with a familiar perturbative (heterotic or type II) description.

\item Temporarily eschewing CHL constructions, we applied our general constraints to the 2-plane preserving $O(\Lambda)$ conjugacy classes which are labeled by Frame shapes that also appear in Umbral moonshine. As before, the procedure yielded a limited set of functions, which we delineated in table \ref{tab:notFourFixing}. Some of these functions coincide with the twining genera of Umbral moonshine, while the other functions do not appear in any existing moonshine-based recipe for generating twining genera. However, the Conway moonshine of \cite{Duncan:2015xoa} was limited to 4-plane preserving conjugacy classes, which directly correspond to SUSY-preserving symmetries of K3 NLSMs. The functions in table \ref{tab:notFourFixing} might suggest an extension of the Conway moonshine recipe for producing Jacobi forms. If so, what is the physical meaning of these Jacobi forms and their relation to (string theory on) K3? It is clearly desirable to have a better understanding of these functions. As a first step, one might try to apply the full constraints summarized in \S\ref{sec:secquant} rather than the slightly weakened Corollary \ref{cor:useful} we employed in computations. Would this eliminate any functions in table \ref{tab:notFourFixing}?

\item Throughout this paper, we have focused on `torsion-free' dyons, i.e. dyons with the discrete T-duality invariant $I \equiv \gcd(Q \wedge P) = 1$. In  several cases, dyons with more general $I$ have been counted \cite{dabholkar2011counting, banerjee2008partition, Banerjee:2008pv}, including analyses for all $I$ in the unorbifolded case. It would be interesting to find the counting functions for all $I$ for our CHL models, which already have a more elaborate structure of (continuous) T-duality orbits that remains to be fully understood. Furthermore, for general $I$ it would be interesting to deduce the properties of the dyon counting functions, explore BKM interpretations thereof, and so on. 

\item The growth rate of the coefficients of ordinary modular and Jacobi forms have been explored extensively in both mathematics and physics, including with recent applications to holography in e.g.  \cite{hartman2014universal, Keller:2014xba,  Belin:2014fna, benjamin2015elliptic, benjamin2015emergent, Belin:2015hwa}. It would be similarly fruitful to derive constraints on Siegel modular forms to obtain growth rates that would guarantee, e.g. an extended regime of validity for Cardy-like growth. Are the constraints in this paper a (perhaps roundabout) way to guarantee `slow growth'? There has been interesting recent work studying Siegel forms obtained via multiplicative lifts and studying putative (subleading) contributions to macroscopic black hole entropy \cite{Belin:2016knb}. It would be educational to extend the analysis of this paper to our generating functions. More modestly, since we have a large new class of CHL dyon counting functions, it would be instructive to check if we reproduce the Bekenstein-Hawking entropy in the limit of large charges, as expected.

\item We have discovered that our spacetime counting functions are determined from minimal data, namely $1/2$ BPS degeneracies on the \textit{worldsheet}, plus information about the location of the walls. Firstly, it would be satisfying to have a deeper explanation for why these intricate functions, which contain much dynamical data, are fixed by such paltry information. Is there a more natural way to constrain the functions than the methods we employ here? For the unorbifolded case, it is known that the $1/4$ BPS counting function is completely determined by Siegel automorphy plus the $1/2$ BPS counting functions, which are manifest as one studies the degeneration limit $z \rightarrow 0$. Is such a phenomenon general? If not, what additional information is required to fix the CHL counting functions for larger $N$ and/or nontrivial $\lambda$? Note also that the constraint of Corollary \ref{cor:useful} was sufficient in practice to fix our Jacobi forms (in the 4-plane fixing cases), though it was strictly weaker than the general constraints to  eliminate unphysical walls that we derived in \S\ref{sec:secquant}. One first step towards understanding  the power of the various constraints discussed might be to understand why this weaker condition is nonetheless so effective.

\end{itemize}

\section*{Acknowledgments}
We are grateful to M. Cheng, S. Harrison, and S. Kachru for interesting discussions on related topics and for helpful comments on earlier drafts of this paper. N.M.P. is supported by an NSF Graduate Research Fellowship under grant DGE-114747 and gratefully acknowledges QMAP at UC Davis for hospitality while this work was in progress. RV is supported by a grant from Programma per Giovani Ricercatori `Rita Levi Montalcini', and thanks SLAC and the Stanford Institute for Theoretical Physics for hospitality. M.Z. is supported by the Mellam Family Fellowship at the Stanford Institute for Theoretical Physics.

\appendix

\section{Basics on modular forms and Jacobi forms}\label{a:basics}
The classical theta functions are Jacobi forms of weight $1/2$ and index $1$ and can be written as follows, using $q:= e^{2 \pi i \tau}, y:= e^{2 \pi i z}$:

\begin{align}
\theta_1(\tau,z) 
&= -i q^{\frac{1}{8}} y^{\half} \prod_{n=1}^{\infty} (1-q^n)(1-y q^n)(1 - y^{-1} q^{n-1}) \\  
&= i \sum_{n=-\infty}^{\infty} (-1)^n q^{\frac{(n-\half)^2}{2}} y^{n-\half} \nonumber \\
\theta_2(\tau,z) 
&=q^{\frac{1}{8}} y^{\half} \prod_{n=1}^{\infty} (1-q^n) (1+y q^n) (1 + y^{-1} q^{n-1}) \\
&= \sum_{n=-\infty}^{\infty} q^{\frac{(n-\half)^2}{2}} y^{n-\half} \nonumber \\
\theta_3(\tau,z) 
&= \prod_{n=1}^{\infty} (1-q^n) (1 + y q^{n-\half}) (1 + y^{-1} q^{n-\half})  \\
&= \sum_{n=-\infty}^{\infty} q^{\frac{n^2}{2}} y^n \nonumber \\
\theta_4(\tau,z)
&= \prod_{n=1}^{\infty} (1-q^n) (1 - y q^{n-\half}) (1 - y^{-1} q^{n-\half}) \\
&= \sum_{n=-\infty}^{\infty} (-1)^n q^{\frac{n^2}{2}} y^n \nonumber.
\end{align}

The usual Dedekind eta function of weight $\half$ is defined to be
\begin{equation}
\eta(\tau) = q^{\frac{1}{24}} \prod_{n=1}^{\infty} (1-q^n)=q^\frac{1}{24}\sum_{n=-\infty}^\infty (-1)^n q^\frac{3n^2-n}{2}=q^{1/24}(1-q-q^2+O(q^3)).
\end{equation}
This is modular for $SL(2,\ZZ)$ with a multiplier system, $v_\eta$: if $\gamma=\twoMatrix{a}{b}{c}{d}$, then
\be \eta\parens{\gamma\tau}=v_\eta(\gamma)(c\tau+d)^{1/2}\eta(\tau) \ , \ee
where $v_\eta(\gamma)$ is a phase. We determine the phase via the following rules:
\begin{align}
v_\eta(\gamma)&=\piecewise{e^{\frac{b i\pi}{12}}}{c=0,d=1}{e^{i\pi\brackets{\frac{a+d}{12c}-s(d,c)-\frac{1}{4}}}}{c>0}\\
s(h,k)&=\sum_{n=1}^{k-1}\frac{n}{k}\parens{\frac{hn}{k}-\floor{\frac{hn}{k}}-\frac{1}{2}}.
\end{align}
Thinking of $\gamma$ as being valued in $PSL(2,\ZZ)$, we can always multiply $\gamma$ by $\pm1$ and end up in one of the cases $(c=0,d=1)$ or $c>0$.

We also write the standard Jacobi forms $\chi_{0, 1}(\tau, z)$ of weight $0$ and index $1$ and $\chi_{-2, 1}(\tau, z)$ of weight $-2$ and index $1$ \cite{eichler_zagier}:
\begin{align*}
\chi_{0, 1}(\tau, z)&= 4\left(\sum_{i=2}^4 \frac{\theta_i(\tau, z)^2}{\theta_i(\tau, 0)^2} \right) = (y^{-1}+10+y)+O(q) \\
\chi_{-2, 1}(\tau, z)&= -\frac{\theta_1(\tau, z)^2}{\eta(\tau)^6} = (y^{-1}-2+y)+O(q).
\end{align*}

\section{Modular forms for congruence subgroups}

\subsection{Introduction}\label{sec:congruence}

In this section we describe some properties of the spaces of weight 2 modular forms for congruence subgroups\cite{steinmodular}. Such modular forms are defined to transform in the usual way, except only under a subgroup of $SL(2,\ZZ)$ which is defined via congruence relations. In order to specify this transformation more explicitly, we introduce the following actions of $GL^+(2,\RR)$ (the $+$ indicates restriction to matrices with positive determinant) on the upper half plane, $\HH=\{x+iy\in\CC|y>0\}$, and the set $\F$ of functions $f:\HH\rightarrow\CC$:
\begin{align}
\alpha\tau&=\frac{a\tau+b}{c\tau+d},\quad \tau\in\HH,\, \\
(f|\alpha)(\tau)&=(\det\alpha)(c\tau+d)^{-2}f(\alpha\tau),\quad \alpha=\twoMatrix{a}{b}{c}{d}\in GL^+(2,\RR)\ .
\end{align}
Modular forms of weight two for the congruence subgroup $\Gamma\subset SL(2,\ZZ)$ are functions $f\in \F$ that satisfy
\be f|\alpha=f,\quad \alpha\in \Gamma\ ,\ee
as well as certain growth conditions at the cusps $\QQ\PP^1=\QQ\cup\{\infty\}$. Defining $\hat\HH=\HH\cup \QQ\PP^1$, we can restate this definition as the requirement that $f(\tau)d\tau$ be a meromorphic $1$-differential on $\hat \HH/\Gamma$ with at most single poles at the cusps and which is holomorphic elsewhere.  We will frequently have cause to modify this definition slightly, by allowing for a multiplier, that is, a phase $\xi(\alpha)$ on the right side of this equation which is independent of $\tau$.

Examples of popular congruence subgroups are the principal congruence subgroup of level $N>0$,
\be \Gamma(N)=\braces{\twoMatrix{a}{b}{c}{d}\in SL(2,\ZZ): a,d\equiv 1\mbox{ (mod $N$)},b,c\equiv 0 \mbox{ (mod $N$)}}\ , \ee
and the Hecke congruence subgroup of level $N$,
\be \Gamma_0(N)=\braces{\twoMatrix{a}{b}{c}{d}\in SL(2,\ZZ): c\equiv 0 \mbox{ (mod $N$)}} \ .\ee
We can define subgroups of $\Gamma_0(N)$ corresponding to any subgroup $G$ of $(\ZZ/N\ZZ)^\times$ as follows:
\be \Gamma_G(N)= \braces{\twoMatrix{a}{b}{c}{d}\in SL(2,\ZZ): a,d\mbox{ (mod $N$)}\in G,c\equiv 0 \mbox{ (mod $N$)}}\ .\ee
We need the cases $G=(\ZZ/N\ZZ)^\times$ (corresponding to $\Gamma_G(N)=\Gamma_0(N)$), $G=\langle -1\mbox{ (mod $N$)}\rangle$ (in which case we write $\Gamma_G(N)=\gmo{N}$), and the case where $G$ is the trivial group (in which case we use the standard notation $\Gamma_G(N)=\Gamma_1(N)$) in the main text. A congruence subgroup is said to be of level $N$ if it contains $\Gamma(N)$ and does not contain $\Gamma(M)$ for any $M<N$. Modular forms for a level $N$ congruence subgroup are themselves also said to be of level $N$.

One important difference between congruence subgroups, $\Gamma$, and the full group $SL(2,\ZZ)$ is that the quotient of the upper half plane $\hat\HH$ by the former will not, in general, identify all cusps $\cc.$ Thus, while modular forms for $SL(2,\ZZ)$ have a single $q$-expansion (say, about $\infty$), modular forms for $\Gamma$ may have inequivalent $q$-expansions about cusps that are not identified by $\Gamma$. Such an expansion about a cusp $\cc$ is a power series in $q_{\cc}=e^{2\pi i\tau_\cc/w_\cc}$, where $\tau_\cc=\gamma\tau$ is the image of $\tau$ under a transformation $\gamma\in PSL(2,\ZZ)$ that maps $\cc$ to $i\infty$, and where $w_{\cc}$, the width of the cusp $\cc$ relative to $\Gamma$, is the smallest positive integer $H$ such that $\gamma^{-1}\twoMatrix{1}{H}{0}{1}\gamma\in \Gamma$. Note that we can replace $\tau_\cc$ by $\tau'_\cc=\tau_\cc+n$ in the last sentence, where $n$ is an arbitrary integer; this means that when $w_\cc\not=1$ there is a certain arbitrariness in the definition of the phase of $q_\cc$, as we may replace $q_\cc$ by $e^{2\pi i n/w_\cc}q_\cc$. Thus, implicit in the notation $\tau_\cc$ is our choice of $\gamma$. However, $w_\cc$ is clearly independent of this choice.

We now describe the complex vector space $M_2(\Gamma)$ of weight $2$ modular forms for $\Gamma$. First, we introduce the subspace $\calS_2(\Gamma)$ of cusp forms whose $q$-expansions about all cusps vanish at order $q^0$. We then define $\N_2(\Gamma)=M_2(\Gamma)/\calS_2(\Gamma)$, so that
\be M_2(\Gamma)=\N_2(\Gamma)\oplus\calS_2(\Gamma)\ .\ee
When $\Gamma_1(N)\subset\Gamma$ (and so, in particular, when $\Gamma$ is of the form $\Gamma_G(N)$), $\N_2(\Gamma)$ is spanned by generalized Eisenstein series, as described in appendix \ref{sec:basis}, and is therefore called the Eisenstein subspace. This subspace has dimension $n-1$, where $n$ is the number of cusps that are not identified by $\Gamma$. The cuspidal subspace has a basis that may be described in terms of certain special cusp forms, called newforms; its dimension equals the genus of $\hat\HH/\Gamma$, which we simply call the genus of $\Gamma$. This follows easily from Hodge theory if we note that weight 2 cusp forms correspond to holomorphic one-forms on $\hat\HH/\Gamma$. The dimension of $M_2(\Gamma)$, and thus of $\N_2(\Gamma)$, follows from the Riemann-Roch theorem.

\subsection{A basis for $M_2(\Gamma)$}\label{sec:basis}

In this section, we introduce the weight 2 modular forms in terms of which we will express the functions $F_{e,g}$.

\subsubsection[Eisenstein series: Was Eisenstein Wrong?]{Eisenstein series}\label{sec:eisenstein}

We begin with Eisenstein series; the interested reader can refer to \cite{miyakemodular,steinmodular} for more on the subject. These are defined by the following $q$-expansion about infinity:
\begin{equation}\label{eq:eisensteinInfty}
E_{k,\chi,\psi}(\tau)=c_0+\sum_{m\ge 1}\parens{\sum_{n|m}\psi(n)\chi(m/n)n^{k-1}}q^m,\quad c_0=\piecewise{0}{k_\chi>1}{-\frac{B_{k,\psi}}{2k}}{k_\chi=1}.
\end{equation}
Here, $k$ is the weight of the Eisenstein series, $\chi$ and $\psi$ are primitive Dirichlet characters with conductors $k_\chi$ and $k_\psi$, respectively, and $B_{k,\psi}$ is a Bernoulli number. (A Dirichlet character mod $M$ induces, in a natural way, a Dirichlet character mod $N$, where $N$ is any multiple of $M$; a primitive character is a character which is not induced in such a way. The modulus of a primitive character is called its conductor. Of particular importance is the primitive principal character, which we denote by $\varepsilon_0$: $\varepsilon_0(n)=1$ for all $n$. This is the unique primitive character with conductor 1. For simplicity, we define $E_k=E_{k,\varepsilon_0,\varepsilon_0}$; we emphasize that our choice of normalization is such that $E_2$ has $q$-expansion $E_2(q)=-\frac{1}{24}+q+O(q^2)$).

We henceforth specialize to the case of weight $k=2$. A basis for the space $\N_2(\Gamma_1(N))$ of Eisenstein forms at level $N$ is given by the functions $E_{k,\chi,\psi}(q^t)$ where $t$ is a positive integer such that $k_\chi k_\psi t|N$, $\chi(-1)=\psi(-1)$, and at least one of $k_\chi$ and $k_\psi$ differs from 1 (that is, at least one of the characters is non-principal), plus the functions $\E_t(q)=-24\brackets{E_2(q)-tE_2(q^t)}$ for all divisors $t>1$ of $N$.  (The fact that Eisenstein series associated to primitive characters span the space of Eisenstein forms is why we restricted our attention to such characters, even though generalizations of Eisenstein series exist for other characters. The restriction $\chi(-1)=\psi(-1)$ simply eliminates trivial functions that vanish identically. We introduced the factor $-24$ in the definition of $\E_t$ so that the $q^0$ term in its $q$-expansion is $1-t$).

\subsubsection{Eta products}\label{sec:etaProds}

Eta products\cite{kohlermodular} are functions of the following form:
$$\eta\brackets{\prod_{t>0}t^{m_t}}(\tau)=\prod_{t>0} \eta^{m_t}(t \tau),$$
where $m_t\in \ZZ$ are non-vanishing only for a finitely many $t$. The formal product on the left hand side is not supposed to be evaluated -- it is to be regarded as a symbol that specifies the eta product under consideration. (Nonzero values of $m_t$ may be either positive or negative; mathematicians sometimes refer to such functions as eta quotients to differentiate from the case where all $m_t$'s are non-negative). We will be interested in a special class of eta products, called holomorphic eta products. These are eta products whose $q$-expansions at all cusps $\cc$ contain no negative powers of $q_\cc$; they are modular forms for $\Gamma_0(N)$ of weight $\sum_t m_t/2$ (generally with a multiplier system -- see \cite{ref:etaSpace} for conditions for the multiplier to be trivial), where $N$ is the least common multiple of the integers in the set $\{t:m_t\not=0\}$. See \cite{kohlermodular} for necessary and sufficient conditions for an eta product to be holomorphic. A sufficient, but not necessary, such condition is $m_t\ge0$ for all $t$; in particular, $\eta(t\tau)$ is a modular form for $\Gamma_0(t)$ with weight $1/2$. Since we are interested in modular forms of weight 2, we will always have $\sum_t m_t=4$.

Eta products will find two uses in the main text. First, we will frequently be able to express the cusp forms that arise in terms of holomorphic eta products. In addition, because eta products can be easily expanded about an arbitrary cusp, while Eisenstein series with non-trivial characters cannot (as we explain in appendix \ref{sec:modularCusp}), we will replace the latter functions with holomorphic eta products. Holomorphic eta products plus the functions $\E_t$ generally do not span the spaces $\N_2(\hat\Gamma_g)$, but nevertheless these functions suffice for us. This is not an accident: the spaces of relevance are not really $\N_2(\hat\Gamma_g)$, but rather the smaller spaces $\N_2^{\xi_{e,g}}(\Gamma_g)$ of modular forms for $\Gamma_g$ with the correct multiplier. Although we have not proven this, it seems likely that holomorphic eta products plus the functions $\E_t$ span $\N_2^{\xi_{e,g}}(\Gamma_g)$ for all $g$.

\subsubsection{Newforms}\label{sec:newforms}

Finally, we define certain newforms for various groups of the form $\Gamma_0(N)$. Strictly speaking, we do not need most of these definitions, since as we show below most of these functions may be (non-canonically) expanded in terms of holomorphic eta products. (We provide these expansions in order to allow one to easily determine the behavior of these functions at arbitrary cusps, using the methods of appendix \ref{sec:modularCusp}. These expansions were determined by slightly modifying the MAGMA code of \cite{ref:etaSpace} in order to output a basis of weight 2 holomorphic eta products at level $N$). We nonetheless introduce these functions, first because they provide a convenient shorthand notation, and second because their use enables easy comparison of our results with those of \cite{Cheng:2014zpa,Persson:2013xpa}.
\begin{align*}
f_{20}(q)&=q - 2q^3 - q^5 + 2q^7 + q^9 + 2q^{13} + 2q^{15} - 6q^{17}+\ldots\\
&=\frac{3}{2}\eta[1^{-6}2^{13}4^{-5}5^{6}10^{-5}20^{1}]-15\eta[2^{-4}4^{8}]+15\eta[10^{-4}20^{8}]+\frac{1}{16}\eta[1^{8}2^{-4}]\\
&\qquad-\frac{15}{2}\eta[1^{1}2^{-4}4^{3}5^{-5}10^{16}20^{-7}]+\frac{95}{16}\eta[5^{8}10^{-4}]\\
f_{21}(q)&=q - q^2 + q^3 - q^4 - 2q^5 - q^6 - q^7 + 3q^8 + q^9 +\ldots\\
&=\frac{2}{9}\eta[1^{3}3^{-1}7^{3}21^{-1}]+18\eta[3^{-1}9^{3}21^{-1}63^{3}]+2\eta[1^{3}3^{-1}21^{-1}63^{3}]-\frac{1}{18}\eta[1^{-3}3^{10}9^{-3}]\\
&\qquad+2\eta[3^{-1}7^{3}9^{3}21^{-1}]-\frac{7}{18}\eta[7^{-3}21^{10}63^{-3}]+\frac{1}{36}\eta[1^{6}3^{-2}]+\frac{7}{36}\eta[7^{6}21^{-2}]-\frac{3}{4}\eta[3^{-2}9^{6}]\\
&\qquad-\frac{21}{4}\eta[21^{-2}63^{6}]\\
f_{23}(q)&=q-q^3-q^4-2q^6+2q^7-q^8+2q^9+2q^{10}+\ldots,\\
f_{24}(q)&=q - q^3 - 2q^5 + q^9 + 4q^{11} - 2q^{13} + 2q^{15} + 2q^{17}+\ldots\\
&=-\frac{41}{9}\eta[2^{-4}4^{8}]+\frac{5}{18}\eta[4^{8}8^{-4}]-\frac{13}{36}\eta[1^{8}2^{-4}]+\frac{80}{9}\eta[8^{-4}16^{8}]+25\eta[6^{-4}12^{8}]+\frac{3}{2}\eta[12^{8}24^{-4}]\\
&\qquad+\frac{5}{4}\eta[3^{8}6^{-4}]+48\eta[24^{-4}48^{8}]+6\eta[1^{2}2^{-4}3^{-6}6^{12}]-\frac{8}{3}\eta[1^{1}3^{1}4^{-5}6^{-2}8^{8}12^{5}24^{-4}]\\
f_{36}(q)&=q - 4q^7 + 2q^{13} + 8q^{19} - 5q^{25} - 4q^{31} - 10q^{37}+\ldots\\
&=-6\eta[1^{-1}4^{1}6^{7}9^{1}12^{-5}18^{-5}36^{6}]+\frac{4}{3}\eta[2^{-4}4^{8}]+6\eta[18^{-4}36^{8}]+\frac{1}{24}\eta[1^{8}2^{-4}]-\frac{21}{8}\eta[9^{8}18^{-4}]\\
&\qquad+\frac{58}{3}\eta[6^{-4}12^{8}]+\frac{11}{12}\eta[3^{8}6^{-4}]+\frac{2}{3}\eta[2^{6}6^{-2}]+3\eta[1^{2}2^{-4}3^{-6}6^{12}]-\frac{5}{3}\eta[3^{-6}6^{12}9^{2}18^{-4}]\\
&\qquad+\frac{8}{3}\eta[4^{6}12^{-2}]\\
f_{40}(q)&=q + q^5 - 4q^7 - 3q^9 + 4q^{11} - 2q^{13} + 2q^{17} +\ldots\\
&=-\frac{560}{3}\eta[40^{-4}80^{8}]+\frac{3}{2}\eta[1^{-6}2^{13}4^{-5}5^{6}10^{-5}20^{1}]-\frac{29}{3}\eta[2^{-4}4^{8}]-3\eta[2^{-6}4^{13}8^{-5}10^{6}20^{-5}40^{1}]\\
&\qquad+\frac{65}{3}\eta[10^{-4}20^{8}]-\frac{1}{6}\eta[4^{8}8^{-4}]-\frac{35}{6}\eta[20^{8}40^{-4}]+2\eta[1^{-1}2^{2}4^{2}5^{1}8^{-1}40^{1}]-\frac{5}{24}\eta[1^{8}2^{-4}]\\
&\qquad+\frac{185}{24}\eta[5^{8}10^{-4}]-\frac{16}{3}\eta[8^{-4}16^{8}]\\
f_{44}(q)&=q+q^3-3q^5+2q^7-2q^9-q^{11}-4q^{13}+\ldots\\
&=-\frac{44}{5}\eta[2^{-4}4^{8}]-\frac{11}{20}\eta[1^{8}2^{-4}]+\frac{44}{5}\eta[22^{-4}44^{8}]+\frac{11}{20}\eta[11^{8}22^{-4}]+\frac{27}{5}\eta[1^{2}11^{2}]+\frac{12}{5}\eta[4^{2}44^{2}]\\
&\qquad+24\eta[2^{2}22^{2}]+24\eta[1^{-2}2^{4}11^{-2}22^{4}]-6\eta[1^{3}4^{-1}11^{-1}44^{3}]\\
f_{48}(q)&=q + q^3 - 2q^5 + q^9 - 4q^{11} - 2q^{13} - 2q^{15} + 2q^{17} + 4q^{19}+\ldots\\
&=\frac{26}{9}\eta[2^{-4}4^{8}]-\frac{25}{36}\eta[4^{8}8^{-4}]+\frac{4}{3}\eta[1^{1}2^{-3}3^{-1}4^{5}6^{4}8^{2}12^{-3}16^{-1}24^{-1}48^{1}]+\frac{4}{9}\eta[1^{8}2^{-4}]\\
&\qquad-\frac{32}{9}\eta[8^{-4}16^{8}]-14\eta[6^{-4}12^{8}]-\frac{1}{4}\eta[12^{8}24^{-4}]+\frac{1}{2}\eta[3^{8}6^{-4}]-32\eta[24^{-4}48^{8}]\\
&\qquad-6\eta[1^{2}2^{-4}3^{-6}6^{12}]+\frac{1}{3}\eta[1^{6}2^{-3}8^{-1}16^{2}]+3\eta[3^{6}6^{-3}24^{-1}48^{2}]-\frac{4}{3}\eta[2^{8}4^{-5}6^{-4}12^{5}16^{1}24^{-2}48^{1}]\\
f_{63,a}(q)&=q + q^2 - q^4 + 2 q^5 - q^7 - 3 q^8 + 2 q^{10} - 4 q^{11} - 2 q^{13}+\ldots\\
&=\frac{4}{3}\eta[1^{3}3^{-1}21^{-1}63^{3}]-\frac{1}{27}\eta[1^{-3}3^{10}9^{-3}]+\frac{4}{3}\eta[3^{-1}7^{3}9^{3}21^{-1}]-\frac{7}{27}\eta[7^{-3}21^{10}63^{-3}]+\frac{1}{27}\eta[1^{6}3^{-2}]\\
&\qquad+\frac{7}{27}\eta[7^{6}21^{-2}]+\eta[3^{-2}9^{6}]+7\eta[21^{-2}63^{6}]\\
f_{63,b}(q)&=q + q^4 + q^7 - 6q^{10} + 2q^{13} - 5q^{16} - 4q^{19} +\ldots\\
&=-\eta[1^{3}3^{-1}21^{-1}63^{3}]+\eta[3^{-1}7^{3}9^{3}21^{-1}]\\
f_{72}(q)&=q + 2q^5 - 4q^{11} - 2q^{13} - 2q^{17} - 4q^{19} + 8q^{23} - q^{25}+\ldots\\
&=4\eta[1^{-1}4^{1}6^{7}9^{1}12^{-5}18^{-5}36^{6}]-\frac{44}{9}\eta[2^{-4}4^{8}]+4\eta[18^{-4}36^{8}]-4\eta[2^{-1}8^{1}12^{7}18^{1}24^{-5}36^{-5}72^{6}]\\
&\qquad+\frac{1}{3}\eta[4^{8}8^{-4}]-\frac{3}{2}\eta[36^{8}72^{-4}]-\frac{29}{72}\eta[1^{8}2^{-4}]+\frac{25}{8}\eta[9^{8}18^{-4}]+\frac{32}{3}\eta[8^{-4}16^{8}]\\
&\qquad-8\eta[1^{1}2^{-1}3^{1}8^{2}12^{-1}18^{-1}36^{1}72^{2}]-48\eta[72^{-4}144^{8}]+\frac{8}{9}\eta[6^{-4}12^{8}]+\frac{7}{6}\eta[12^{8}24^{-4}]\\
&\qquad+\frac{5}{18}\eta[3^{8}6^{-4}]+\frac{112}{3}\eta[24^{-4}48^{8}]-\frac{2}{3}\eta[2^{6}6^{-2}]+\frac{8}{9}\eta[8^{6}24^{-2}]+3\eta[1^{2}2^{-4}3^{-6}6^{12}]\\
&\qquad+\frac{17}{9}\eta[3^{-6}6^{12}9^{2}18^{-4}]-\frac{22}{9}\eta[4^{6}12^{-2}]-\frac{8}{3}\eta[1^{1}3^{1}4^{-5}6^{-2}8^{8}12^{5}24^{-4}]\\
f_{80,a}(q)&=q + q^5 + 4q^7 - 3q^9 - 4q^{11} - 2q^{13} + 2q^{17} +\ldots\\
&=\frac{4160}{9}\eta[40^{-4}80^{8}]-\frac{128}{3}\eta[4^{-5}8^{13}16^{-6}20^{1}40^{-5}80^{6}]+\frac{1}{6}\eta[1^{-6}2^{13}4^{-5}5^{6}10^{-5}20^{1}]\\
&\qquad+\frac{8}{3}\eta[1^{1}2^{-3}4^{6}5^{1}8^{-3}16^{1}80^{1}]-2\eta[2^{-4}4^{8}]-\frac{1}{3}\eta[2^{-6}4^{13}8^{-5}10^{6}20^{-5}40^{1}]-\frac{290}{9}\eta[10^{-4}20^{8}]\\
&\qquad-\frac{5}{3}\eta[1^{2}2^{3}4^{-2}8^{-1}16^{2}]-\frac{83}{12}\eta[4^{8}8^{-4}]+\frac{5}{3}\eta[5^{2}10^{3}20^{-2}40^{-1}80^{2}]+2\eta[1^{-1}2^{6}4^{-3}5^{1}20^{1}]\\
&\qquad-\frac{95}{36}\eta[20^{8}40^{-4}]-\frac{2}{3}\eta[1^{-1}2^{2}4^{2}5^{1}8^{-1}40^{1}]-\frac{15}{8}\eta[1^{8}2^{-4}]+\frac{40}{3}\eta[1^{1}2^{-4}4^{3}5^{-5}10^{16}20^{-7}]\\
&\qquad-\frac{125}{72}\eta[5^{8}10^{-4}]-\frac{8}{3}\eta[2^{-1}4^{2}8^{2}10^{1}16^{-1}80^{1}]-\frac{256}{3}\eta[8^{-4}16^{8}]\\
f_{80,b}(q)&=q^3 - q^5 - 3q^7 + 2q^9 + 2q^{11} + 2q^{13} - q^{15} +\ldots\\
&=-\frac{730}{9}\eta[40^{-4}80^{8}]+\frac{58}{3}\eta[4^{-5}8^{13}16^{-6}20^{1}40^{-5}80^{6}]+\frac{1}{24}\eta[1^{-6}2^{13}4^{-5}5^{6}10^{-5}20^{1}]\\
&\qquad+\frac{2}{3}\eta[1^{1}2^{-3}4^{6}5^{1}8^{-3}16^{1}80^{1}]-\frac{11}{3}\eta[2^{-4}4^{8}]+\frac{13}{6}\eta[2^{-6}4^{13}8^{-5}10^{6}20^{-5}40^{1}]+\frac{100}{9}\eta[10^{-4}20^{8}]\\
&\qquad+\frac{5}{6}\eta[1^{2}2^{3}4^{-2}8^{-1}16^{2}]+\frac{79}{24}\eta[4^{8}8^{-4}]-\frac{5}{6}\eta[5^{2}10^{3}20^{-2}40^{-1}80^{2}]-\eta[1^{-1}2^{6}4^{-3}5^{1}20^{1}]\\
&\qquad+\frac{365}{72}\eta[20^{8}40^{-4}]-\frac{1}{6}\eta[1^{-1}2^{2}4^{2}5^{1}8^{-1}40^{1}]+\frac{97}{96}\eta[1^{8}2^{-4}]-\frac{35}{3}\eta[1^{1}2^{-4}4^{3}5^{-5}10^{16}20^{-7}]\\
&\qquad+\frac{25}{288}\eta[5^{8}10^{-4}]+\frac{7}{3}\eta[2^{-1}4^{2}8^{2}10^{1}16^{-1}80^{1}]+\frac{118}{3}\eta[8^{-4}16^{8}]\\
f_{144,a}(q)&=q + 4q^7 + 2q^{13} - 8q^{19} - 5q^{25} + 4q^{31} - 10q^{37} - 8q^{43}+\ldots\\
&=9\eta[6^{-4}12^{5}18^{8}24^{-2}36^{-5}48^{1}144^{1}]+9\eta[3^{-1}6^{4}9^{1}12^{-3}18^{-3}24^{-1}36^{5}48^{1}72^{2}144^{-1}]\\
&\qquad+6\eta[1^{-1}4^{1}6^{7}9^{1}12^{-5}18^{-5}36^{6}]+\frac{2}{3}\eta[2^{-4}4^{8}]-12\eta[4^{1}12^{-5}16^{-1}24^{7}36^{6}72^{-5}144^{1}]\\
&\qquad+18\eta[18^{-4}36^{8}]+\frac{5}{6}\eta[4^{8}8^{-4}]-\eta[1^{1}2^{-3}3^{-1}4^{5}6^{4}8^{2}12^{-3}16^{-1}24^{-1}48^{1}]+\frac{15}{2}\eta[36^{8}72^{-4}]\\
&\qquad+\frac{1}{12}\eta[1^{8}2^{-4}]+\frac{9}{2}\eta[9^{8}18^{-4}]+\frac{32}{3}\eta[8^{-4}16^{8}]-48\eta[72^{-4}144^{8}]-\frac{194}{3}\eta[6^{-4}12^{8}]\\
&\qquad-\frac{9}{2}\eta[12^{8}24^{-4}]-\frac{61}{12}\eta[3^{8}6^{-4}]-80\eta[24^{-4}48^{8}]-\frac{2}{3}\eta[2^{6}6^{-2}]-\frac{8}{3}\eta[8^{6}24^{-2}]\\
&\qquad-6\eta[1^{2}2^{-4}3^{-6}6^{12}]+\frac{10}{3}\eta[3^{-6}6^{12}9^{2}18^{-4}]-\frac{16}{3}\eta[4^{6}12^{-2}]-4\eta[3^{6}6^{-3}24^{-1}48^{2}]\\
&\qquad+2\eta[1^{1}3^{1}4^{-5}6^{-2}8^{8}12^{5}24^{-4}]+18\eta[3^{1}6^{-2}9^{1}12^{5}24^{-4}36^{-5}72^{8}]+\eta[2^{8}4^{-5}6^{-4}12^{5}16^{1}24^{-2}48^{1}]\\
&\qquad+4\eta[3^{2}6^{-1}8^{-1}16^{-1}24^{3}48^{4}72^{-1}144^{-1}]\\
f_{144,b}(q)&=q^5 - 2q^7 + 2q^{11} - 2q^{13} - q^{17} + 6q^{19} - 4q^{23} + 2q^{25}+\ldots\\
&=\frac{21}{2}\eta[6^{-4}12^{5}18^{8}24^{-2}36^{-5}48^{1}144^{1}]+\frac{3}{2}\eta[3^{-1}6^{4}9^{1}12^{-3}18^{-3}24^{-1}36^{5}48^{1}72^{2}144^{-1}]\\
&\qquad-6\eta[1^{-1}4^{1}6^{7}9^{1}12^{-5}18^{-5}36^{6}]-\frac{35}{6}\eta[2^{-4}4^{8}]+12\eta[4^{1}12^{-5}16^{-1}24^{7}36^{6}72^{-5}144^{1}]\\
&\qquad-\frac{191}{2}\eta[18^{-4}36^{8}]+\frac{1}{9}\eta[4^{8}8^{-4}]-\frac{1}{6}\eta[1^{1}2^{-3}3^{-1}4^{5}6^{4}8^{2}12^{-3}16^{-1}24^{-1}48^{1}]+\frac{23}{4}\eta[36^{8}72^{-4}]\\
&\qquad-\frac{11}{16}\eta[1^{8}2^{-4}]-\frac{197}{16}\eta[9^{8}18^{-4}]-\frac{28}{9}\eta[8^{-4}16^{8}]+2\eta[1^{1}2^{-1}3^{1}8^{2}12^{-1}18^{-1}36^{1}72^{2}]\\
&\qquad+172\eta[72^{-4}144^{8}]-\eta[2^{2}8^{-1}12^{-1}16^{1}18^{2}36^{1}48^{1}72^{-1}]-\frac{1}{18}\eta[1^{6}3^{-2}]+\frac{403}{3}\eta[6^{-4}12^{8}]\\
&\qquad-\frac{8}{9}\eta[16^{6}48^{-2}]+\frac{13}{3}\eta[12^{8}24^{-4}]+\frac{71}{12}\eta[3^{8}6^{-4}]+\frac{512}{3}\eta[24^{-4}48^{8}]+\frac{13}{18}\eta[2^{6}6^{-2}]\\
&\qquad+\frac{26}{9}\eta[8^{6}24^{-2}]+16\eta[1^{2}2^{-4}3^{-6}6^{12}]-8\eta[3^{-6}6^{12}9^{2}18^{-4}]+\frac{59}{9}\eta[4^{6}12^{-2}]-\frac{1}{6}\eta[1^{6}2^{-3}8^{-1}16^{2}]\\
&\qquad-\frac{3}{2}\eta[9^{6}18^{-3}72^{-1}144^{2}]-\eta[3^{6}6^{-3}24^{-1}48^{2}]-\frac{7}{3}\eta[1^{1}3^{1}4^{-5}6^{-2}8^{8}12^{5}24^{-4}]\\
&\qquad-27\eta[3^{1}6^{-2}9^{1}12^{5}24^{-4}36^{-5}72^{8}]+\frac{5}{6}\eta[2^{8}4^{-5}6^{-4}12^{5}16^{1}24^{-2}48^{1}]\\
&\qquad-2\eta[3^{2}6^{-1}8^{-1}16^{-1}24^{3}48^{4}72^{-1}144^{-1}]-2\eta[2^{1}3^{-1}6^{4}9^{1}12^{-3}16^{1}18^{-4}24^{-1}36^{5}48^{1}72^{2}144^{-2}]\\
&\qquad-2\eta[1^{1}3^{1}6^{-1}8^{1}9^{-2}12^{-3}18^{2}24^{4}36^{5}48^{-1}72^{-4}144^{1}]
\end{align*}

\subsection{Fourier expanding our basis at various cusps}\label{sec:modularCusp}

In this section, we explain how to expand the modular forms defined in appendix \ref{sec:basis} about various cusps. In the main text, we have used physical arguments to determine the behavior of modular forms at arbitrary cusps. In order to use this data to expand these modular forms in terms of the basis described in appendix \ref{sec:basis}, we need to know how to expand the elements of this basis that are not cusp forms about all cusps. In addition, the constraints from wall crossing require us to be able to expand our cusp forms about $\tau=0$.

\subsubsection{Eisenstein series}

Our strategy for determining the values of Eisenstein series at arbitrary cusps will be to determine the transformation properties of these forms under $SL(2,\ZZ)$ transformations, which allow us to map any cusp to infinity, where we know the function's $q$-expansion. The Eisenstein series transform trivially under the $T$ operation that maps $\tau$ to $\tau+1$, as is obvious from $q=e^{2\pi i\tau}$. Therefore, we only need to determine the series' behavior under $S:\tau\mapsto -1/\tau$.\footnote{Readers may safely skip to the conclusion of this argument, equation \eqref{eq:etwo}.} (Actually, while this reasoning does end up working for the functions $\E_t$, we will find that knowing the $S$ and $T$ transformations of the other Eisenstein series is not sufficient to determine their general $SL(2,\ZZ)$ transformations. Hence, in the main text we use holomorphic eta products instead of the Eisenstein series other than $\E_t$. Nevertheless, since it requires little extra work and illustrates why we are modifying our basis -- and because the result for non-principal characters is, to the extent of our knowledge, unpublished -- we will determine the $S$ transformation of all of the Eisenstein series $E_{2,\chi,\psi}$). We specify the value of a character $\chi$ at $-1$ via the notation $\chi(-1)=(-1)^{a_\chi}$, where $a_\chi\in\{0,1\}$; recall that for each Eisenstein series $E_{2,\chi,\psi}$ we have $a_\chi=a_\psi$.

Our method, due to Hecke, is described in the proof of Theorem 4.3.5 in \cite{miyakemodular}. Define coefficients $c_m(\chi,\psi)$ as follows:
$$E_{2,\chi,\psi}=\sum_{m\ge 0}c_m(\chi,\psi)q^m.$$
We have $c_m(\varepsilon_0,\varepsilon_0)=\sigma_1(m)$ for all $m\ge 1$, where $\sigma_1(m)=\sum_{d|m}d=O(m^2).$ Since (for $m\ge 1$) $|c_m(\chi,\psi)|\le c_m(\varepsilon_0,\varepsilon_0)$ for all $\chi,\psi$, this shows that $c_m(\chi,\psi)=O(m^2)$ in this case as well. Using these coefficients, we introduce two new functions:
$$g(s;\chi,\psi)=\frac{\Gamma(s)}{(2\pi)^s}\sum_{m\ge 1}\frac{c_m(\chi,\psi)}{m^s},$$
and
\begin{equation}\label{eq:h}
h(y;\chi,\psi)=\sum_{n\ge 1}c_n(\chi,\psi) e^{-2\pi n y}=E_{2,\chi,\psi}(iy)-c_0(\chi,\psi).
\end{equation}
The former converges absolutely for $\Real(s)>3$, while the latter converges for all positive real $y$. The integral representation of the $\Gamma$ function lets us relate these (when $\Real(s)>3$):
$$g(s;\chi,\psi)=\int_0^\infty dt\, h(t;\chi,\psi)t^{s-1}.$$
Substituting $t=e^x$ shows that $g(c+2\pi i z;\chi,\psi)$ is the Fourier transform of $h(e^x;\chi,\psi)e^{cx}$ when $c>3$. An inverse Fourier transform then yields
$$h(y;\chi,\psi)=\frac{1}{2\pi i}\int_{c-i\infty}^{c+i\infty}ds\, y^{-s}g(s;\chi,\psi)\quad (c>3,y>0).$$
To make use of this equation, we find another expression for $g$. Note that
$$\sum_{m\ge 1}\frac{c_m(\chi,\psi)}{m^s}=\sum_{m\ge 1,n|m}\frac{\psi(n)\chi(m/n)n}{m^s}=\sum_{r,n\ge 1}\frac{\psi(n)}{n^{s-1}}\cdot\frac{\chi(r)}{r^s}=L(s-1,\psi)L(s,\chi).$$
The last equation introduced the $L$-function $L(s,\chi)$ associated to a Dirichlet character $\chi$, which may be analytically continued to an entire function, unless $\chi$ is principal, in which case the $L$-function is the Riemann zeta function, which has a single pole at $s=1$. $L(s,\chi)$ has a simple zero at all negative even/odd integers if $a_\chi$ is even/odd, so $g(s;\chi,\psi)=\frac{\Gamma(s)}{(2\pi)^s}L(s,\chi)L(s-1,\psi)$ has no poles at the negative integers, even though such poles are present in $\Gamma(s)$. These facts allow us to determine the residues we pick up as we move the integration contour:
\begin{align*}
h(y;\chi,\psi)&=\frac{1}{2\pi i}\int_{4-i\infty}^{4+i\infty}ds\, y^{-s}g(s;\chi,\psi)\\
&=\text{Residues at a subset of $s=0,1,2$} + \frac{1}{2\pi i}\int_{-2-i\infty}^{-2+i\infty}ds\,  y^{-s}g(s;\chi,\psi)\\
&=\text{Residues}+\frac{1}{2\pi i}\int_{4-i\infty}^{4+i\infty}ds' y^{-2+s'}g(2-s',\chi,\psi).
\end{align*}
We now reap another benefit of our having expressed $g$ in terms of $L$-functions: we may take advantage of the functional equation (valid when $\chi$ is primitive)
$$\Lambda(s,\chi)=\parens{\frac{k_\chi}{\pi}}^{s/2}\Gamma\parens{\frac{s+a_\chi}{2}}L(s,\chi)\Rightarrow \Lambda(1-s,\bar\chi)=\frac{i^{a_\chi}k_\chi^{1/2}}{\tau(\chi)}\Lambda(s,\chi).$$
In this equation, $\tau(\chi)=\sum_{m=1}^k \chi(m)e^{2\pi i m/k_\chi}$ is the Gauss sum associated to $\chi$, which satisfies $k_\chi=|\tau(\chi)|^2$.
This identity relates $g(2-s,\chi,\psi)$ to $g(s;\bar\psi,\bar\chi)$, yielding
\begin{align*}
h(y;\chi,\psi)&=\text{Residues}-\frac{1}{2\pi i y^2 k_\chi}\sqrt{\frac{\tau(\chi)\tau(\psi)}{\tau(\bar\chi)\tau(\bar\psi)k_\chi k_\psi}}\int_{4-i\infty}^{4+i\infty}ds'\, (yk_\chi k_\psi)^{s'}g(s';\bar\psi,\bar\chi)\\
&=\text{Residues}-\frac{h(1/(y k_\chi k_\psi);\bar\psi,\bar\chi)}{y^2k_\chi}\sqrt{\frac{\tau(\chi)\tau(\psi)}{\tau(\bar\chi)\tau(\bar\psi)k_\chi k_\psi}}.
\end{align*}
There are four different cases that we need to analyze, as they have different residues: $\chi=\psi=\varepsilon_0$, $\chi=\varepsilon_0\not=\psi$, $\chi\not=\varepsilon_0=\psi$, and $\chi\not=\varepsilon_0$ and $\psi\not=\varepsilon_0$. Plugging in these residues and using \eqref{eq:h} to replace $h$ with an Eisenstein series yields the following transformation rules:
\begin{align*}
E_2(iy)&=-\frac{1}{4\pi y}-\frac{E_2(i/y)}{y^2},\\
E_{2,\chi,\psi}(iy)&=-\frac{E_{2,\bar\psi,\chi}(i/(yk_\chi k_\psi))}{y^2 k_\chi}\sqrt{\frac{\tau(\chi)\tau(\psi)}{\tau(\bar\chi)\tau(\bar\psi) k_\chi k_\psi}}\quad (\chi\not=\varepsilon_0\mbox{ or }\psi\not=\varepsilon_0).
\end{align*}
Since both sides of these equations are holomorphic functions of $\tau=iy$ on the upper half plane (away from cusps), these equations may be extended from the positive imaginary axis to the whole upper half plane:
\begin{align}
E_2(\tau)&=\frac{1}{4\pi i\tau}+\frac{E_2(-1/\tau)}{\tau^2}\label{eq:bothprincipal}\\
E_{2,\chi,\psi}(\tau)&=\frac{E_{2,\bar\psi,\chi}(-1/(\tau k_\chi k_\psi))}{\tau^2 k_\chi}\sqrt{\frac{\tau(\chi)\tau(\psi)}{\tau(\bar\chi)\tau(\bar\psi) k_\chi k_\psi}}\quad (\chi\not=\varepsilon_0\mbox{ or }\psi\not=\varepsilon_0).\label{eq:notbothprincipal}
\end{align}

With the $S$ and $T$ transformations of $E_2$ in hand, a simple inductive argument proves that
\begin{equation}\label{eq:etwo}
E_2(\gamma\tau)=(c\tau+d)^2 E_2(\tau)-\frac{c}{4\pi i}(c\tau+d),
\end{equation}
for any $\gamma=\twoMatrix{a}{b}{c}{d}\in SL(2,\ZZ)$. (We assume that this result holds for some $\gamma$ and then prove that it holds for $S\gamma,T\gamma,S^{-1}\gamma$, and $T^{-1}\gamma$. Since \eqref{eq:etwo} obviously holds for the base case where $\gamma$ is the identity matrix, it follows that it holds for an arbitrary $\gamma\in SL(2,\ZZ)$. Note that $S=S^{-1}$ within the group $PSL(2,\ZZ)$ that acts on $\tau$, so we do not need to do extra work to determine the $S^{-1}$ transformation of $E_2$). Note that this reasoning does not work in the case $k_\chi k_\psi\not=1$ -- the extra $1/k_\chi k_\psi$ in the argument of the Eisenstein series is problematic.\footnote{There is another method, called Hecke's trick, that is more commonly employed to determine the $SL(2,\ZZ)$ transformation of $E_2$. This method involves relating $E_2$ to the analytic continuation to $s=0$ of a non-holomorphic function that almost transforms under $SL(2,\ZZ)$ as a modular form of weight 2. Unfortunately, this method also does not seem well-suited to more general characters, since the functions that we analytically continue in these cases transform nicely only under a smaller group, $\Gamma_0(k_\chi,k_\psi)\subset SL(2,\ZZ).$} 

We now use \eqref{eq:etwo} to determine the behavior of the functions $\E_t$ near an arbitrary cusp, $\cc\in\QQ\PP^1$. Our reasoning is similar to that used in the proof of Proposition 2.1 in \cite{kohlermodular}. If $\cc=i\infty$, then we already know the answer: \eqref{eq:eisensteinInfty}. In particular, $\E_t(i\infty)=1-t$. Now, we assume $\cc\in\QQ$. Write $\cc=m/n$, where $m,n\in\ZZ$, $n>0$, and $\gcd(m,n)=1$, so that there exist $r,s\in\ZZ$ such that $sn-rm=1$. Then, $\gamma=\twoMatrix{r}{-s}{n}{-m}$ maps $\cc$ to $\infty$. Define $\cc'=t\cc=m'/n'$ with $m',n'\in\ZZ$, $n'>0$, and $\gcd(m',n')=1$, and find $r',s'\in\ZZ$ such that $s'n'-r'm'=1$. Define $\tau'=t\tau$ and $\gamma'=\twoMatrix{r'}{-s'}{n'}{-m'}$. We then have:
\begin{align}
E_2(\gamma\tau)&=(n\tau-m)^2 E_2(\tau)-\frac{n}{4\pi i}(n\tau-m)\nonumber\\
E_2(\gamma'\tau')&=(n'\tau'-m')^2 E_2(\tau')-\frac{n'}{4\pi i}(n'\tau'-m'\nonumber)\\
&=(n'^2t/n^2)\left[(n\tau-m)^2tE_2(t\tau)-\frac{n}{4\pi i}(n\tau-m)\right]\nonumber\\\
E_2(\gamma\tau)-\frac{n^2}{n'^2t}E_2(\gamma'\tau')&=(n\tau-m)^2\left[E_2(\tau)-tE_2(t\tau)\right]=-\frac{1}{24}(n\tau-m)^2 \E_t(\tau)\nonumber\\
\E_t(\tau)&=\frac{-24}{(n\tau-m)^2}\brackets{E_2(\gamma\tau)-\frac{n^2}{n'^2t}E_2(\gamma'\tau')}\label{eq:bigetrans}.
\end{align}
To get the $q$-expansion about $\cc$, we multiply by $(n\tau-m)^2$. The constant term can be read off easily, since $E_2(i\infty)=-1/24$: $a_0(\cc;t)=1-\frac{n^2}{n'^2t}$. That is,
$$a_0(\cc;t)=1-\frac{g^2}{t},$$
where we have defined
$$g=\gcd(t,\mbox{denominator($\cc$)})=\frac{n}{n'}.$$
Higher-order terms in the $q$-series are only a bit harder to obtain. The case $\cc=0$ enjoys a nice simplification, as $\gamma=\gamma'=\twoMatrix{0}{-1}{1}{0}$ and $\gamma(t\tau)=(\gamma\tau)/t$. Recalling that the expansion parameter $q_0$ about the cusp $\cc=0$ is $q_0=e^{2\pi i \tau_0/w_0}$, where $w_0$ is the width of the cusp $\cc=0$ and $\tau_0=\gamma\tau$, we find
$$\E_{t;0}(\tau_0)=-24\brackets{E_2(q_0^{w_0})-\frac{1}{t}E_2(q_0^{w_0/t})};$$
the 0 subscript labels the cusp about which we are expanding, as in \eqref{eq:Fc}:
\be \E_t(\tau)d\tau=\E_{t;\cc}(\tau_\cc)d\tau_\cc \ .\ee
More generally, define
$$\alpha=\gamma'\twoMatrix{t}{0}{0}{1}\gamma^{-1}=\twoMatrix{\mu}{\nu}{\rho}{\sigma}.$$
This maps $\gamma\tau$ to $\gamma'\tau'$; in particular, it fixes $i\infty$. Therefore, $\rho=0$. Multiplying these matrices out, we also find that $\mu=g$. Using $\mu\sigma=\det\alpha=t\not=0$, we find that $\mu/\sigma=\mu^2/t$ and $\nu/\sigma=\mu\nu/t$, so that
\begin{equation}\label{eq:generalEisenstein}
\E_{t;\cc}(\tau_\cc)=-24\brackets{E_2(q_\cc^{w_\cc})-\frac{g^2}{t}E_2(e^{2\pi i g\nu/t}q_\cc^{g^2 w_\cc/t})}.
\end{equation}



\subsubsection{Eta products}

In order to determine the $q$-expansion of an eta product at an arbitrary cusp, we employ a technique similar to the one we employed in deriving \eqref{eq:generalEisenstein}. 
Fix some positive integer $t$. Definine $m,n,r,s,\gamma,$ and their primed counterparts as above \eqref{eq:bigetrans}. Also, as above, define $\tau'=t\tau$ and $g=\gcd(t,\mbox{denominator($\cc$)})=n/n'$. We then have
\begin{align}
\eta(\gamma'\tau')&=v_\eta(\gamma')(n'\tau'-m')^{1/2}\eta(\tau')=v_\eta(\gamma')\parens{\frac{n't}{n}}^{1/2}(n\tau-m)^{1/2}\eta(t\tau)\nonumber\\
&=v_\eta(\gamma')\parens{\frac{n't}{n}}^{1/2}(n\tau-m)^{1/2}\eta(t\tau)\\
\Rightarrow \eta(t\tau)&=v_\eta(\gamma')^{-1}\parens{\frac{n}{n't}}^{1/2}(n\tau-m)^{-1/2}\eta(\gamma'\tau')\label{eq:etaTrans}.
\end{align}
As in the derivation of \eqref{eq:generalEisenstein}, we can now find $\nu\in\ZZ$ such that $\gamma'\tau'=(g^2/t)\gamma\tau + g\nu/t=(g^2/t)\tau_\cc+g\nu/t$. Via a slight abuse of notation (since $\eta[t^1](\tau)$ is not a weight 2 modular form) we define
\begin{equation}\label{eq:generalEta}
\eta[t^1]_\cc(\tau_\cc)=v_\eta(\gamma')^{-1}\parens{\frac{g}{t}}^{1/2}\eta\parens{\frac{g\nu}{t}+\frac{g^2 \tau_\cc}{t}}.
\end{equation}
The $q$-expansion of the eta product $\eta[\prod_t t^{m_t}]$ about the cusp $\cc$ is then obtained by raising the functions \eqref{eq:generalEta} to the powers $m_t$ and multiplying them together; note that there will be a $\gamma'$, $g$, and $\nu$ for each $t$. 
Since we are interested in holomorphic eta products, the $q_\cc^0$ coefficient in such a $q$-expansion comes from the leading terms in each of the functions \eqref{eq:generalEta}:
\begin{equation}\label{eq:etaZero}
a_0(\cc;\{m_t\})=\piecewise{\prod_{t>0}\brackets{v_\eta(\gamma'_t)^{-1}\parens{\frac{g_t}{t}}^{1/2}e^{\pi i g_t\nu_t/12t}}^{m_t}}{\sum_t \frac{g_t^2 m_t}{t}=0}{0}{\mbox{else}}.
\end{equation}
Here, $\gamma'_t$ is the $\gamma'$ matrix corresponding to $t$ -- that is, $\gamma'_t$ maps $t\cc$ to $i\infty$. We also denote the $g$ and $\nu$ values corresponding to a given $t$ by $g_t$ and $\nu_t$, respectively.

\subsubsection{Arbitrary cusp forms}

We now explain how to determine the expansions of arbitrary cusp forms for $\Gamma_0(N)$ about 0. (In most cases of interest to us, this is easily done using the techniques of the previous subsection, since we can write most of our cusp forms in terms of eta products. However, we present a method that works for all of our cusp forms). The tool that enables this is the Fricke involution $W=\twoMatrix{0}{-1}{N}{0},$ which has three nice properties: it maps $\infty$ to 0 (when acting on $\hat\HH$), it linearly maps the cuspidal subspace of $\Gamma_0(N)$ into itself (when acting on $\F$), and it squares to $N\gamma$ for some $\gamma\in\Gamma_0(N)$. In terms of its action on cusp forms for $\Gamma_0(N)$, this last property means that $W^2=1$. We may therefore decompose an arbitrary cusp form $f$ into a sum of cusp forms $f_+ + f_-$, where the $W$ eigenforms $f_{\pm}$ reside in the eigenspaces with eigenvalues $\pm 1$. (Software packages such as MAGMA enable one to determine bases of Fricke eigenforms. In particular, we note that a holomorphic eta product $\eta[\prod_t t^{m_t}]$ where $t|N$ whenever $m_t\not=0$ is a Fricke eigenform iff $m_{N/t}=m_t$ for all $t|N$, and in this case the Fricke eigenvalue is always $-1$\cite{kohlermodular}). More explicitly, we have
$$ f_{\pm}(\tau)=\pm\frac{N}{(N\tau)^2}f_{\pm}(-1/N\tau)=\pm\frac{1}{N\tau^2}f_{\pm}(\tau_0/N) \ ,$$
where $\tau_0=-1/\tau$ approaches $\infty$ as $\tau\to 0$. As usual, in order to determine the $q$-expansion of these forms about 0, we strip off the factor of $1/\tau^2$:
\be f_{\pm;0}(\tau_0)=\pm\frac{1}{N}f_{\pm}(\tau_0/N)=\pm\frac{1}{N}f_{\pm}(q_0^{w_0/N})=\pm\frac{1}{N}f_{\pm}(q_0) \ ,\ee
where the last equality follows from the fact that we always have $w_0=N$.

We can similarly expand about cusps of the form $e/N$ where $e|N$ and $\gcd(e,N/e)=1$ by replacing Fricke involutions in this argument with Atkin-Lehner involutions $W_e=\twoMatrix{e}{b}{N}{de}$ with integers $b,d$ such that $de-b\frac{N}{e}=1$, but in general this reasoning does not allow us to expand about arbitrary cusps. (Since 1 is always $\Gamma_0(N)$ equivalent to 0, the argument with the Fricke involution is really the case $e=N$). More explicitly, combining the logic of the previous paragraph with that of the previous sections, we define the $SL(2,\ZZ)$ matrix $\gamma_e=\twoMatrix{de}{-b}{-N/e}{1}$ that maps $e/N$ to $\infty$ and the $GL^+(2,\RR)$ matrix $\alpha=W_e^{-1}\gamma_e^{-1}=\twoMatrix{1/e}{0}{0}{1}$ that fixes $\infty$ in order to obtain
\be f_{e,\pm;e/N}(\tau_{e/N})=\pm\frac{1}{e}f_{e,\pm}(\alpha\tau_{e/N})=\pm\frac{1}{e}f_{e,\pm}(\tau_{e/N}/e)=\pm\frac{1}{e}f_{e,\pm}(q_{e/N}^{w_{e/N}/e})=\pm\frac{1}{e}f_{e,\pm}(q_{e/N})\ ,\ee
where $f_{e,\pm}$ are eigenforms for $W_e$, which satisfy
$$f_{e,\pm}(\tau)=\pm e(N\tau+de)^{-2} f_{e,\pm}(W_e\tau)\ ,$$
and $\tau_{e/N}=\gamma_e\tau$.

As a useful aside, we note that although the functions $\E_t$ are not cusp forms, they are nonetheless eigenforms with eigenvalue $-1$ for the Fricke operator defined with $N=t$ \cite{kohlermodular}. Thus, the same arguments as above imply that
\be \E_{t;0}(\tau_0)=-\frac{1}{t}\E_t(q_0^{w_0/t})\ .\ee
(We leave $w_0$ arbitrary, as it depends on the $SL(2,\ZZ)$ subgroup for which we are viewing $\E_t$ as being a modular form).


\section{Charges in the case $\lambda>1$}\label{a:mess}

In this section, we describe the quantization of the electric-magnetic charges of a CHL model associated to a symmetry $(\delta,g)$, where $g$ is a symmetry of an NLSM on K3 that does not satisfy the level-matching condition (see section \ref{sec:levelfail} and \cite{PV}). Furthermore, we derive, for this class of models, the possible channels of decay of a 1/4 BPS dyon into a pair of 1/2 BPS states and compare the expected domain walls with the poles of the corresponding function $1/\Phi_{g,e}$.

The lattice of electric-magnetic charges $(\hat m\ m'\ \hat w\ w' | \hat W\ W'\ \hat M\ M')^t$ was derived in \cite{PV} and is given by the $\ZZ$-span of the following $8$ vectors 
\be 
\begin{matrix}
  1 \\
 0\\
  0\\
 0\\
 \hline
   0 \\
 0\\
 0\\
 0
\end{matrix}\qquad
\begin{matrix}
  0\\
  1/N \\
 0\\
 0\\
 \hline
   0 \\
 0\\
 0\\
 0
\end{matrix}\qquad
\begin{matrix}
  0 \\
 0\\
  1\\
 0\\
 \hline
   0 \\
 0\\
 0\\
 0
\end{matrix}\qquad
\begin{matrix}
  0\\
  \E/N\lambda \\
 0\\
 1\\
 \hline
   0 \\
 0\\
 0\\
 0
\end{matrix}\qquad
\begin{matrix}
  0\\
  -\E'/N\lambda \\
 0\\
 0\\
 \hline
   1 \\
 0\\
 0\\
 0
\end{matrix}\qquad
\begin{matrix}
  0\\
  0 \\
 0\\
 0\\
 \hline
   0 \\
 1\\
 0\\
 0
\end{matrix}\qquad
\begin{matrix}
  0\\
  \E\E'/N\lambda \\
 0\\
 0\\
 \hline
   0 \\
 0\\
 1\\
 0
\end{matrix}\qquad
\begin{matrix}
  \E\E'/\lambda\\
  y \\
 -\E'/\lambda\\
 0\\
 \hline
   0 \\
 -\E/\lambda\\
 0\\
N
\end{matrix}
 \ee
(we adapted the results of \cite{PV} to the conventions of this paper). Here, $\E'$ and $\lambda$ determine the failure of the level matching condition for the $g$-twisted sector of the K3 NLSM (see eq.\eqref{levelfail}), while $\E$ plays the same role for the level matching condition of the $g$-twisted sector in the heterotic frame
\be (L_0-\bar L_0)_{\rvert \Hh_g}\in \frac{\E}{N\lambda}+\frac{1}{N}\ZZ \qquad \qquad \text{(heterotic)}\ .
\ee As described in section \ref{sec:secquant}, the ground level of the heterotic $g$-twisted sector is given by the constant $-A/(24N\lambda)$. Observing that for all $g$ with $\lambda>1$ one has $A=24$, we find
\be \E\equiv-1\modu \lambda\qquad\qquad \text{for }\lambda>1\ .
\ee

For these CHL models, a D1-D5 system analogous to the one considered in section \ref{sec:BPScount} has charges
\be \begin{pmatrix}
 \hat m \\
 m'\\
 \hat w\\
 w'
\end{pmatrix}=\begin{pmatrix}
 0 \\
\frac{ -n\lambda +1 -m\E'+\E'}{N\lambda}\\
 0\\
 -1
\end{pmatrix}\ ,\qquad 
\begin{pmatrix}
 \hat W \\
 W'\\
 \hat M\\
 M'
\end{pmatrix}=\begin{pmatrix}
 m \\
 -l\\
 1\\
 0
\end{pmatrix}\ ,
\ee $l,m,n\in \ZZ$,
so that
\be Q^2=2\frac{n\lambda -1 -\E'(m-1)}{N\lambda}\ ,\qquad P^2=2m\ ,\qquad P\cdot Q=l\ .
\ee Comparing with the $\lambda=1$ case, the only difference is the complicated quantization of the momentum $m'$ along $S^1$. This quantization is necessary in order to get a non-zero multiplicity $d(Q^2/2,P^2/2,P\cdot Q)$, as follows from the condition that the exponents $\hat c^g_{m,n}(\frac{mn}{N\lambda},l)$ in the infinite product \eqref{Phige} vanish unless $n-\E'm\equiv 0\modu \lambda$. In practice, we can simply consider the set of charges
\be \begin{pmatrix}
 \hat m \\
 m'\\
 \hat w\\
 w'
\end{pmatrix}=\begin{pmatrix}
 0 \\
 -n'/N\lambda\\
 0\\
 -1
\end{pmatrix}\ ,\qquad 
\begin{pmatrix}
  \hat W \\
 W'\\
 \hat M\\
 M'
\end{pmatrix}=\begin{pmatrix}
 m \\
-l\\
 1\\
 0
\end{pmatrix}\ee for all $l,m,n'\in \ZZ$;
the corresponding multiplicity will be zero unless $n'$ is of the form $n\lambda -1 -\E'(m-1)$ for some $n\in \ZZ$.

\medskip

Let us now determine the possible decay channels of 1/4 BPS dyon of charges $\column{Q}{P}$ into a pair of 1/2 BPS states of charges $\column{Q_1}{P_1}$ and $\column{Q_2}{P_2}$. Following the same reasoning as in section \ref{sec:chl}, we obtain
\be  \column{Q_1}{P_1}=\begin{pmatrix}
adQ+dbP\\ \hline -caQ-cbP
\end{pmatrix}=\begin{pmatrix}
 dbm \\
 -adn'/N\lambda-dbl\\
 db\\
 -ad \\ \hline
 -cbm\\
 can'/N\lambda+cbl\\
 -cb\\
 ac
\end{pmatrix}
\ee
\be \column{Q_2}{P_2}=\begin{pmatrix}
-bcQ-bdP\\ \hline acQ+adP
\end{pmatrix}=\begin{pmatrix}
  -dbm \\
 bcn'/N\lambda+dbl\\
 -bd\\
 bc \\ \hline
  adm\\
 -acn'/N\lambda-adl\\
 ad\\
 -ac
\end{pmatrix}\ .
\ee
The condition that $\smcolumn{Q_1}{P_1}$ and  $\smcolumn{Q_2}{P_2}$ are contained in the lattice of electric-magnetic charges gives 
\be ad\in \ZZ\qquad bc\in \ZZ\qquad ac\in N\ZZ\qquad bd\in \frac{1}{\lambda	}\ZZ\ ,
\qquad \frac{ac \E'}{N\lambda}+bd\in \ZZ\ .
\ee
We can use the rescaling
\be \begin{pmatrix}
a & b\\ c & d
\end{pmatrix}\to \begin{pmatrix}
xa & xb\\ c/x & d/x
\end{pmatrix}\ee to make $a$ and $b$ integral and coprime  (or equal to $0$ and $\pm 1$, in case one of the two vanishes). This implies that also $c$ and $\lambda d$ are integral.

For each given decomposition labeled by $a,b,c,d$ as above, the location of the corresponding wall can be found as described in section \ref{sec:contour}.\footnote{For this derivation to hold in the case $\lambda>1$, it is crucial  that the covariance or invariance properties of the scalar product, the central charge vector and the BPS mass hold for the whole real group $SL(2,\RR)$ and not just for $SL(2,\ZZ)$.} As a final result, the walls of marginal stability are given by the equation $(\alpha,\Z)=0$, where $\alpha$ is still given by \eqref{eq:alphaMat} and $a,b,c,d$ satisfy
\be a,b,c\in \ZZ, \ d\in \frac{1}{\lambda}\ZZ \qquad ad-bc=1\qquad ac\in N\ZZ\qquad \frac{ac \E'}{N\lambda}+bd\in \ZZ\ .
\label{lambdawalls}\ee
These conditions are not what one would naively expect just from replacing $N$ by $N\lambda$ in the $\lambda=1$ case.

\bigskip

The zeroes and poles of $1/\Phi_{g,e}$ are located at
\be m\sigma+n\frac{\tau}{N\lambda}+lz=k 
\ee for $m,n,l,k\in \ZZ$ with $4\frac{mn}{N\lambda}-l^2<0$ and have multiplicity $\hat c^g_{m,n,l}(4\frac{mn}{N\lambda}-l^2)$. Noting that for $m\equiv 0\modu {N\lambda}$ the only pole is given by $\hat c_{0,0,1}(-1)=2$ and using the isomorphisms \eqref{isomodule}, we find that there is a special set of poles
\be Nr\sigma+\frac{s}{\lambda}\tau+l z=k
\ee where $r,s,l,k\in \ZZ$ with $4Nr\frac{s}{\lambda}-l^2=-1$ and $s\equiv -\E'r\modu \lambda$ with multiplicity $\hat c^g_{Nr,-Nr\E',1}(-1)=2$. These poles occur for all $g$ of order $N$ and multiplier $\lambda$ and correspond to walls of equation
\be (\Z,\twoMatrix{2\frac{s}{\lambda}}{-l}{-l}{2Nr})=0\qquad\qquad r,s,l\in \ZZ,\quad 4Nr\frac{s}{\lambda}-l^2=-1,\quad s\equiv -\E'r\modu \lambda\ .\label{lambdapoles}
\ee Let us show that the walls \eqref{lambdapoles} are in one to one correspondence with the expected physical walls. Given a `physical' wall labeled by $a,b,c,d$ satisfying \eqref{lambdawalls}, it is easy to see that one obtains a  wall of the form \eqref{lambdapoles} by setting
\be r:=\frac{ac}{N}\qquad l:=ad+bc\qquad s:=\lambda bd\ .
\ee 
Vice versa, consider a wall of the form \eqref{lambdapoles}. Let us first assume that $s\neq 0$, and set
\be t:=\frac{l-1}{2},\qquad b:=\gcd\left(\frac{s}{\gcd(\lambda,s)},t\right),\qquad d:=\frac{\gcd(\lambda,s)}{\lambda}\gcd\left(\frac{s}{\gcd(\lambda,s)},t+1\right),\ee
as well as
\be  a:=\frac{t+1}{d},\qquad c:=\frac{t}{b}.
\ee Then, it is clear that $a,b,c,\lambda d\in \ZZ$ and $ad-bc=1$. Furthermore,
\be bd= \frac{\gcd(\lambda,s)}{\lambda}\gcd\left(\frac{s}{\gcd(\lambda,s)},t(t+1)\right)=\frac{\gcd(\lambda,s)}{\lambda}\gcd\left(\frac{s}{\gcd(\lambda,s)},\frac{Nrs}{\lambda}\right)=\frac{s}{\lambda}\ ,
\ee where the first equality follows because $t$ and $t+1$ are coprime, the second from the relation $\frac{Nrs}{\lambda}=\frac{l^2-1}{4}$ and the last because $\frac{s}{\gcd(\lambda,s)}$ is a divisor of $\frac{Nrs}{\lambda}$. Finally,
\be ac=\frac{abcd}{bd}= \frac{t(t+1)}{s/\lambda}=Nr\ ,
\ee which shows that $a,b,c,d$ satisfy all congruences in \eqref{lambdawalls} and therefore label a `physical' wall equivalent to \eqref{lambdapoles}. 
When $s=0$, the relations \eqref{lambdapoles} imply $l=\pm 1$. For $l=1$, we set
\be a=d=1,\qquad b=0\qquad c=Nr, 
\ee while for $l=-1$ we set
\be a=Nr,\qquad  b=-1,\qquad c=1,\qquad d=0,
\ee and in both cases $a,b,c,d$ label a physical wall equivalent to \eqref{lambdapoles}.

\bibliographystyle{utphys}

\bibliography{Refs}

\end{document}